\documentclass[%
     reprint,
     superscriptaddress,
     amsmath,amssymb,amsfonts
     aps,
     pra,
     floatfix,
    ]{revtex4-1}

\usepackage{acronym}
\usepackage{float}
\usepackage{xfrac}
\usepackage{braket}
\usepackage{graphicx} \graphicspath{{PIC}}
\usepackage{verbatim}
\usepackage{etoolbox}
\usepackage[utf8]{inputenc}
\usepackage{mathtools}
\usepackage{soul,xcolor,colortbl}
\usepackage{hyperref} \hypersetup{colorlinks=true,citecolor=blue,linkcolor=blue,urlcolor=blue}
\usepackage{csquotes}
\usepackage{multirow}
\usepackage{cellspace}
\usepackage{resizegather}
\usepackage{dcolumn}
\usepackage[normalem]{ulem}
\usepackage{siunitx}
\usepackage{tabularx}
\usepackage{pdfpages}

\usepackage{bm}
\usepackage{array}
\newcolumntype{C}[1]{>{\centering\arraybackslash}p{#1}}\usepackage{soul}
\definecolor{Gray}{gray}{0.85}

\usepackage{color}
\usepackage{morefloats}

\definecolor{Gray}{gray}{0.9}
\definecolor{LightCyan}{rgb}{0.88,1,1}
\definecolor{green}{rgb}{0.5451,0.2706,0.0745}

\def\DFT{{\small  DFT}}

\def\PBE{{\small  PBE}}
\def\SCAN{{\small  SCAN}}

\def\VASP{{\small VASP}}

\makeatletter
\AtBeginDocument{\let\LS@rot\@undefined}
\makeatother

\begin{document}

\title{Non-van der Waals Heterostructures}

\author{Anastasiia Nihei}
\affiliation{Theoretical Chemistry, Technische Universit\"at Dresden, 01062 Dresden, Germany}
\affiliation{Institute of Ion Beam Physics and Materials Research, Helmholtz-Zentrum Dresden-Rossendorf, 01328 Dresden, Germany}
\author{Tom Barnowsky}
\affiliation{Theoretical Chemistry, Technische Universit\"at Dresden, 01062 Dresden, Germany}
\affiliation{Institute of Ion Beam Physics and Materials Research, Helmholtz-Zentrum Dresden-Rossendorf, 01328 Dresden, Germany}
\author{Rico Friedrich}
\email{rico.friedrich@tu-dresden.de}
\affiliation{Theoretical Chemistry, Technische Universit\"at Dresden, 01062 Dresden, Germany}
\affiliation{Institute of Ion Beam Physics and Materials Research, Helmholtz-Zentrum Dresden-Rossendorf, 01328 Dresden, Germany}

\date{\today}

\begin{abstract}

Beyond the study of individual materials, their interfaces and arising functionality are crucial enablers of fundamental science and technological {progress}.
Recently, the arena of two-dimensional (2D) materials was extended to so-called non-van der Waals (non-vdW) compounds derived from strongly bonded non-layered bulk crystals.
{These nanosystems have advantageous active surfaces and thus {now complement}} -- through their chemically bonded heterostructure (HS) interfaces -- the established space of weakly interacting vdW HSs.
Here, we study a wide range of 55 candidate non-vdW HSs with {extensive} density functional calculations and find that the resulting interfacial chemical hybridization and bonding is the key factor determining their electronic and magnetic properties.
It gives rise to the formation of {unique hybrid interface bands}, strong magnetic
coupling, and substantial electronic as well as magnetic moir{\'e} surface property modulations upon twisting.
Our work thus provides a {valuable platform for} charting the new interface class of non-vdW HSs.

\vspace{0.2cm}

\end{abstract}

\maketitle

\section*{Introduction} \label{intro}

Interfaces of materials are at the heart of technologies by creating decisive functionalities such as in electronics, catalysis, batteries, and data storage media -- or as Herbert Kroemer put it: ``the interface is the device" \cite{Kroemer_RMP_2001}.
In addition to lateral junctions \cite{Levendorf_Nature_2012,Mehlich_2DMat_2024}, stacked two-dimensional (2D) materials in the form of both homobilayers \cite{Pakdel_NCom_2024} and heterostructures (HSs) have emerged as a powerful platform for functional interface design.
They can host unique excitonic effects, special (tunneling) transport behavior as well as optical, plasmonic, magnetic, topological, and optoelectronic properties \cite{Geim_Nature_2013,Liu_NatRevMat_2016,Novoselov_Science_2016,Gibertini_NNano_2019,Pham_ChemRev_2022}.
Particular progress has been made {toward the} reliable and scalable manufacturing of HSs {with} atomically sharp interfaces {using both top-down methods as well as bottom-up} growth approaches \cite{Geim_Nature_2013,Wang_Nanoscale_2014,Kim_NanoLett_2016,Liu_NatRevMat_2016,Li_MatTod_2016,Castellanos-Gomez_NatRev_Meth_Prim_2022}.

The twist angle between stacked sheets has been pointed out as an important additional degree of freedom.
The interfacing can then bring about qualitatively new, and at first sight unexpected, features for which the observation of unconventional superconductivity in twisted bilayer graphene is a striking example \cite{Cao_Nature_2018}.
Beyond that, long-range moir{\'e} patterns in large twisted bilayers of MoS$_2$ can, for instance, initiate emerging features in the band structures resembling signatures of graphene or hexagonal boron nitride albeit at a much smaller dispersive energy scale \cite{Arnold_MatTod_2024}.

The zoo of traditional 2D systems derived from layered crystals bonded by weak van der Waals (vdW) forces \cite{Lebegue_prx_2013,Cheon_NanoLett_2017,Mounet_AiiDA2D_NNano_2018} provides a range of possibilities for the associated vdW HSs.
Nevertheless, the weak dispersive interactions between the layers in the stack pose also a {significant} restriction as the corresponding interlayer interaction energy is intrinsically low.

The recently discovered new class of non-van der Waals 2D materials \cite{Balan_MatTod_2022,Kaur_AdvMat_2022} derived from strongly bonded non-layered bulk crystals offers qualitatively new opportunities.
A decent amount of representatives has already been found by both top-down exfoliation as well as more recently also bottom-up growth approaches \cite{Puthirath_Balan_NNANO_2018,Puthirath_Balan_CoM_2018,Guan_AdvMat_2017,Yadav_AdvMatInt_2018,Liu_JMCA_2019,Puthirath_Small_2020,Kaur_ACSNano_2020,Puthirath_JPCC_2021,Moinuddin_AdvElMat_2021,Hu_two-dimensional_2021,Yousaf_JPCC_2021,Guo_Nanoscale_2021,Gibaja_AdvMat_2021,Xu_ChemSci_2021,Peng_NChem_2021,Homkar_ACSApplNanoMat_2021,Chahal_ACSML_2021,Zhang_NMAT_2021,Chen_FlatChem_2022,Toksumakov_NPJ2DM_2022,Puthirath_Balan_OxOpMatSci_2023,Jiang_NatSyn_2023,Zhao_Nature_2025}
-- supplemented by data-driven computational studies suggesting a few dozen additional candidates \cite{Friedrich_NanoLett_2022,Barnowsky_AdvElMats_2023,Bagheri_JPCL_2023}.

\begin{figure*}[ht!]
    \includegraphics[width=\textwidth]{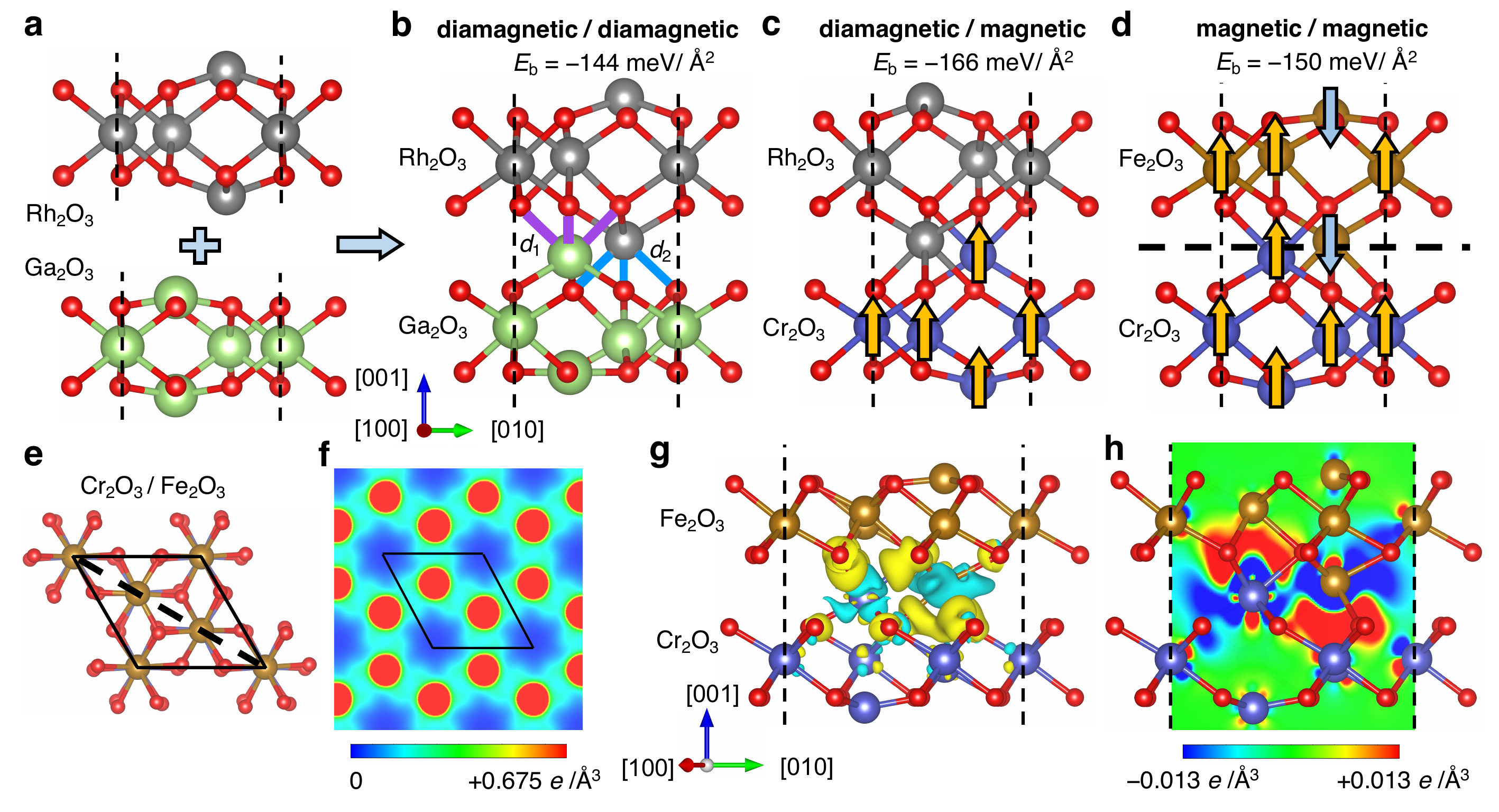}
    \caption{\small \textbf{Interfacial bonding in non-vdW HSs.} HSs are build from stacking and matching non-vdW 2D sheets (\textbf{a}) leading to interfacial bonding (highlighted in purple and light blue) in optimized structures for combinations of two diamagnetic (\textbf{b}), a diamagnetic and a magnetic (\textbf{c}), and two magnetic (\textbf{d}) systems.
    Distances $d_1$ and $d_2$ of interfacial bonds are marked in (\textbf{b}).
    Arrows indicate moments in magnetic layers.
    Binding energies $E_{\mathrm{b}}$ are also indicated.
    The compass indicating crystal directions refers to all top row subfigures.
    (\textbf{e}) Top view of the representative Cr$_2$O$_3$/Fe$_2$O$_3$ HS.
    (\textbf{f}) Top view of the charge density in an extended region on the plane indicated by the black dashed line in (\textbf{d}).
    The unit cell is given by the black frame.
    (\textbf{g}) Isosurface plot of the charge density difference for Cr$_2$O$_3$/Fe$_2$O$_3$ with charge gain/loss indicated by yellow/bluish parts (isosurface value: $\pm$0.027~$e$/\AA$^3$).
    (\textbf{h}) Cut through the charge density difference on the (110) plane indicated by the black dashed line in (\textbf{e}).
    The compass in (\textbf{g}) refers also to (\textbf{h}).
    Dashed vertical black lines represent unit cell boundaries.
    Colors: O, red; Ga, light green; Rh, gray; Cr, blue; Fe, gold. }
    \label{fig1}
\end{figure*}

The advent of these systems came as a surprise as it was intuitively thought that chemically bonded materials cannot be exfoliated.
Their active surfaces \cite{Barnowsky_NanoLett_2024,Jin_JPCL_2024}, resulting from dangling bonds and surface states generated during exfoliation, {now give} rise to behavior {that differs from that of} traditional 2D systems.
As such, they have already been demonstrated to show exceptional catalytic activity \cite{Puthirath_Balan_NNANO_2018} as well as substantial surface magnetism with almost perfect spin polarization \cite{Friedrich_NanoLett_2022}.
{These (re-)active surfaces of non-vdW 2D materials foreshadow a qualitatively new and unique behavior for their HS interfaces: interfacial chemical bonding between the 2D components, resulting from hybridization of electronic states, can give rise to strong chemical coupling inaccessible in vdW HSs.}

{Signatures of strong interfacial interactions are already known from experiments on grown thin film oxide stacks giving rise to emerging metallicity \cite{Chen_NanoLett_2011}, superconductivity \cite{Reyren_Science_2007}, and magnetism \cite{Brinkman_NMAT_2007} due to interfacial 2D electron gases \cite{Chen_NCom_2013,Chen_NatMat_2015,Cossu_NPJCM_2022,Erlandsen_NanoLett_2022,Bjorlig_ACSAELM_2022,Li_ACSNano_2022}.
Several recent reports of experimentally realized stacked and twisted oxides even outline that specific rotations between the films can be achieved \cite{Li_AdvMat_2022,Shen_ACSAMI_2022,Wang_AdvMat_2024,Schmidbauer_AplPhysLett_2025}.
Well defined chemical bonds were detected at the interface and even signatures of emerging moir{\'e} patterns were identified.
Studies employing twisted oxide bilayers have very recently demonstrated novel ferroelectric vortex states~\cite{Sanchez_Nature_2024,Sha_NatComm_2024}.
While these findings were obtained with films of several ten nanometer thickness, the interfacial interactions are still strongly confined to the 2D }

Here, we perform an exhaustive computational investigation of a large set of over 50 non-vdW HS candidates.
Based on density functional theory (\DFT) we demonstrate that their properties and potential active electronic and magnetic functionality is determined by the strong interfacial interactions.
Study of the optimized structures, densities, and interlayer binding energies clearly indicates substantial interfacial bonding for most systems steming from ionic and covalent contributions.
These strong interactions give rise to the formation of unique hybrid interface bands with sizable contributions from both 2D components.
{Furthermore, interfacial chemical bonding mediates} strong magnetic coupling that stabilizes specific magnetic configurations {and can enable multiferroic behavior}.
Eventually, also twisted non-vdW HSs can be generated exhibiting characteristic local atomic moir{\'e} patterns that lead to long-range signatures in the surface charge and magnetization densities.
Our large-scale screening showcases the rich variety of features to be expected from non-vdW HSs.
It thus uncovers the potential of this new interface class and offers the prospect of converging major concepts of 2D materials and oxide thin films research.

\section*{Methods} \label{methods}

\textbf{Hetbuilder:}
HSs are generated by a custom ``hetbuilder" implementation of the coincidence lattice method
following the work of Koda \emph{et al.} \cite{Koda_jpcc_2016,Kempt_hetbuilder}.
The initial structures for the input 2D sheets are sourced from the dataset of previously proposed non-van der Waals 2D materials \cite{Friedrich_NanoLett_2022,Barnowsky_AdvElMats_2023}.
The components are initially placed 2~\AA\ apart in the vertical direction perpendicular to the sheets and duplicate structure removal is based on structure comparison and energy evaluation due to force fields.
The mismatch criterion (tolerance) is set to 0.2~\AA\ and the maximum number of unit cell repetitions is limited to five ensuring that the HS cell contains less than 500 atoms.
The hetbuilder tool automatically calculates the nominal strain for the resulting HSs.

For a HS out of two 2D components, the strain is calculated as the sum of the relative deformations in both individual systems:

\[
\text{strain} = 2 \left( \left| 1 - \frac{a_{\text{HS}}}{a_{\text{1}}} \right| + \left|  1 - \frac{a_{\text{HS}}}{a_{\text{2}}} \right| \right),
\]

where \(a_{\text{HS}}\) is the lattice constant of the initial or relaxed HS and \(a_{\text{1/2}}\) is the lattice constant of the relaxed first/second 2D component.
The factor of two accounts for the two equivalent hexagonal directions.

\textbf{Density functional calculations:}
For all HSs, \DFT\ calculations including structural relaxations as well as evaluation of band structure and DOS are performed.
The computations for the exchange-correlation functional PBE+$U$ \cite{PBE,Anisimov_Mott_insulators_PRB1991,Liechtenstein1995,Dudarev1998} are executed with AFLOW \cite{Oses_CMS_2023,Esters_CMS_2023,Divilov_HighEntAlloyMat_2025} commanding the Vienna Ab-initio Simulation Package (\VASP) \cite{vasp_prb1996,kresse_vasp_paw} with settings according to the AFLOW standard \cite{curtarolo:art104},
and internal \VASP\ precision set to ACCURATE.
Electrostatic dipole corrections for energies, potentials, and forces are included \cite{Neugebauer_PRB_1992}, along with non-spherical contributions from gradient corrections inside projector-augmented wave (PAW) spheres.
For systems with an absolute binding energy smaller than 20 meV/\AA$^2$ from plain PBE+$U$, the D3 dispersion correction \cite{Grimme_JCP_2010} is applied.
The AFLOW internal automatic determination of $k$-point sets is used in conjunction with an extension for 2D systems enabling only in-plane sampling to be released in a future version.
{Convergence tests with respect to $k$-point density were performed for forces, lattice constants, total energy, and electronic structure (band gaps). Based on this, the number of $k$-points per reciprocal atom (2D KPPRA) is set to 2500 (12×12×1) for static calculations which is reduced to 400 (5×5×1) for structural relaxations. The chosen grids ensure convergence of the lattice constants within 0.1 m\AA, total energies within 0.1 meV/cell,  and band gaps within 5 meV (see Section~I. in the SI). For checking the forces, we note that for 80\% of the systems, the residual forces on all atoms were below 0.04 eV/\AA\ in the static calculation; for the remaining systems, the maximum force on any atom did not exceed 0.075 eV/\AA.}

The charge density difference is computed as:

\begin{equation}
    n_{\mathrm{diff}}(\mathbf{r})= n_{\mathrm{HS}}(\mathbf{r}) - n_{\mathrm{1}}^{\mathrm{static}}(\mathbf{r}) - n_{\mathrm{2}}^{\mathrm{static}}(\mathbf{r}),
\end{equation}

where $n_{\mathrm{HS}}(\mathbf{r})$ is the charge density of the relaxed HS and $n_{\mathrm{1/2}}^{\mathrm{static}}(\mathbf{r})$ is the density of 2D component 1/2 with the same atomic positions as in the HS obtained from a single self-consistent electronic calculation.

The binding energy $E_\mathrm{b}$ per unit area is calculated as:
\begin{equation}
    E_\mathrm{b} = \frac{E_\mathrm{HS} - E_1 - E_2}{A},
\end{equation}
where $E_\mathrm{HS}$ is the total energy of the HS, $E_1$/$E_2$ is the energy of the individual 2D component 1/2, and $A$ is the area of the relaxed HS.
The graphene homobilayer binding energy is computed with both D3 dispersion correction and \SCAN$+$rVV10 vdW functional \cite{Perdew_SCAN_PRL_2015,Peng_PRX_2016} to $-$16~meV/\AA$^2$.

{For the shifted cells, constrained calculations were performed.
Based on the initial commensurate structure obtained from the hetbuilder tool, the shifted cells were generated by displacing the lower 2D component.
The out-of-plane coordinates were allowed to relax while the in-plane positions were kept fixed as otherwise the structure would relax back to the unshifted geometry.
This systematic shifting created a homogeneous grid of configurations over the unit cell, enabling the evaluation of the total energy for each point on the grid.
}

{All HSE06~\cite{HSE06} (static and band structure) calculations were carried out on the geometries relaxed at the PBE+$U$ level, with $k$-point grids identical to the relaxation step and all other computational parameters kept consistent with the respective PBE+$U$ calculations.}

{The dynamical stability of the promising candidates for experimental realization was evaluated via phonon calculations using the machine-learning interatomic potential (MLIP) mace-mh-1 \cite{batatia2025crosslearning}. Prior to the phonon analysis, all structures were relaxed using the MLIP. Phonon spectra were computed using the finite displacement method as implemented in the Atomic Simulation Environment (ASE) \cite{Larsen_JPhysCM_2017}. To validate the MLIP-relaxed structures, we compared them with DFT-relaxed geometries using AFLOW-XtalFinder \cite{curtarolo:art170}, obtaining a structural misfit parameter $\varepsilon$ below 0.2. It indicates good agreement between the two structures and suggests a consistent description of the local potential energy surface relevant for phonon calculations.}

All systems are rigorously checked for magnetism.
For systems formed from diamagnetic components, both spin-polarized and non spin-polarized calculations are conducted.
However, none of the systems developed a magnetic ground state significantly lower in energy than the diamagnetic one.
In HSs involving one magnetic 2D component, the magnetic configuration is initialized as known from our previous studies \cite{Friedrich_NanoLett_2022,Barnowsky_AdvElMats_2023} of the individual sheets and still trying for five different sizes of all magnetic moments from 1 to 5 $\mu_{\mathrm{B}}$ in the spirit of the procedure introduced in Ref.~\cite{Friedrich_CCE_2019}.
For systems with two magnetic components, the magnetic configurations are guided by the individual cases and {are initialised} in two different ways, considering both parallel (ferromagnetic) and antiparallel (antiferromagnetic) spin alignment at the interface.
The resulting coupling energy $E_\mathrm{c}$ is calculated as the difference between the corresponding anitferromagnetic ($E_\mathrm{AFM}$) and ferromagnetic ($E_\mathrm{FM}$) states:
\begin{equation}
    E_\mathrm{c} = E_\mathrm{AFM} - E_\mathrm{FM}.
\end{equation}
A positive coupling energy thus indicates that the ferromagnetic alignment is energetically preferred.

\begin{figure*}[ht!]
    \centering
    \includegraphics[width=\textwidth]{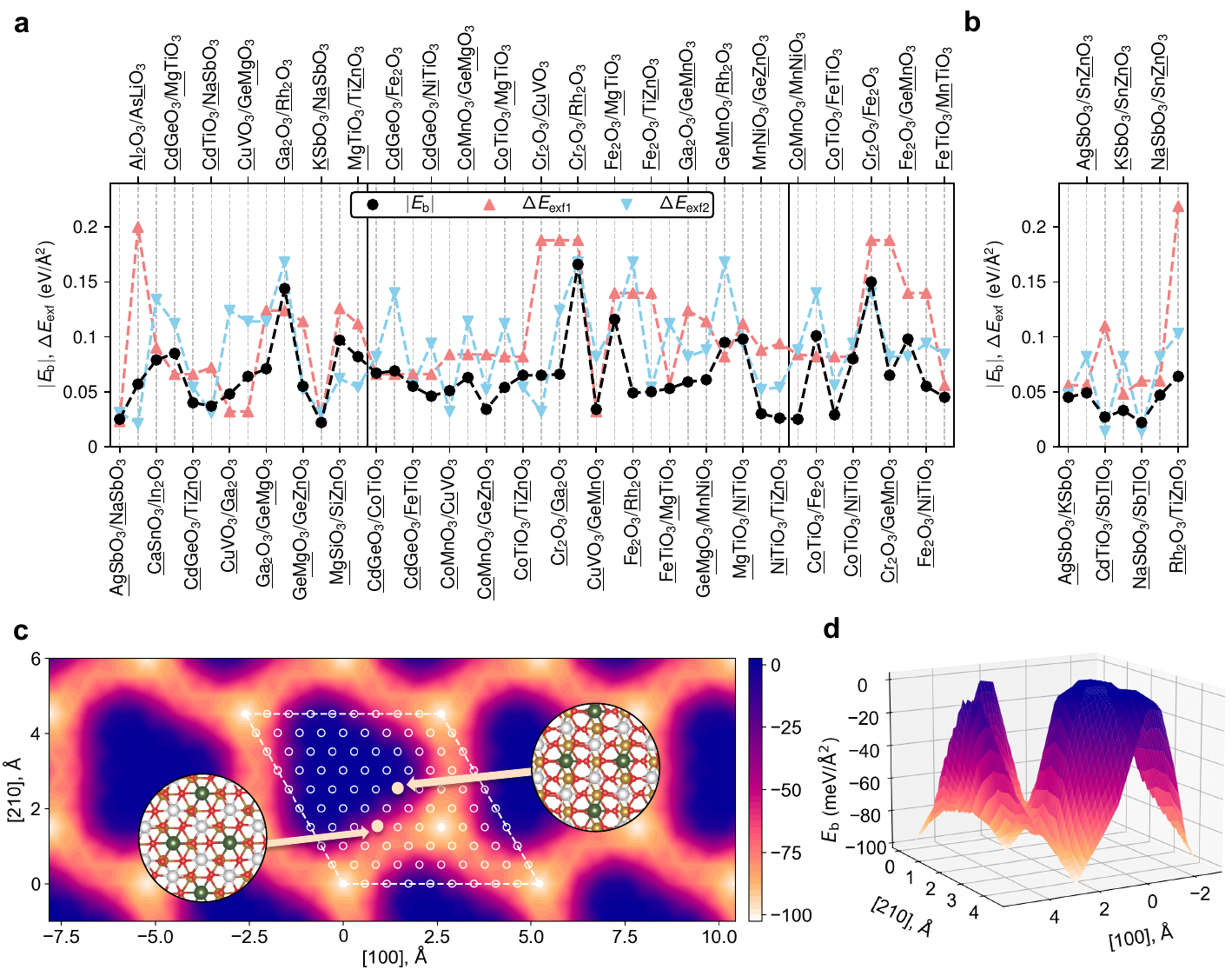}
    \caption{\small \textbf{Binding energies.} Comparison of HS binding energies and exfoliation energies of the individual 2D components calculated with plain \PBE$+U$ (\textbf{a}) and with D3 dispersion correction (\textbf{b}).
    $\Delta E_{\mathrm{exf1/2}}$ refers to the exfoliation energy of the first/second component in the formula of the HS.
    Vertical black lines separate results for the different groups.
    As a guide to the reader, elements at the surface of the isolated 2D sheets are underlined.
    These are then on one side located at the interface of the respective HS.
    {
    (\textbf{c}) Energy landscape for shifted Fe$_2$O$_3$/MgTiO$_3$ HS cells.
    The white open circles indicate the relative positions of the interfacial cations with the one from the upper 2D component placed at ($\frac{1}{3}$,$\frac{2}{3}$) unit cell fractional coordinates and the one from the lower 2D component shifted according to the circles.
    For the two highlighted cases, a top view of the shifted structures is given in the insets.
    Dashed white lines indicated unit cell boundaries.
    The color map is a linear interpolation between the values sampled at the circles and is extended beyod the unit cell boundaries to signify the overall (triangular) symmetry.
    The magnitude of the computed binding energies is indicated by the color bar.
    (\textbf{d}) Perspective view of the linearly interpolated shift energy landscape clearly indicating pronounced minima at preferential orientation of the interfacial cations.}
    }
    \label{fig2}
\end{figure*}

\section*{Results and Discussion} \label{results_discussion}

The HSs are build out of 36 non-van der Waals 2D materials outlined previously \cite{Friedrich_NanoLett_2022,Barnowsky_AdvElMats_2023} (see Fig.~\ref{fig1}(\textbf{a})).
The entries (V$_2$O$_3$ and Ti$_2$O$_3$) exhibiting broken symmetry are excluded so that all 2D materials have the same hexagonal structure as the experimentally realized hematene \cite{Puthirath_Balan_NNANO_2018} and ilmenene \cite{Puthirath_Balan_CoM_2018} systems.
Out of the remaining 34 single layers, a total of 561 combinations can be formed.
During the search for matching cell sizes at different twist angles (for details see Methods section), the total strain is limited to at max 2~\% as larger values are considered generally unfeasible, resulting in the generation of in total 1217 HSs.

Here, we are interested in outlining the fundamental effects arising from the strong interfacial interactions in these non-vdW HSs.
Hence we focus first on the smallest cells possible with a zero degree twist angle, aligned in such a way that cations do not face each other, yielding a total of 53 systems.
These HSs are further categorized into three groups: combinations of (i) two diamagnetic layers (21 HSs, example structure in Fig.~\ref{fig1}(\textbf{b})), (ii) a diamagnetic and a magnetic layer with a sizeable magnetic moment for at least one cation species (23 HSs, Fig.~\ref{fig1}(\textbf{c})), and (iii) two magnetic layers (9 HSs, Fig.~\ref{fig1}(\textbf{d})).
A representative top view of Cr$_2$O$_3$/Fe$_2$O$_3$ is {shown} in Fig.~\ref{fig1}(\textbf{e}).
Table~\ref{tab1} lists the nominal strain for all systems as obtained from the initial HS setup and the ``relaxed strain'' of the optimized structure.
The second varies from the nominal value due to specific relaxations for each system albeit not strongly in most cases.
Note that the tabulated total strain is the sum of the (tensile and compressive) strains on both 2D materials.
Analysis of the relaxed strain on each individual 2D component showed that the maximum value per spatial direction is 1.25\% for TiZnO$_3$ in Rh$_2$O$_3$/TiZnO$_3$.

\onecolumngrid

{
{
\begin{table*}[h!]
    \caption{ \textbf{Key properties of non-vdW HSs.} List of all HSs with nominal (relaxed) strain, binding energy $E_\mathrm{b}$ per area and per bond, {exfoliation energies $E_\mathrm{exf}$ of the individual (first/second) 2D components}, bond lengths at the interface ($d_1$/$d_2$ is the interfacial cation-O bond length of the first/second component in the formula of the HS), and interfacial magnetic coupling energies.
    For the entries marked by `*', the properties computed for the structure optimized with D3 dispersion correction are reported.
    Elements at the surface of the isolated 2D sheets are underlined.
    These are then on one side located at the interface of the respective HS.}
    \begin{tabular}{|>{\centering\arraybackslash}m{3cm}|>{\centering\arraybackslash}m{2.0cm}|>{\centering\arraybackslash}m{2.0cm}|>{\centering\arraybackslash}m{2.0cm}|>{\centering\arraybackslash}m{2.2cm}|>{\centering\arraybackslash}m{1.9cm}|>{\centering\arraybackslash}m{1.9cm}|>{\centering\arraybackslash}m{2.0cm}|}
            \hline
        \textbf{System} & \textbf{Strain (\%) (Relaxed)} & \textbf{$E_\mathrm{b}$ per area (meV/\AA$^2$)} & \textbf{$E_\mathrm{b}$ per bond (meV)} & {\textbf{$E_\mathrm{exf}$ per area (meV/\AA$^2$)}} &  \textbf{$d_1$ (\AA)} & \textbf{$d_2$ (\AA)} & \textbf{$E_\mathrm{c}$ (meV)} \\
        \hline
        \multicolumn{7}{|c|}{\textbf{diamagnetic/diamagnetic combinations}} \\
        \hline
        \underline{Ag}SbO$_3$\slash \underline{K}SbO$_3$* & 0.06 (0.71) & --45 & --190 & {56 \slash 48} & 2.86 & 2.82      & --- \\
        \hline
        \underline{Ag}SbO$_3$\slash \underline{Na}SbO$_3$ & 1.92 (1.81) & --25 & --107 & {23 \slash 31} & 2.73 & 2.73      & --- \\
        \hline
        \underline{Ag}SbO$_3$\slash Sn\underline{Zn}O$_3$* & 0.04 (1.59) & --49 & --210 & {56 \slash 82} & 2.65 & 2.98     & --- \\
        \hline
        \underline{Al}$_2$O$_3$\slash As\underline{Li}O$_3$ & 0.29 (1.18) & --57 & --200 & {200 \slash 21} & 2.45 & 2.16    & --- \\
        \hline
        \underline{Ca}SnO$_3$\slash \underline{In}$_2$O$_3$ & 0.07 (2.18) & --79 & --363 & {90 \slash 134} & 2.43 & 2.52    & --- \\
        \hline
        \underline{Cd}GeO$_3$\slash \underline{Mg}TiO$_3$ & 0.51 (0.85) & --85 & --337 & {66 \slash 112} & 2.44 & 2.31      & --- \\
        \hline
        \underline{Cd}GeO$_3$\slash Ti\underline{Zn}O$_3$ & 0.91 (2.09) & --40 & --155 & {66 \slash 54} & 2.52 & 2.48      & --- \\
        \hline
        \underline{Cd}TiO$_3$\slash \underline{Na}SbO$_3$ & 1.10 (1.76) & --37 & --153 & {72 \slash 31} & 2.80 & 2.53      & --- \\
        \hline
        \underline{Cd}TiO$_3$\slash Sb\underline{Tl}O$_3$* & 0.39 (0.81) & --27 & --112 & {110 \slash 14} & 4.13 & 3.32     & --- \\
        \hline
        \underline{Cu}VO$_3$\slash \underline{Ga}$_2$O$_3$ & 1.12 (2.12) & --48 & --178 & {32 \slash 124} & 2.25 & 2.36     & --- \\
        \hline
        \underline{Cu}VO$_3$\slash Ge\underline{Mg}O$_3$ & 0.64 (2.52) & --64 & --238 & {32 \slash 114} & 2.32 & 2.30       & --- \\
        \hline
        \underline{Ga}$_2$O$_3$\slash Ge\underline{Mg}O$_3$ & 1.76 (2.47) & --71 & --263 & {124 \slash 114} & 2.41 & 2.23    & --- \\
        \hline
        \underline{Ga}$_2$O$_3$\slash \underline{Rh}$_2$O$_3$ & 0.93 (0.93) & --144 & --547 & {124 \slash 168} & 2.08 & 2.12 & --- \\
        \hline
        Ge\underline{Mg}O$_3$\slash Ge\underline{Zn}O$_3$ & 1.50 (2.31) & --55 & --200 & {114 \slash 52} & 2.33 & 2.37      & --- \\
        \hline
        \underline{K}SbO$_3$\slash \underline{Na}SbO$_3$ & 1.92 (1.93) & --22 & --93 & {24 \slash 31} & 2.85 & 2.93        & --- \\
        \hline
        \underline{K}SbO$_3$\slash Sn\underline{Zn}O$_3$* & 0.03 (0.70) & --33 & --143 & {48 \slash 82} & 2.85 & 3.36      & --- \\
        \hline
        \underline{Mg}SiO$_3$\slash Si\underline{Zn}O$_3$ & 1.43 (1.43) & --97 & --325 & {126 \slash 62} & 2.22 & 2.25      & --- \\
        \hline
        \underline{Mg}TiO$_3$\slash Ti\underline{Zn}O$_3$ & 1.42 (2.58) & --82 & --320 & {112 \slash 54} & 2.23 & 2.26      & --- \\
        \hline
        \underline{Na}SbO$_3$\slash Sb\underline{Tl}O$_3$* & 0.71 (0.84) & --22 & --94 & {60 \slash 14} & 3.56 & 3.17      & --- \\
        \hline
        \underline{Na}SbO$_3$\slash Sn\underline{Zn}O$_3$* & 1.88 (2.18) & --47 & --197 & {60 \slash 82} & 2.63 & 2.99     & --- \\
        \hline
        \underline{Rh}$_2$O$_3$\slash Ti\underline{Zn}O$_3$* & 1.99 (2.50) & --64 & --244 & {219 \slash 103} & 2.45 & 2.34   & --- \\
        \hline
        \multicolumn{7}{|c|}{\textbf{diamagnetic/magnetic combinations}} \\
        \hline
        \underline{Cd}GeO$_3$\slash \underline{Co}TiO$_3$ & 0.68 (1.52) & --67 & --262 & {66 \slash 82} & 2.45 & 2.37      & --- \\
        \hline
        \underline{Cd}GeO$_3$\slash \underline{Fe}$_2$O$_3$ & 1.30 (1.30) & --69 & --271 & {66 \slash 140} & 3.86 & 2.39    & --- \\
        \hline
        \underline{Cd}GeO$_3$\slash \underline{Fe}TiO$_3$ & 1.78 (1.78) & --55 & --217 & {66 \slash 56} & 2.46 & 2.40      & --- \\
        \hline
        \underline{Cd}GeO$_3$\slash \underline{Ni}TiO$_3$ & 0.53 (2.44) & --46 & --177 & {66 \slash 94} & 2.51 & 2.46      & --- \\
        \hline
        \underline{Co}MnO$_3$\slash \underline{Cu}VO$_3$ & 1.15 (2.63) & --51 & --186 & {84 \slash 32} & 2.35 & 2.28       & --- \\
        \hline
        \underline{Co}MnO$_3$\slash Ge\underline{Mg}O$_3$ & 0.51 (2.15) & --63 & --234 & {84 \slash 114} & 2.44 & 2.37      & --- \\
        \hline
        \underline{Co}MnO$_3$\slash Ge\underline{Zn}O$_3$ & 0.99 (2.13) & --34 & --126 & {84 \slash 52} & 2.51 & 2.48      & --- \\
        \hline
        \underline{Co}TiO$_3$\slash \underline{Mg}TiO$_3$ & 0.18 (3.12) & --54 & --210 & {82 \slash 112} & 2.48 & 2.40      & --- \\
        \hline
        \underline{Co}TiO$_3$\slash Ti\underline{Zn}O$_3$ & 1.60 (2.78) & --65 & --252 & {82 \slash 54} & 2.26 & 2.25      & --- \\
        \hline
        \underline{Cr}$_2$O$_3$\slash \underline{Cu}VO$_3$ & 1.72 (3.08) & --65 & --241 & {188 \slash 32} & 2.24 & 2.21     & --- \\
        \hline
        \underline{Cr}$_2$O$_3$\slash \underline{Ga}$_2$O$_3$ & 0.60 (4.00) & --66 & --248 & {188 \slash 124} & 2.19 & 2.19  & --- \\
        \hline
        \underline{Cr}$_2$O$_3$\slash \underline{Rh}$_2$O$_3$ & 0.33 (0.97) & --166 & --631 & {188 \slash 168} & 2.08 & 2.11 & --- \\
        \hline
        \underline{Cu}VO$_3$\slash Ge\underline{Mn}O$_3$ & 1.76 (2.02) & --34 & --126 & {32 \slash 82} & 2.45 & 2.53       & --- \\
        \hline
        \underline{Fe}$_2$O$_3$\slash \underline{Mg}TiO$_3$ & 1.81 (1.80) & --116 & --454 & {140 \slash 112} & 2.27 & 2.17   & --- \\
        \hline
        \underline{Fe}$_2$O$_3$\slash \underline{Rh}$_2$O$_3$ & 1.60 (1.77) & --49 & --186 & {140 \slash 168} & 2.18 & 2.28  & --- \\
        \hline
        \underline{Fe}$_2$O$_3$\slash Ti\underline{Zn}O$_3$ & 0.39 (3.00) & --50 & --192 & {140 \slash 54} & 2.34 & 2.26    & --- \\
        \hline
        \underline{Fe}TiO$_3$\slash \underline{Mg}TiO$_3$ & 1.28 (3.25) & --53 & --208 & {56 \slash 112} & 2.35 & 2.30      & --- \\
        \hline
        \underline{Ga}$_2$O$_3$\slash Ge\underline{Mn}O$_3$ & 0.64 (2.09) & --59 & --223 & {124 \slash 82} & 2.46 & 2.33    & --- \\
        \hline
        Ge\underline{Mg}O$_3$\slash Mn\underline{Ni}O$_3$ & 1.85 (2.52) & --61 & --221 & {114 \slash 88} & 2.31 & 2.37      & --- \\
        \hline
        Ge\underline{Mn}O$_3$\slash \underline{Rh}$_2$O$_3$ & 0.30 (0.47) & --95 & --361 & {82 \slash 168} & 2.21 & 2.26    & --- \\
        \hline
        \underline{Mg}TiO$_3$\slash \underline{Ni}TiO$_3$ & 1.04 (2.99) & --98 & --381 & {112 \slash 94} & 2.20 & 2.22      & --- \\
        \hline
        Mn\underline{Ni}O$_3$\slash Ge\underline{Zn}O$_3$ & 0.35 (2.27) & --30 & --110 & {88 \slash 52} & 2.44 & 2.41      & --- \\
        \hline
        \underline{Ni}TiO$_3$\slash Ti\underline{Zn}O$_3$ & 0.38 (3.33) & --26 & --100 & {94 \slash 54} & 2.44 & 2.44      & --- \\
        \hline
        \multicolumn{7}{|c|}{\textbf{magnetic/magnetic combinations}} \\
        \hline
        \underline{Co}MnO$_3$\slash Mn\underline{Ni}O$_3$ & 1.34 (1.34) & --25 & --94 & {84 \slash 88} & 2.46 & 2.31       & $-$72 \\
        \hline
        \underline{Co}TiO$_3$\slash \underline{Fe}$_2$O$_3$ & 1.99 (2.09) & --101 & --392 & {82 \slash 140} & 2.19 & 2.25   & +30 \\
        \hline
        \underline{Co}TiO$_3$\slash \underline{Fe}TiO$_3$ & 1.10 (2.95) & --29 & --112 & {82 \slash 56} & 2.51 & 2.49      & $-$7 \\
        \hline
        \underline{Co}TiO$_3$\slash \underline{Ni}TiO$_3$ & 1.22 (3.15) & --80 & --310 & {82 \slash 94} & 2.23 & 2.21      & +13 \\
        \hline
        \underline{Cr}$_2$O$_3$\slash \underline{Fe}$_2$O$_3$ & 1.93 (1.92) & --150 & --575 & {188 \slash 140} & 2.10 & 2.13 & $-$74 \\
        \hline
        \underline{Cr}$_2$O$_3$\slash Ge\underline{Mn}O$_3$ & 0.06 (2.91) & --65 & --244 & {188 \slash 82} & 2.39 & 2.31    & $-$61 \\
        \hline
        \underline{Fe}$_2$O$_3$\slash Ge\underline{Mn}O$_3$ & 1.90 (1.89) & --98 & --376 & {140 \slash 82} & 2.32 & 2.26    & +10 \\
        \hline
        \underline{Fe}$_2$O$_3$\slash \underline{Ni}TiO$_3$ & 0.77 (3.27) & --55 & --212 & {140 \slash 94} & 2.36 & 2.32    & +15 \\
        \hline
        \underline{Fe}TiO$_3$\slash \underline{Mn}TiO$_3$ & 1.09 (3.26) & --45 & --177 & {56 \slash 84} & 2.39 & 2.38      & $-$27 \\
        \hline
    \end{tabular}
    \label{tab1}
\end{table*}

}

}

\twocolumngrid

\textbf{Interfacial bonding:}
Upon relaxation of the structures, the individual 2D layers bind strongly (highlighted interfacial bonds in Fig.~\ref{fig1}(\textbf{b})).
This is clearly visible in all representative optimized geometries in Fig.~\ref{fig1}(\textbf{b}-\textbf{d}).
The bond formation at the interface indicates chemical compatibility and enhances the overall stability of the HS, {crucially determining its properties beyond those of traditional vdW stacks.}
This chemical bonding can already be detected for the example of Cr$_2$O$_3$/Fe$_2$O$_3$ from an inspection of the charge density on the interfacial plane,
{as} provided in Fig.~\ref{fig1}(\textbf{f}).
While the main charge is concentrated at the transition metals, there is also significant density in the regions between these centers indicative of electronic delocalization and coupling.
An even more insightful picture emerges from the charge density difference computed as the change of the electronic density between the HS and its isolated 2D components with the same atomic positions (see Methods section for details).
The isosurface/cut through the density difference in Fig.~\ref{fig1}(\textbf{g})/(\textbf{h}) clearly indicates that, upon interfacial interaction, charge is transferred from around the central cations towards the neighboring oxygens of the other 2D sheet with which bonding occurs.
A comparison of the charge density differences for all the three HSs in Fig.~\ref{fig1}(\textbf{b}-\textbf{d}) is presented in section~II. of the supporting information (SI).

\textbf{Binding energies:}
To quantify the interfacial interaction strength, the binding energy $E_{\mathrm{b}}$ is computed as the difference in total energy between the HS and the isolated relaxed 2D systems normalized per area.
The binding energy has a negative sign, as energy is gained upon bonding the individual sheets.
The values of all systems are included in Table~\ref{tab1} and the absolute numbers in comparison to the exfoliation energies of the respective 2D component sheets {(also included in Table~\ref{tab1})} \cite{Friedrich_NanoLett_2022,Barnowsky_AdvElMats_2023} are presented in Fig.~\ref{fig2}{(\textbf{a}) and (\textbf{b})} with a separation according to the three groups.
The exfoliation energy $\Delta E_{\mathrm{exf}}$ is the (positive) energy required to delaminate a 2D material of given thickness from the surface of its bulk parent material which has also been proven to coincide with the interlayer binding energy of the bulk \cite{Jung_NanoLett_2018}.

The energy gain resulting from the bonding within the HS ranges from only a few meV/\AA$^2$ in a small number of cases to $-$166 meV/\AA$^2$ (Cr$_2$O$_3$/Rh$_2$O$_3$), \emph{i.e.}, for most systems (around two third), a large binding energy of at least around $-$50~meV/\AA$^2$ is observed.
Five systems even have an absolute $E_{\mathrm{b}}$ exceeding 100~meV/\AA$^2$: Ga$_2$O$_3$/Rh$_2$O$_3$, Cr$_2$O$_3$/Rh$_2$O$_3$, Fe$_2$O$_3$/MgTiO$_3$ CoTiO$_3$/Fe$_2$O$_3$, and Cr$_2$O$_3$/Fe$_2$O$_3$.
This energy scale is substantially larger as for any vdW bonded layers, underscoring the unique strength of interfacial bonding in non-vdW heterostructures.
For example, the absolute binding energy of a vdW bonded graphene homobilayer is computed to 16~meV/\AA$^2$ which is in the same well-known energy range as typical exfoliation energies of vdW bonded (layered) materials \cite{Mounet_AiiDA2D_NNano_2018,Lebegue_prx_2013,Bjorkman_PRL_2012}.
An exfoliation or absolute binding energy of $\sim$20~meV/\AA$^2$ was even pointed out to be a universal value for layered systems \cite{Bjorkman_PRL_2012}.

The large $E_{\mathrm{b}}$ of non-vdW HSs stem from ionic and covalent interactions at the interface between the 2D sheets.
To underscore this point, we also compute average binding energies per bond at the interface.
As depicted in Figs.~\ref{fig1}(\textbf{b})-(\textbf{d}), for essentially all structures, six new cation-oxygen bonds are formed between the surface cation of one component and O$^{2-}$ anions of the other component in the HS.
As outlined in Table~\ref{tab1}, these values per bond are in almost all cases on the order of several hundred meV reaching the highest energy gain again for Cr$_2$O$_3$/Rh$_2$O$_3$ with $-631$~meV indicative of ionic and/or covalent interactions.

Also the individual interfacial bond lengths $d_1$ and $d_2$ labelled in Fig.~\ref{fig1}(\textbf{b}) for each cation-O pair included in Table~\ref{tab1} corroborate this picture.
These are almost exclusively between 2 and 3~\AA\ and hence typical for covalent and ionic bonds for the elements involved.

When compared to the exfoliation energies of their 2D components in Fig.~\ref{fig2}{(\textbf{a}) and (\textbf{b})}, the absolute binding energies of the HSs are generally on the same order of magnitude with a slight tendency towards lower values.
Thus, most HSs are bound about as strongly as each of the components is bound to its respective bulk parent compound.

There are, however, a few cases where the coupling in the HS is significantly smaller for which Fe$_2$O$_3$/Rh$_2$O$_3$ is a representative example with its absolute $E_{\mathrm{b}}$ of 49~meV/\AA$^2$ being about 90~meV/\AA$^2$ smaller than the lower of the two exfoliation energies (140~meV/\AA$^2$ for Fe$_2$O$_3$).
This signifies that for strong bonding proper electronic compatibility is required.

There are a few HSs in our set with small absolute $E_{\mathrm{b}}$.
For these cases and a few others with higher binding energies, the role of including dispersive interactions via established extensions to DFT (D3 \cite{Grimme_JCP_2010}, \SCAN$+$rVV10 \cite{Perdew_SCAN_PRL_2015,Peng_PRX_2016}) is validated with details discussed in section~III in the SI.
Based on this analysis, the binding energies of the HSs marked with an asterisk in Table~\ref{tab1} (the cases where plain \PBE$+U$ yields $|E_{\mathrm{b}}|<$ 20 meV/\AA$^2$) are presented with the dispersion correction as only then proper negative values are obtained.
Their binding energies compared to the respective exfoliation energies are presented in Fig.~\ref{fig2}(\textbf{b}).
Also here the HS shows a tendency towards weaker bonding compared to exfoliation from the parent compounds.
In all other cases, negative $E_{\mathrm{b}}$ exceeding the typical energy scale of dispersive interactions \cite{Bjorkman_PRL_2012} is already achieved with plain DFT.
For HSs with high binding energies (more negative than $-$50~meV/\AA$^2$), which we are mostly
interested in, the effect of the dispersion on the energetics and, in particular, on bond lengths is minor, as expected, and properties can be {reliably} computed from {structures optimized with} plain \PBE$+U$.

\begin{figure*}[ht!]
    \includegraphics[width=0.9\textwidth]{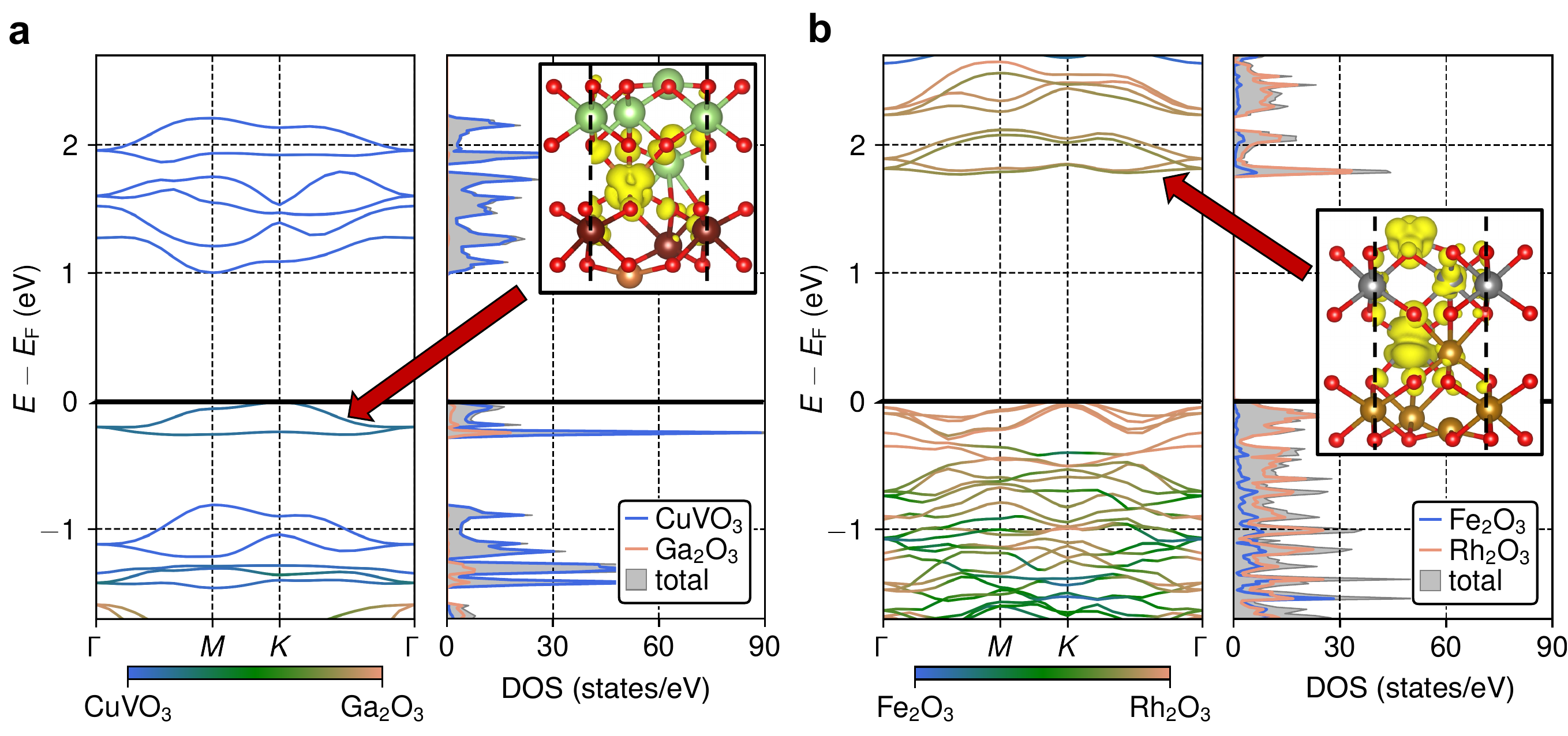}
    \caption{\small {\textbf{Hybrid interface bands.}
    Representative band structures with projection contributions onto the atoms of each 2D sheet of CuVO$_3$/Ga$_2$O$_3$ (\textbf{a}) and Fe$_2$O$_3$/Rh$_2$O$_3$ (\textbf{b}) exhibiting hybrid interface bands close to the gap computed with HSE06.
    The insets show isosurfaces of the partial charge density of the hybrid interface bands marked by the respective arrows (isosurface value: 0.027~$e$/\AA$^3$).
    Colors: same as in Fig.~\ref{fig1} and V, brown; Cu, orange.}
    }
    \label{fig3}
\end{figure*}

The systems including SbTlO$_3$ deserve a special note.
This compound has been classified as a non-vdW 2D material due to its structural similarity to all the other systems in the set with the Tl cations at the surface of the sheet.
The bulk form of SbTlO$_3$ is, however, a layered crystal held together mostly by vdW interactions.
{Therefore,} this system was pointed out to be in the transition region between vdW and non-vdW materials \cite{Barnowsky_AdvElMats_2023}.
Consequently, also in HSs involving SbTlO$_3$, interfacial interactions are expected to be weaker than in the other cases with mostly dispersive contributions as corroborated by the small $E_{\mathrm{b}}$ values.

{We also verify that the studied commensurate HS cells are relevant for experimental investigations.
Firstly, the strain energy of the 2D components in the HS is very small.
Structures with minimal strain are {deliberately} studied
, reaching maximally 1.25\% per direction as mentioned above.
From calculations on selected individual 2D systems (including TiZnO$_3$) strained to the lattice constants {of} the HS, it is found that the energy penalty due to strain is at most a few meV/\AA$^2$ and thus negligibly small compared to the interfacial binding energies, which are on the order of up to 100~meV/\AA$^2$.}

{Secondly, the dominant factor for the interaction in the non-vdW HSs is the directional bonding between the 2D surface cations at the interface and {therefore} their relative alignment.
This is indicated in the relaxed structures in Figs.~\ref{fig1}(\textbf{b}-\textbf{c}) where the cations properly avoid each other to bind to the oxygen anions of the other 2D component.
The consequence of less preferential alignment can be estimated from constructing shifted cells where the 2D components and thus their surface cations are displaced with respect to each other (within the minimal unit cell).
Fig.~\ref{fig2}(\textbf{c}) shows the resulting energy landscape of these shifted cells for the example of Fe$_2$O$_3$/MgTiO$_3$ (In section~IV. in the SI a large scale depiction of the energy and interlayer distance landscapes is included.).
Each point of the investigated 9$\times$9 grid represents a different relative alignment within the HS as underlined by top views in the insets for two exemplary cases.
The energy becomes minimal for the above studied commensurate HS with the upper component cation at ($\frac{1}{3}$,$\frac{2}{3}$) unit cell fractional coordinates and the lower component cation at ($\frac{2}{3}$,$\frac{1}{3}$).
An equal energy minimum from a symmetry equivalent structure is present when the lower component cation is placed at the origin.
The maximum with zero binding energy is achieved with both cations at ($\frac{1}{3}$,$\frac{2}{3}$), thus facing each other at the interface.
The perspective view of the shift energy landscape in Fig.~\ref{fig2}(\textbf{d}) clearly shows the dominant minimum for preferred cation alignment.
This preferential interaction is maximized in the minimal commensurate cells, thus promoting them as favorable HSs to study.
}

{To further assess the thermodynamic stability of the proposed HSs, we compare the binding energy per atom with respect to isolated 2D components ($E_{\mathrm{b}}$) and the energy relative to bulk (pre-exfoliation) phases ($E_{\mathrm{B}}$). As shown in section~V. in the SI, all HSs exhibit negative $E_{\mathrm{b}}$ values ($-27$ to $-189$ meV/atom), indicating that the formation of the interface from isolated layers is energetically favorable. In contrast, the corresponding $E_{\mathrm{B}}$ values are positive (35 to 295 meV/atom), reflecting that the bulk phases remain thermodynamically more stable than the HSs.}

{Such behavior is expected, particularly for non-vdW HSs, where the formation of an interface competes with strong bulk bonding. Therefore, positive $E_{\mathrm{B}}$ values indicate metastability with respect to phase separation into bulk components.
However, since HSs are often build from previously obtained 2D components, the binding energy with respect to these is generally the more relevant quantity.
Moreover, the proposed HSs are expected to be kinetically accessible and therefore can be experimentally realized, particularly via layer-by-layer assembly.}

{\textbf{Hybrid interface bands}:}
{While the electronic properties of HSs are often governed by the specific band alignment at the interface, for the present non-vdW stacks, the interfacial bonding creates new hybrid interface bands that can appear in or around the edges of the original band gap.
To illustrate these unique features of the new interface class of non-vdW HSs, we focus on two representative examples: CuVO$_3$/Ga$_2$O$_3$ and Fe$_2$O$_3$/Rh$_2$O$_3$.
{
In Fig.~\ref{fig3}(\textbf{a},\textbf{b}) this is visible in the band structures with color-coded projection contributions onto the atoms of the individual 2D components.
While in the first case, the highest occupied bands show strongly mixed character with respect to contributions from the individual 2D components, the second case depicts a hybridization of the hematene surface state with Rh$_2$O$_3$ bands in the conduction band region.}

{To further underline the mixed interfacial character, the insets of Fig.~\ref{fig3}(\textbf{b},\textbf{c}) show isosurfaces of the partial charge density of the hybrid bands marked by the red arrows.
There, the $d$-states of the transition metal of one 2D component hybridize with the oxygen $p$-states of the other, contributing to the interfacial bonding.
In the second case, contributions at both the interfacial Rh and Fe ions are visible, together with {contributions from} the Rh at the top of the structure.
These hybrid interface bands signify that the electronic properties of the HSs are not just a simple superposition of their 2D components but showcase qualitatively new hybrid features that will also govern their transport properties.}

{When studying electronic properties, band gaps from standard DFT are usually unreliable.
We therefore decided to present band structures as computed on the hybrid functional level (HSE06~\cite{HSE06}) while all (binding) energies, densities, and structural properties are studied with PBE$+U$.
In section~VI. in the SI, the different approaches are benchmarked with respect to each other for several systems showing that the hybrid functional essentially only increases the band gap but the dispersion and hybrid character of the bands is already well captured within simpler PBE+$U$.
On this basis, a
comprehensive comparison of band structures and densities of states (DOSs) for all HSs together with a direct comparison to the bands of the component 2D systems is provided
in section~VII. in the SI.
Band structures with projection contributions for all systems are also included in section~VII. in the SI.}

\begin{figure*}[ht!]
    \includegraphics[width=0.9\textwidth]{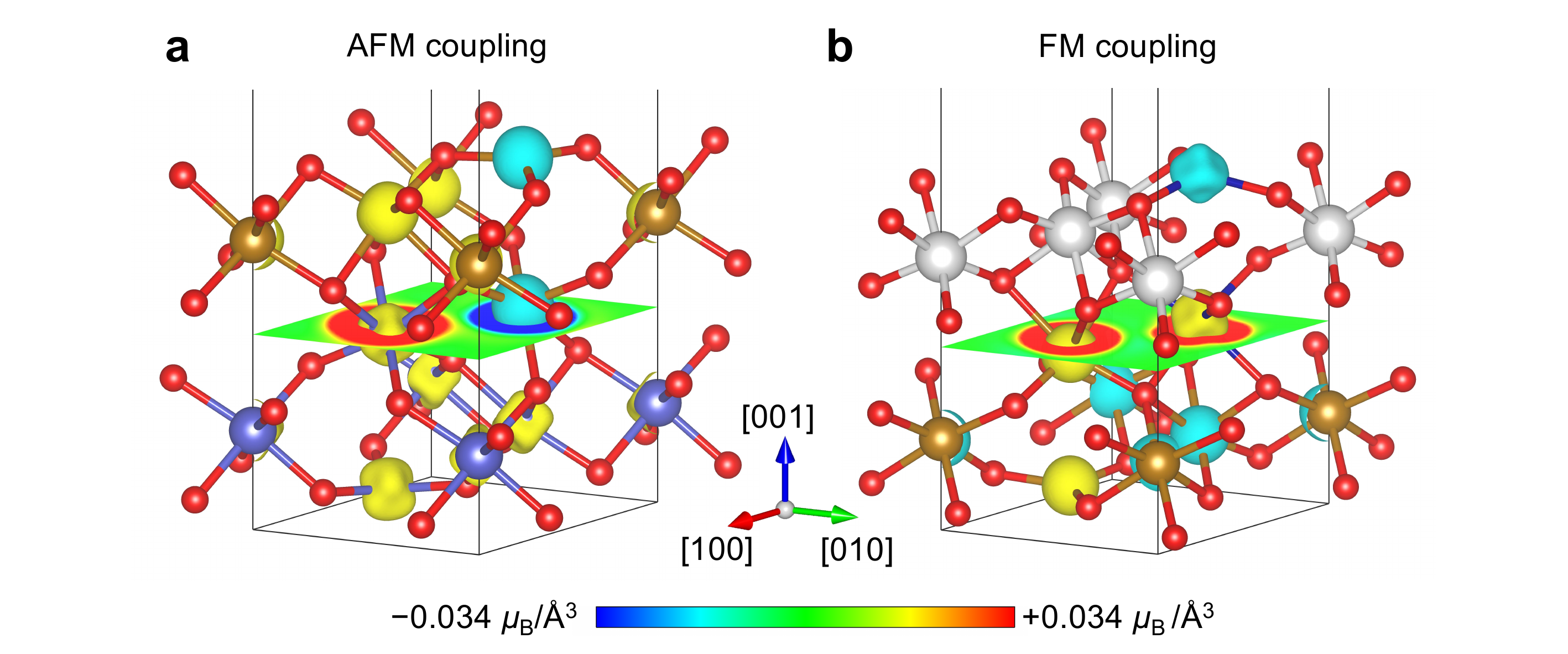}
    \caption{\small \textbf{Interfacial magnetic coupling.}
    Perspective view of Cr$_2$O$_3$/Fe$_2$O$_3$ (\textbf{a}) and CoTiO$_3$/Fe$_2$O$_3$ (\textbf{b}) with magnetization density isosurfaces as well as a visualization of this quantity on the interfacial plane indicating AFM and FM interfacial coupling, respectively (isosurface value: 0.88~$\mu_{\mathrm{B}}$/\AA$^3$).
    Yellow/bluish parts of the isosurface correspond to positive/negative values.
    The black line indicates unit cell boundaries.
    Colors: same as in Fig.~\ref{fig1} and Co, dark blue; Ti, light gray.
    }
    \label{fig4}
\end{figure*}

\textbf{Interfacial magnetic coupling}:
The chemical bonding {now also} translates into strong interfacial magnetic coupling between the 2D components.
Several coupling paths for superexchange between the transition metal ions mediated via oxygen anions are initiated through the interfacial bonds.
In {group three} HSs composed of two magnetic 2D systems, the relative alignment of the magnetic moments at the interface is a key factor {in} determining the magnetic characteristics of the hybrid material.
The moments themselves change by less than 0.2~$\mu_{\mathrm{B}}$ upon HS formation.

The moments can be aligned or antialigned, \emph{i.e.} the interfacial magnetic coupling can be ferromagnetic (FM) or antiferromagnetic (AFM).
As such, calculations with {(i)} the interfacial moments in both component systems aligned {(FM)} and {(ii)} for all moments of the second component flipped {(AFM)} can quantify the strength of the magnetic interactions initiated upon interfacial bonding.
Understanding these interactions is crucial, as they directly influence the stability and potential functionality of magnetic HSs.
The corresponding coupling energies $E_{\mathrm{c}}$ included in Table~\ref{tab1} computed as the energy difference between {anti-aligned (AFM) and aligned (FM)} states are substantial on the order of several 10~meV/cell {and thus preferably large compared to vdW HSs}.
Positive/negative values indicate preference of the FM/AFM state.
We are mainly interested in the interfacial magnetic coupling and thus assume that the magnetic alignment within each 2D component is not affected by formation of the HS.
Among the studied nine combinations, four exhibit a preference for FM alignment (CoTiO$_3$/Fe$_2$O$_3$, CoTiO$_3$/NiTiO$_3$, Fe$_2$O$_3$/GeMnO$_3$, Fe$_2$O$_3$/NiTiO$_3$) whereas the five others (CoMnO$_3$/MnNiO$_3$, CoTiO$_3$/FeTiO$_3$, Cr$_2$O$_3$/Fe$_2$O$_3$, Cr$_2$O$_3$/GeMnO$_3$, FeTiO$_3$/MnTiO$_3$) favor an AFM state.

For the cases with strongest AFM and FM interfacial couplings (Cr$_2$O$_3$/Fe$_2$O$_3$ and CoTiO$_3$/Fe$_2$O$_3$) represented by the most negative and positive coupling energies of $-74$ and $+30$~meV/cell, the magnetization densities are visualized in Fig.~\ref{fig4}.
These clearly indicate the localized positive and negative moments originating from the transition metal $d$-states.
The cut through the magnetisation density at the interfacial planes in Fig.~\ref{fig4} highlights once again the type of magnetic coupling and signifies the strong interfacial spin polarization.
Remarkably, these two HSs also show the largest binding energies among all group three systems more negative than $-$100~meV/\AA$^2$.

\begin{figure*}[ht!]
    \centering
    \includegraphics[width=\textwidth]{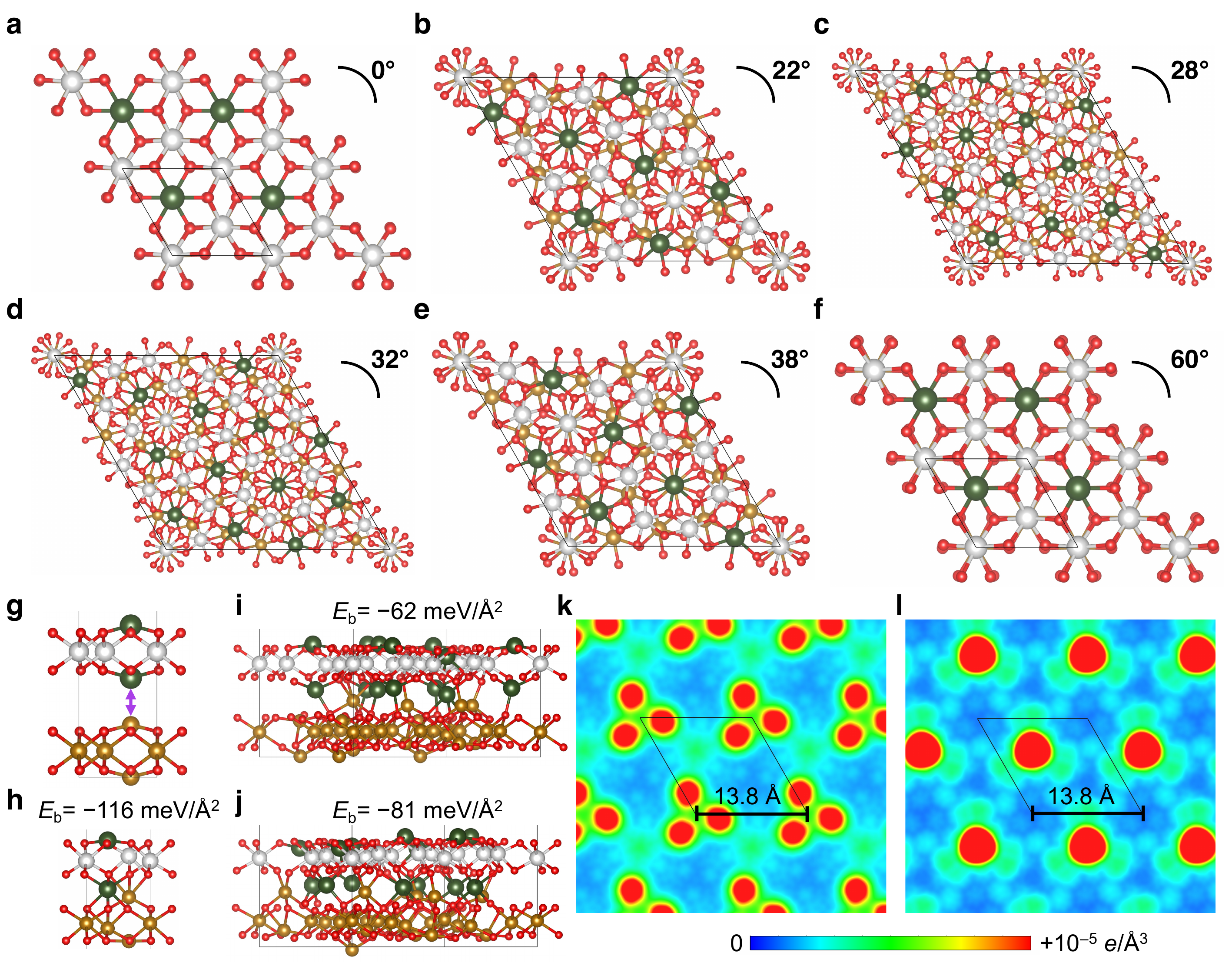}
    \caption{\small \textbf{Non-vdW HSs at different twist angles.} (\textbf{a}-\textbf{f}) Top views of Fe$_2$O$_3$/MgTiO$_3$ HSs obtained at different twist angles showcasing different local atomic moir{\'e} patterns for the cases different from 0 and 60\textdegree.
    (\textbf{g}/\textbf{h}) Side view of the initial/relaxed structures with 0/60\textdegree\ twist.
    (\textbf{i}/\textbf{j}) Side view of the relaxed structures with 22/38\textdegree\ twist.
    (\textbf{k}/\textbf{l}) Top view of the charge density variation on a plane 3.69/3.15~\AA\ below the bottom surface (below Fe$_2$O$_3$) for 22/38\textdegree.
    The unit cell is indicated by the black frame.
    Colors: same as in Fig.~\ref{fig1} and \ref{fig4} and Mg, dark green.}
    \label{fig5}
\end{figure*}

Note that deducing a magnetic stability of the interfacial coupling with respect to temperature by comparing these effective coupling energies through the interface to the thermal energy scale is complicated by the fact that several new interfacial couplings are initiated.
Moreover, the couplings inside the individual 2D components can be weak such that their magnetization will not switch as one entity \cite{Barnowsky_AdvElMats_2023,Barnowsky_NanoLett_2024,Hess_NJP_2017}.
Nevertheless, the sizeable $E_{\mathrm{c}}$ values highlight the robustness of the interfacial magnetic interactions in non-vdW HSs, which are critical for stabilizing magnetic configurations.

\textbf{Twisted non-vdW HSs:}
In addition to minimal commensurate
cells, especially the study of twisted HSs and homobilayers of vdW materials has attracted strong interest as they can give rise to novel and enhanced properties such as superconductivity.
Such twisted 2D sheets lead to moir{\'e} patterns: large-scale interference effects resulting from different local arrangements of the underlying lattices.
For 2D nanoscale systems, these are associated with characteristic density and magnetisation density variations that can, for instance, be measured via high resolution electron and scanning tunneling microscopy.
Thus, twisted HSs emerge as an appealing platform for discovering and engineering functional materials properties (twistronics) which we anticipate to be even more pronounced in non-vdW HSs due to the significant interfacial chemical interaction.
{Their study is particularly motivated by several recent experimental realizations of twisted oxide thin film interfaces from freestanding membranes also hinting at moir{\'e} signatures~\cite{Li_AdvMat_2022,Shen_ACSAMI_2022,Wang_AdvMat_2024,Schmidbauer_AplPhysLett_2025}.}

Introducing specific twist angles between non-vdW materials can create interfacial structures with both bonded (spatially separated cations) and non-bonded (confronting cations) regions, leading to a combination of weak and strong interfacial interactions.
However, these systems often require supercells with thousands of atoms, making them challenging to handle computationally with \emph{ab initio} methods.
Here, we give an outlook onto twisted non-vdW HSs by focusing on carefully chosen systems of moderate size generated from the hetbuilder tool that can still be treated within \DFT.

In Fig. \ref{fig5}(\textbf{a}-\textbf{f}) representative Fe$_2$O$_3$/MgTiO$_3$ HSs (with $\sim$1.81\% total nominal strain) obtained from a systematic twist angle scan are depicted.
The systems at angles different from 0 and 60\textdegree\ are modelled by larger unit cells containing 140 (22 and 38\textdegree) or 260 (28 and 32\textdegree) atoms.
From a visual inspection of the atomic structures it is already clear that the twisted HSs can give rise to moir{\'e} patterns due to the different local stacking arrangements.
Note that although the structures in Fig.~\ref{fig5}(\textbf{a}) and (\textbf{f}) appear equivalent from their top views, they differ in the relative arrangement of the 2D components.
While for 0\textdegree\ (side view in Fig.~\ref{fig5}(\textbf{g})) cations would face each other leading to a repulsive setup, for 60\textdegree\ (Fig.~\ref{fig5}(\textbf{h})), the arrangement is preferable and bonding can be achieved in a commensurate structure as investigated above.

We explicitly consider the cases in Fig.~\ref{fig5}(\textbf{b}) and (\textbf{e}) as electronic structure calculations were performed for them.
The side views of the relaxed geometries at 22 and 38\textdegree\ twist in Fig.~\ref{fig5}(\textbf{g}) and (\textbf{h}) clearly indicate substantially different interfacial interactions and bonding patterns (the relaxed strain amounts to 1.81 and 2.79\%).
This strong change of the atomic structure upon twisting is also mirrored by the binding energies.
While the commensurate HS at 60\textdegree\ has an $E_{\mathrm{b}}$ of $-$116~meV/\AA$^2$ (see Table~\ref{tab1}), this value reduces to $-$62~meV/\AA$^2$ at 22\textdegree.
At 38\textdegree\ the binding is reinforced to $-$81~meV/\AA$^2$ hinting at more preferential inter-layer bonds between the 2D components as is also visible when comparing the relaxed side views in Fig.~\ref{fig5}(\textbf{i}) and (\textbf{j}).
In general, reduced binding energies for larger twisted structures compared to the minimal commensurate cells are found as always less preferential interfacial arrangements need to be accommodated.

To underscore the strong (local) long-range property variation for the different angles, the charge density on a plane below the bottom surface of the HS is visualized in Fig.~\ref{fig5}(\textbf{k}/\textbf{l}) for 22/38\textdegree.
We focus on the bottom surface density as it depicts the clearest effects.
An equivalent analysis for the top surface (above MgTiO$_3$) can be found {in section~VIII. in the SI}.
There are distinct angle specific patterns in the densities that correspond to the long-range height variation of the transition metal Fe ions.
The pronounced red features can be associated with the Fe atoms sticking out of the HS surface.
These substantial moir{\'e} like density variations observed are thus strongly structure related and should as such be readily observable with experimental techniques such as electron microscopy and scanning tunneling microscopy.
The associated magnetisation density also shows strong long-range angle specific variations as provided {in section~IX. in the SI}.
These findings prove that non-vdW HSs can also host -- once more based on their interfacial interaction -- particularly strong local and twist dependent property modulations.
Such modulations also underscore the potential of these systems for experimental exploration.

\enlargethispage{10pt}

{\textbf{HSs for experimental realization:} Based on the combined analysis of electronic structure, interfacial hybridization, and magnetic interactions, several HSs emerge as particularly promising candidates for experimental realization. In particular, CuVO$_3$/Ga$_2$O$_3$ and Fe$_2$O$_3$/Rh$_2$O$_3$ exhibit distinctive hybrid interface bands, while Cr$_2$O$_3$/Fe$_2$O$_3$ and CoTiO$_3$/Fe$_2$O$_3$ show strong interfacial magnetic coupling. In addition, twisted Fe$_2$O$_3$/MgTiO$_3$ HSs display characteristic moiré patterns, suggesting further tunability of properties.}

{Five out of the seven investigated HSs exhibit no imaginary phonon modes throughout the Brillouin zone, indicating dynamical stability in their freestanding form. In contrast, two systems (CuVO$_3$/Ga$_2$O$_3$ and Fe$_2$O$_3$/Rh$_2$O$_3$) display small imaginary frequencies, indicating dynamical instability (the phonon dispersions are provided in Section~X. in the SI.). Such behavior is not uncommon in low-dimensional systems and may be mitigated under experimental conditions, for example, by substrate-induced constraints.}
{Overall, these results support the experimental viability of the majority of the identified HSs.}

\section*{Conclusions} \label{conclusions}

We have presented an extensive study of the new class of non-vdW HSs by comprehensive computational screening of a large set of 55 representatives.
A characteristic signature of these novel interfaces is the strong chemical bonding between the 2D components -- {a unique feature distinguishing them from weakly interacting vdW HSs}.
Based on investigation of the binding energies, the interactions clearly reveal covalent and/or ionic character reaching values of {over a} hundred meV/\AA$^2$.
The HSs are thus bound almost as strongly as each individual 2D component is to the bulk surface of its parent compound.
The strong interfacial interaction leaves also clear signatures in the electronic structure.
While all HSs remain insulating as their individual 2D components, in most cases a notable narrowing of the band gap is found.
The emergence of {unique} hybrid interface bands is demonstrated as an imprint of the interfacial bonding in the electronic properties.
Inspection of corresponding partial charge densities of these bands unambiguously reveals the delocalized character of these features exhibiting contributions on both 2D components.

The interfacial bonding { then also translates} into substantial magnetic coupling between the 2D components.
We observe both ferromagnetic and antiferromagnetic interfacial coupling with the details depending on the individual HS.
{The interplay between the electronic and magnetic interactions can be a particularly appealing tool to invoke strong multiferroic effects as required for magnetoelectric behavior.}

Finally, also an outlook towards twisted non-vdW HSs is provided as such systems can host new and enhanced effects.
Keeping the sizes tractable for density functional electronic structure calculations, a strong dependence of the binding energy and electronic as well as magnetic properties on the twist angle are found.
Different local atomic arrangements reveal moir{\'e} patterns due to stacking and twisting and thus lead to characteristic long-range variations of the (magnetisation) density above the HS surfaces -- readily observable with different experimental techniques.

{We
specifically suggest synthesis and experimental study of {(i) Cr$_2$O$_3$/Fe$_2$O$_3$ and CoTiO$_3$/Fe$_2$O$_3$ for their strong interfacial magnetic coupling, as well as (ii) Fe$_2$O$_3$/MgTiO$_3$ for its distinctive moir{\'e} patterns when twisted.}}

For unlocking and leveraging the versatility of non-vdW HSs, the present work goes beyond the study and design of individual (nanoscale) materials and moves towards the systematic data-driven design of functional interfaces.
Non-vdW HSs are a prime example of such a new class of interfaces that -- due to their unique interfacial bonding characteristics -- can host a rich variety of appealing effects and phenomena ranging from charge delocalisation over interfacial magnetic and multiferroic coupling down to correlated flatband physics.

\section*{Data Availability} \label{data_avail}
The primary research data of this study are available from the Rossendorf Data Repository (RODARE) via \verb|https://doi.org/10.14278/rodare.3621| \cite{Nihei_Rodare_2025}.

\section*{Acknowledgements}

The authors thank Agnieszka Kuc and Mani Lokamani for fruitful interactions.
R.F. acknowledges funding from the German Research Foundation (DFG), project FR 4545/2-1, the collaborative research center ``Chemistry of Synthetic Two-Dimensional Materials'' SFB-1415-417590517, and for the ``Autonomous Materials Thermodynamics'' (AutoMaT) project by Technische Universit\"at Dresden and Helmholtz-Zentrum Dresden-Rossendorf within the DRESDEN-concept alliance.
The authors thank the HZDR Computing Center, HLRS Stuttgart (HAWK cluster), the Paderborn Center for Parallel Computing (PC2, Noctua 2 cluster), and TU Dresden ZIH (Taurus cluster) for generous grants of CPU time.

\section*{Author contributions} \label{contribs}

R.F. initiated and supervised the project.
A.N. built the HSs and performed all \DFT\ calculations.
A.N., T.B., and R.F. analyzed the data and performed all post-processing.
A.N. wrote the initial draft of the manuscript which was then refined by R.F.
All authors contributed to the writing and discussion of the final manuscript.

\clearpage
\begin{figure}[H]
\centering
\includegraphics[width=0.7\columnwidth]{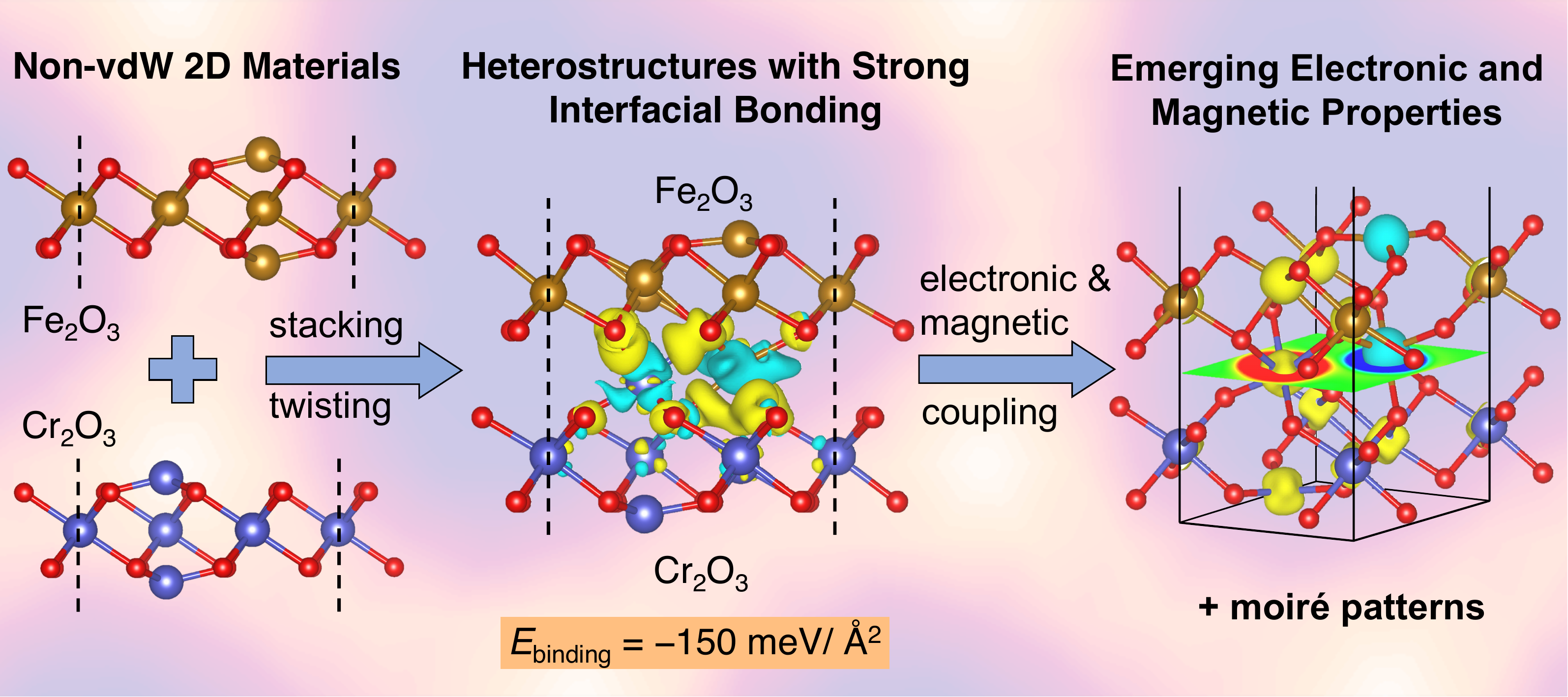}
\caption{\small \textbf{Graphical abstract.}}
\end{figure}
\clearpage


\section*{Supplementary material}

# Non-van der Waals Heterostructures
# Supporting Information

Anastasiia Nihei,[1,2] Tom Barnowsky,[1,2] and Rico Friedrich[1,2,3,*]

[1]*Theoretical Chemistry, Technische Universität Dresden, 01062 Dresden, Germany*
[2]*Institute of Ion Beam Physics and Materials Research,*
*Helmholtz-Zentrum Dresden-Rossendorf, 01328 Dresden, Germany*
[3]*Center for Extreme Materials, Duke University, Durham, NC 27708, USA*
(Dated: March 14, 2025)

## I. Charge density differences for non-vdW HSs

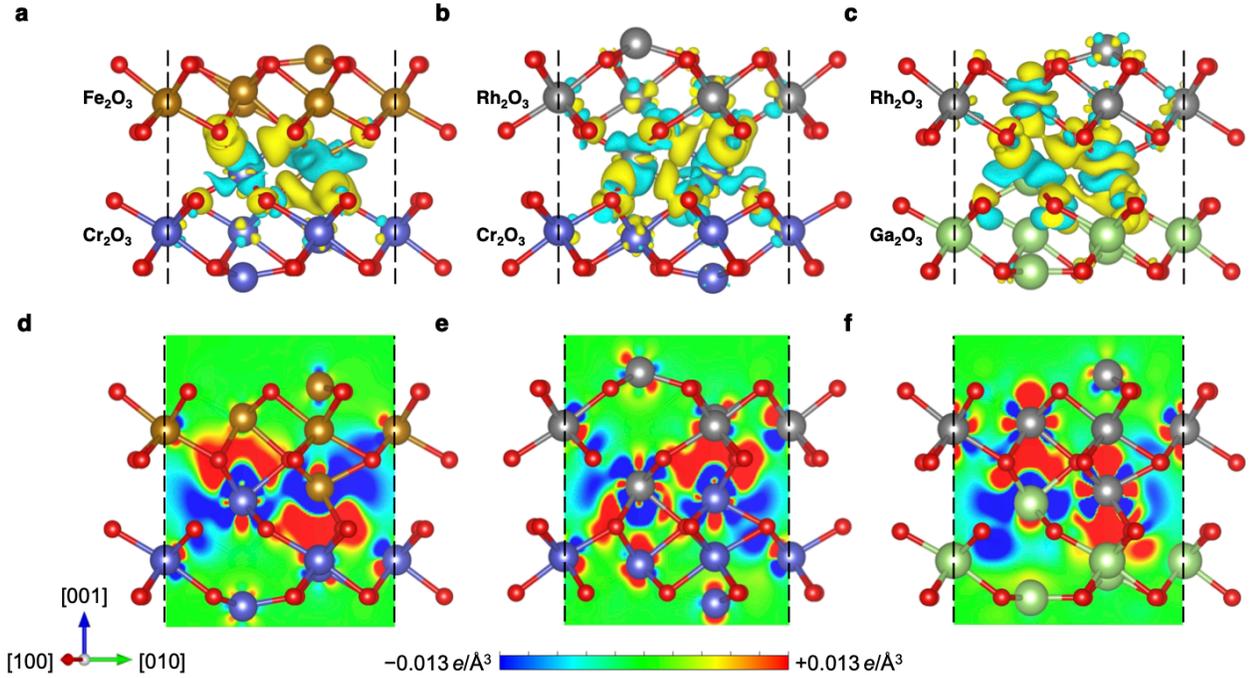

FIG. S1: Isosurface plots of the charge density difference for $Cr_2O_3/Fe_2O_3$ (**a**), $Cr_2O_3/Rh_2O_3$ (**b**), and $Ga_2O_3/Rh_2O_3$ (**c**) with charge gain/loss indicated by yellow/bluish parts (isosurface value: $\pm 0.027\ e/\text{Å}^3$). (**d-f**) Cut through the charge density difference on the (110) plane. The compass indicating crystal directions applies to all subfigures.

## II. Assessing the role of dispersion contributions

The influence of including a D3 [1] dispersion correction to PBE+$U$ [2–5] or employing the SCAN+rVV10 vdW functional [6, 7] on the optimized interfacial bond lengths and binding energies is benchmarked for thirteen HSs (Table I). As expected, for systems with a high binding energy from plain PBE+$U$ more negative than $-70$ meV/Å$^2$, the effect of dispersion is very small as they are already strongly bonded with changes in bond lengths of maximally 0.04 Å being negligible. The increase in binding energy appears to be rather large of the order of $-30$ to $-50$ meV/Å$^2$. This dispersion-induced contribution stems, however, from the fact that the 2D layers are not atomically thin as graphene and then the vdW interactions between several atoms add up but the net effect on the structure and even electronic properties is very minor.

On the other hand, slightly larger (relative) changes are found for systems with binding energies between $-20$ meV/Å$^2$ and $-70$ meV/Å$^2$ where both D3 inclusion and SCAN+rVV10 consistently cause bonds to shorten by typically more than $\sim$0.05 Å and binding energies increase by close to a factor of two. However, we still see no drastic change in the electronic properties.

Dispersion should be included for systems where plain DFT yields absolute binding energies below 20 meV/Å$^2$ – the typical energy scale of vdW interactions [8]. This is clearly the case for the seven HSs with the smallest absolute $E_b$ at the top of the table. For these systems, we also note that the absolute binding energy from plain PBE+$U$ increases by more than 100% for both dispersion approaches setting them apart from all other systems where the influence is weaker. The bond lengths shorten in almost all cases drastically by up to 0.68 Å for both vdW approaches and also the binding energies reach a proper negative dispersive energy range. A particularly special case is the Rh$_2$O$_3$/TiZnO$_3$ system where the small binding energy of $-10$ meV/Å$^2$ increases by a factor of six upon dispersion correction.

TABLE I: Comparison of interfacial bond lengths $d_1$ and $d_2$ as well as binding energies $E_b$ for different systems with and without dispersion correction (D3) as well as with SCAN+rVV10 functional.

| System | PBE(+$U$) | | | PBE(+$U$)+D3 | | | SCAN+rVV10 | | |
|---|---|---|---|---|---|---|---|---|---|
| | $d_1$ (Å) | $d_2$ (Å) | $E_b$ per Area (meV/Å$^2$) | $d_1$ (Å) | $d_2$ (Å) | $E_b$ per Area (meV/Å$^2$) | $d_1$ (Å) | $d_2$ (Å) | $E_b$ per Area (meV/Å$^2$) |
| CdTiO$_3$/SbTlO$_3$ | 4.32 | 3.50 | $-3$ | 4.13 | 3.32 | $-27$ | 3.64 | 3.02 | $-22$ |
| NaSbO$_3$/SbTlO$_3$ | 4.06 | 3.52 | $-4$ | 3.56 | 3.17 | $-22$ | 3.74 | 3.26 | $-15$ |
| KSbO$_3$/SnZnO$_3$ | 3.02 | 3.73 | $-10$ | 2.85 | 3.36 | $-33$ | 2.81 | 3.37 | $-29$ |
| Rh$_2$O$_3$/TiZnO$_3$ | 2.60 | 2.53 | $-10$ | 2.45 | 2.34 | $-64$ | 2.42 | 2.32 | $-60$ |
| AgSbO$_3$/SnZnO$_3$ | 2.78 | 3.19 | $-13$ | 2.65 | 2.98 | $-49$ | 2.66 | 2.96 | $-44$ |
| NaSbO$_3$/SnZnO$_3$ | 2.70 | 3.10 | $-16$ | 2.63 | 2.99 | $-47$ | 2.54 | 2.87 | $-40$ |
| AgSbO$_3$/KSbO$_3$ | 2.90 | 2.86 | $-19$ | 2.86 | 2.82 | $-45$ | 2.78 | 2.77 | $-42$ |
| KSbO$_3$/NaSbO$_3$ | 2.85 | 2.93 | $-22$ | 2.81 | 2.85 | $-46$ | 2.75 | 2.78 | $-38$ |
| AgSbO$_3$/NaSbO$_3$ | 2.73 | 2.73 | $-25$ | 2.69 | 2.68 | $-54$ | 2.65 | 2.64 | $-47$ |
| CdTiO$_3$/NaSbO$_3$ | 2.80 | 2.53 | $-37$ | 2.74 | 2.48 | $-72$ | 2.67 | 2.41 | $-67$ |
| Ga$_2$O$_3$/GeMgO$_3$ | 2.41 | 2.23 | $-71$ | 2.39 | 2.21 | $-108$ | 2.38 | 2.21 | $-110$ |
| CaSnO$_3$/In$_2$O$_3$ | 2.43 | 2.52 | $-79$ | 2.41 | 2.50 | $-108$ | 2.40 | 2.48 | $-119$ |
| MgSiO$_3$/SiZnO$_3$ | 2.22 | 2.25 | $-97$ | 2.19 | 2.22 | $-145$ | 2.20 | 2.25 | $-144$ |

## III. Band structures and densities of states of HSs in comparison to their 2D components as well as band structures with projection contributions

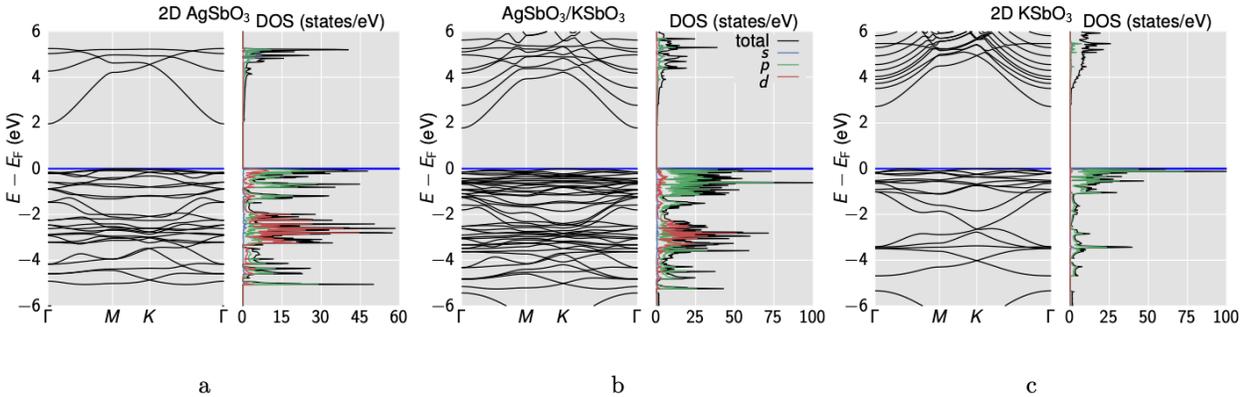

FIG. S2: Band structure and DOS for the individual 2D (a) AgSbO$_3$, (c) KSbO$_3$ systems, and (b) the HS. The energies are aligned at the respective Fermi energy $E_F$. Note that the band structures are computed for geometries optimized with D3 dispersion correction.

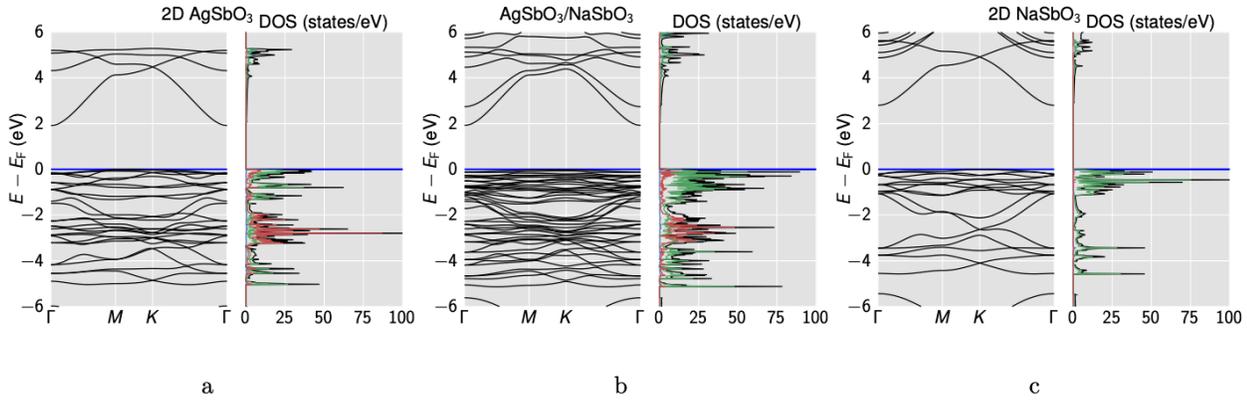

FIG. S3: Band structure and DOS for the individual 2D (a) AgSbO$_3$, (c) NaSbO$_3$ systems, and (b) the HS. The energies are aligned at the respective Fermi energy $E_{\mathrm{F}}$.

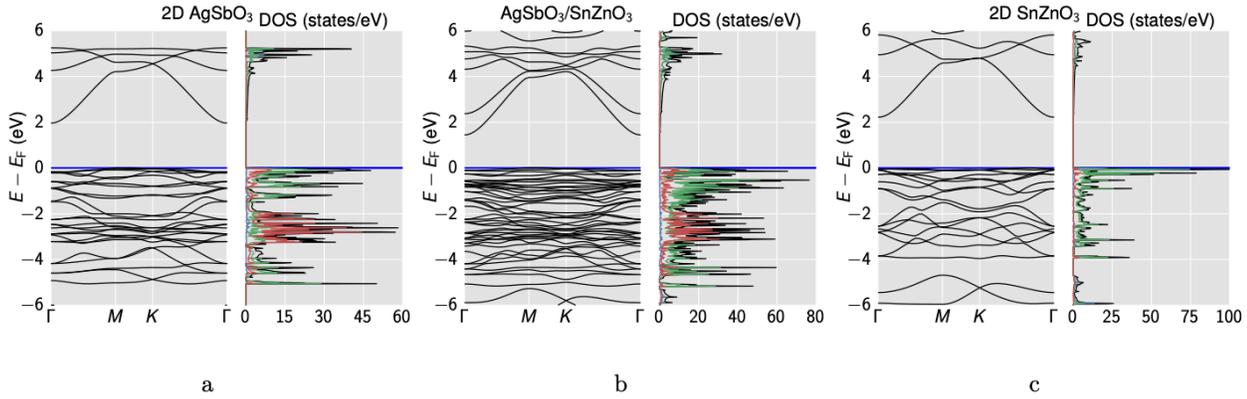

FIG. S4: Band structure and DOS for the individual 2D (a) AgSbO$_3$, (c) SnZnO$_3$ systems, and (b) the HS. The energies are aligned at the respective Fermi energy $E_{\mathrm{F}}$. Note that the band structures are computed for geometries optimized with D3 dispersion correction.

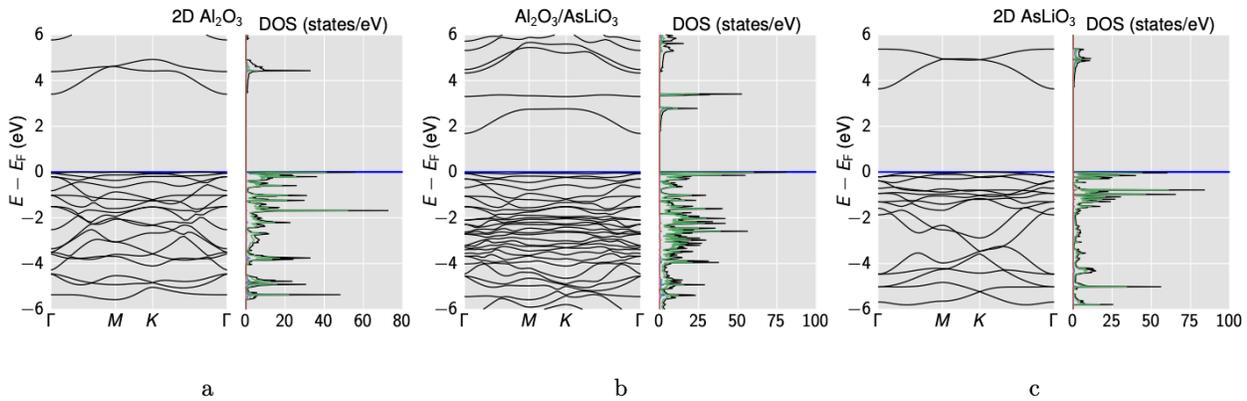

FIG. S5: Band structure and DOS for the individual 2D (a) Al$_2$O$_3$, (c) AsLiO$_3$ systems, and (b) the HS. The energies are aligned at the respective Fermi energy $E_{\mathrm{F}}$.

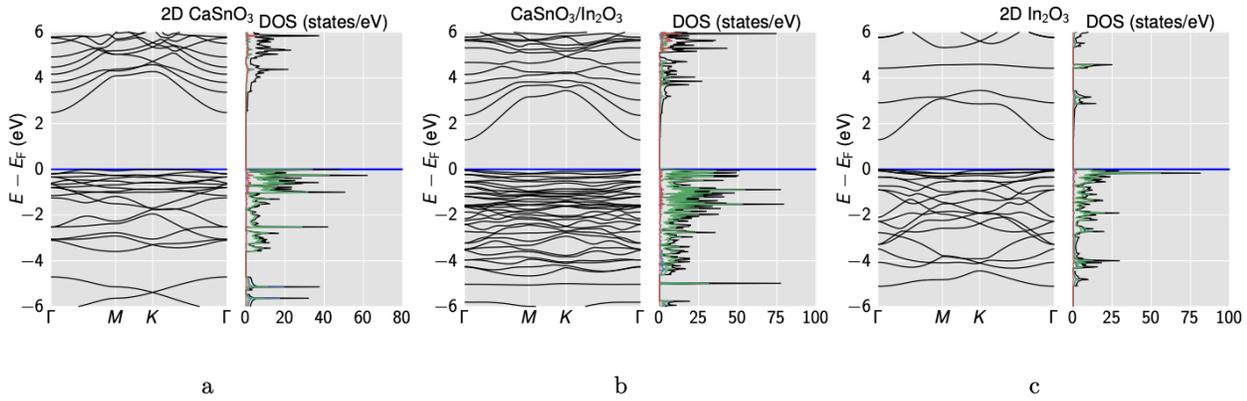

FIG. S6: Band structure and DOS for the individual 2D (a) CaSnO$_3$, (c) In$_2$O$_3$ systems, and (b) the HS. The energies are aligned at the respective Fermi energy $E_F$.

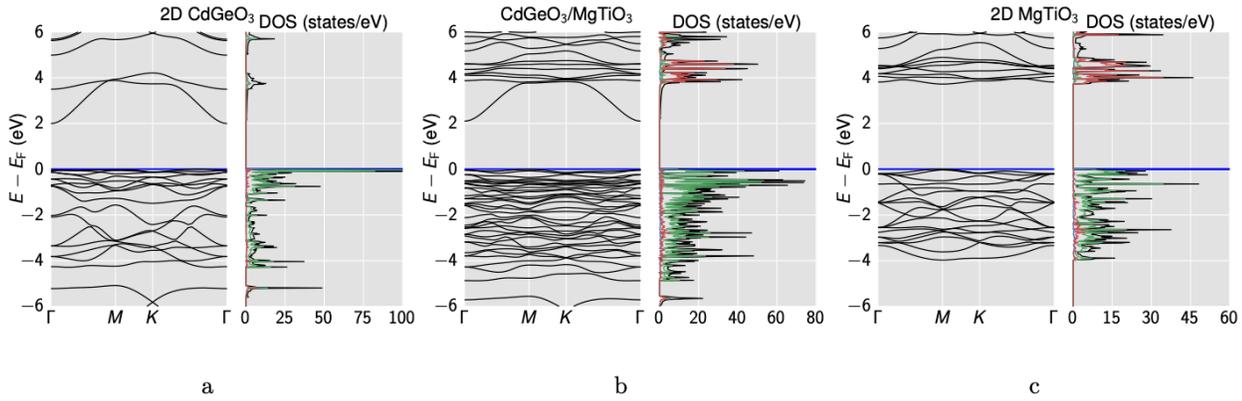

FIG. S7: Band structure and DOS for the individual 2D (a) CdGeO$_3$, (c) MgTiO$_3$ systems, and (b) the HS. The energies are aligned at the respective Fermi energy $E_F$.

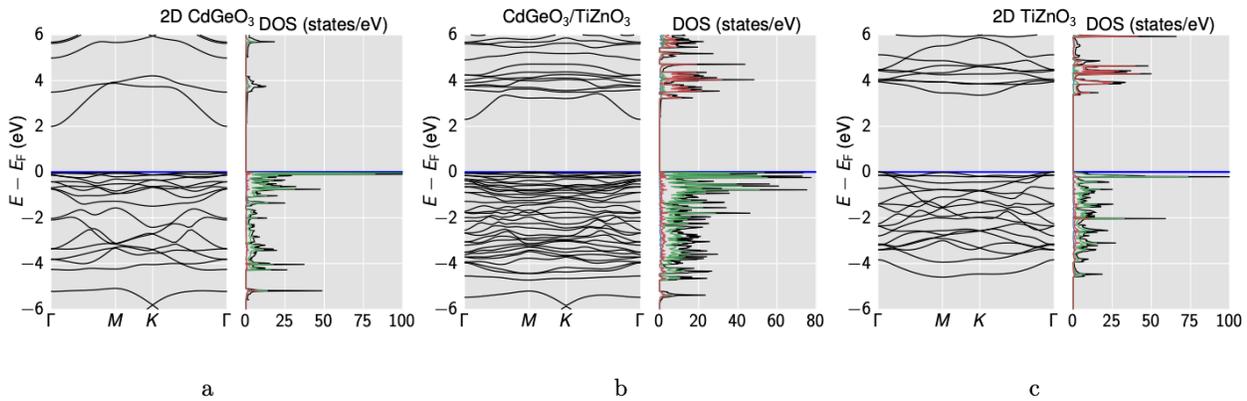

FIG. S8: Band structure and DOS for the individual 2D (a) CdGeO$_3$, (c) TiZnO$_3$ systems, and (b) the HS. The energies are aligned at the respective Fermi energy $E_F$.

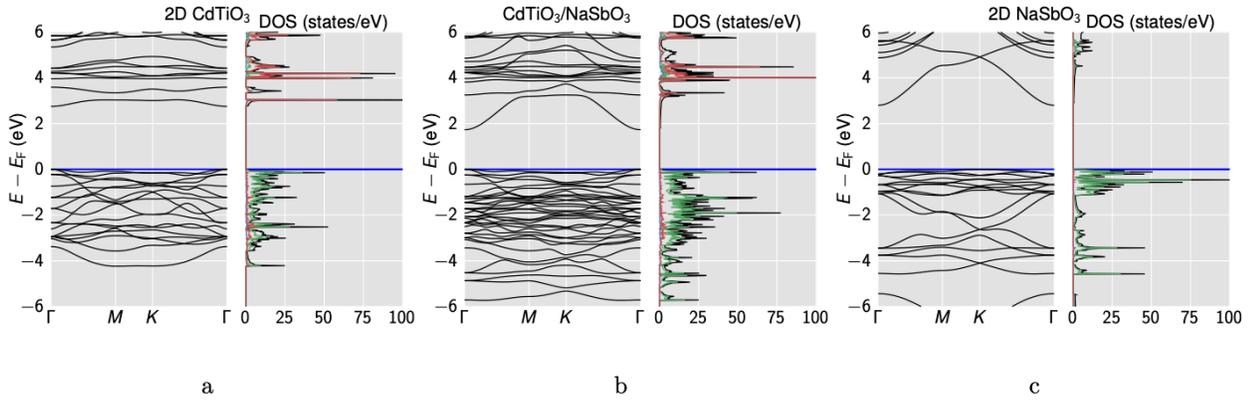

FIG. S9: Band structure and DOS for the individual 2D (a) CdTiO$_3$, (c) NaSbO$_3$ systems, and (b) the HS. The energies are aligned at the respective Fermi energy $E_\mathrm{F}$.

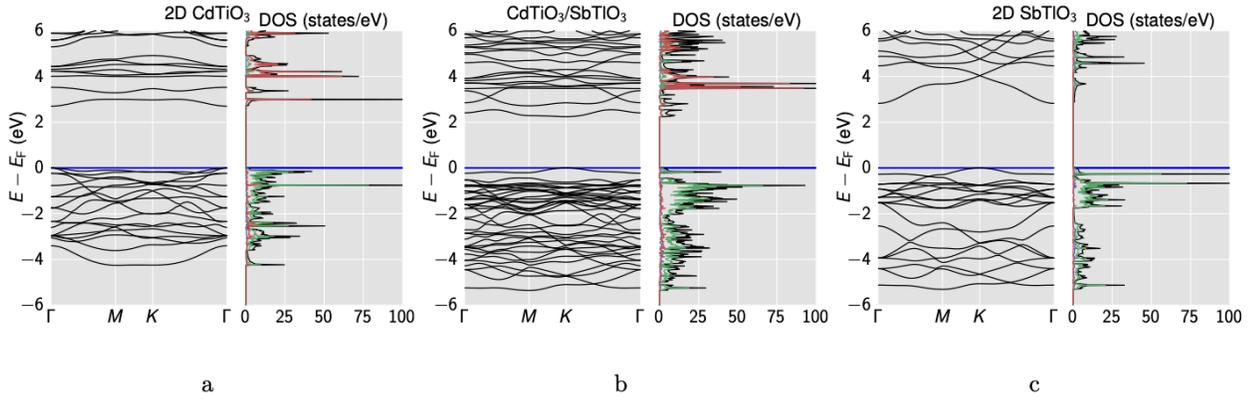

FIG. S10: Band structure and DOS for the individual 2D (a) CdTiO$_3$, (c) SbTlO$_3$ systems, and (b) the HS. The energies are aligned at the respective Fermi energy $E_\mathrm{F}$. Note that the band structures are computed for geometries optimized with D3 dispersion correction.

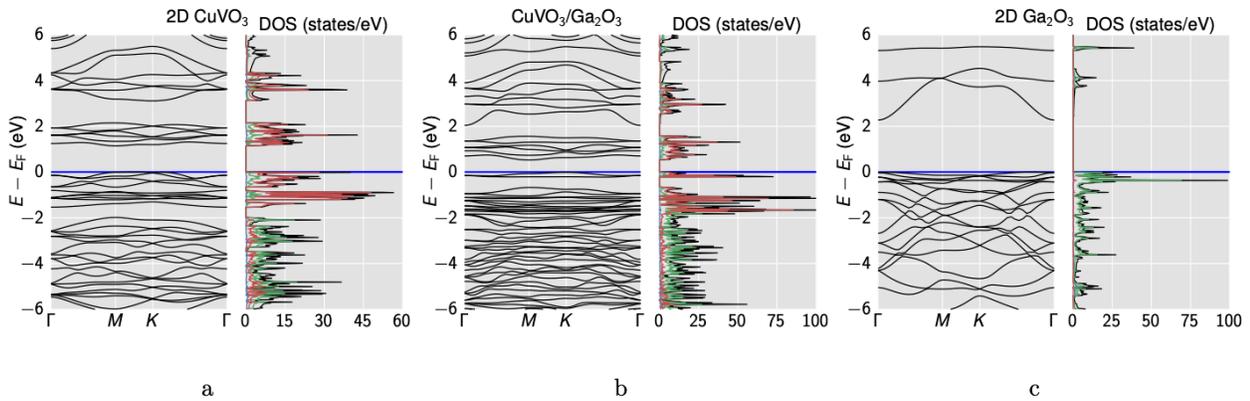

FIG. S11: Band structure and DOS for the individual 2D (a) CuVO$_3$, (c) Ga$_2$O$_3$ systems, and (b) the HS. The energies are aligned at the respective Fermi energy $E_\mathrm{F}$.

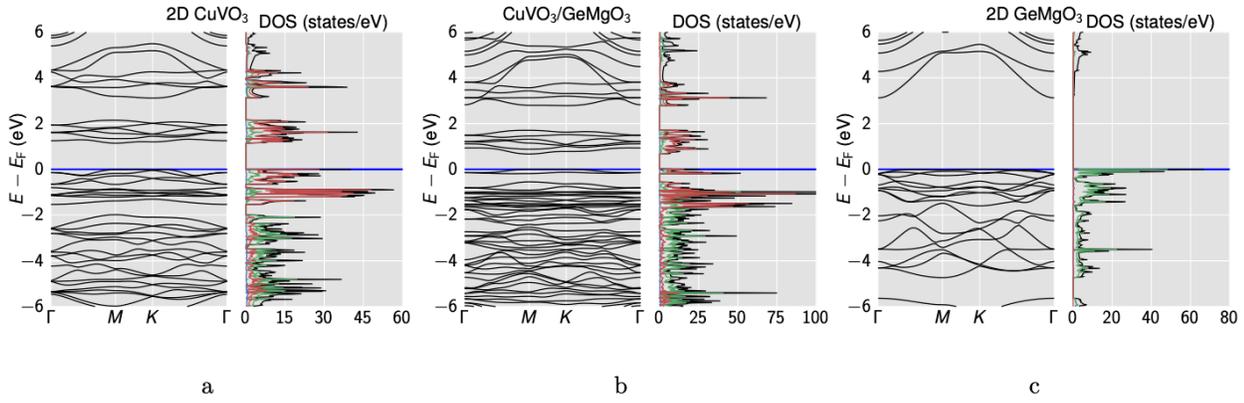

FIG. S12: Band structure and DOS for the individual 2D (a) CuVO$_3$, (c) GeMgO$_3$ systems, and (b) the HS. The energies are aligned at the respective Fermi energy $E_{\mathrm{F}}$.

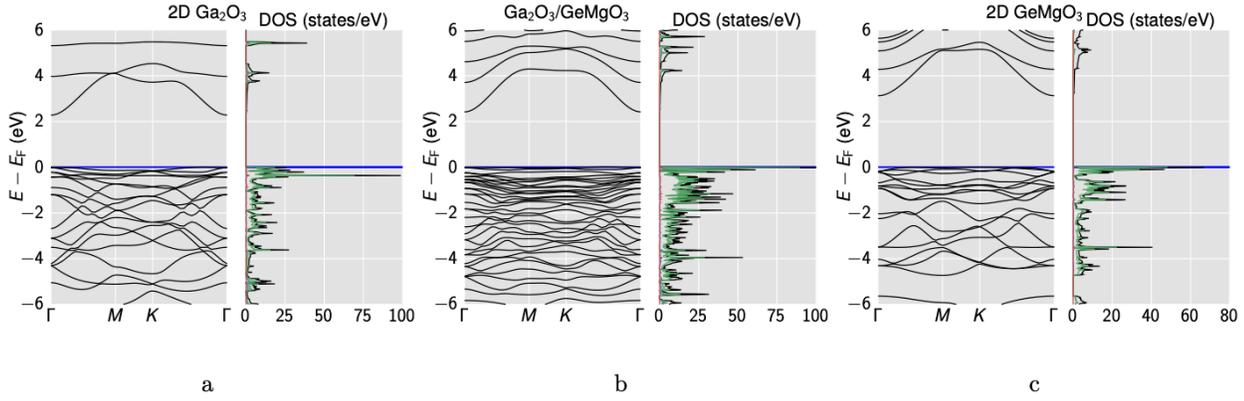

FIG. S13: Band structure and DOS for the individual 2D (a) Ga$_2$O$_3$, (c) GeMgO$_3$ systems, and (b) the HS. The energies are aligned at the respective Fermi energy $E_{\mathrm{F}}$.

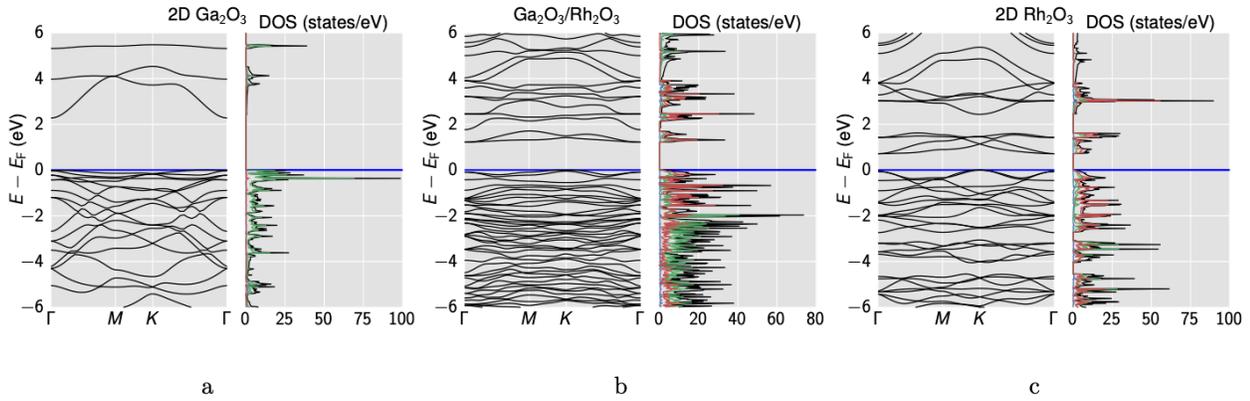

FIG. S14: Band structure and DOS for the individual 2D (a) Ga$_2$O$_3$, (c) Rh$_2$O$_3$ systems, and (b) the HS. The energies are aligned at the respective Fermi energy $E_{\mathrm{F}}$.

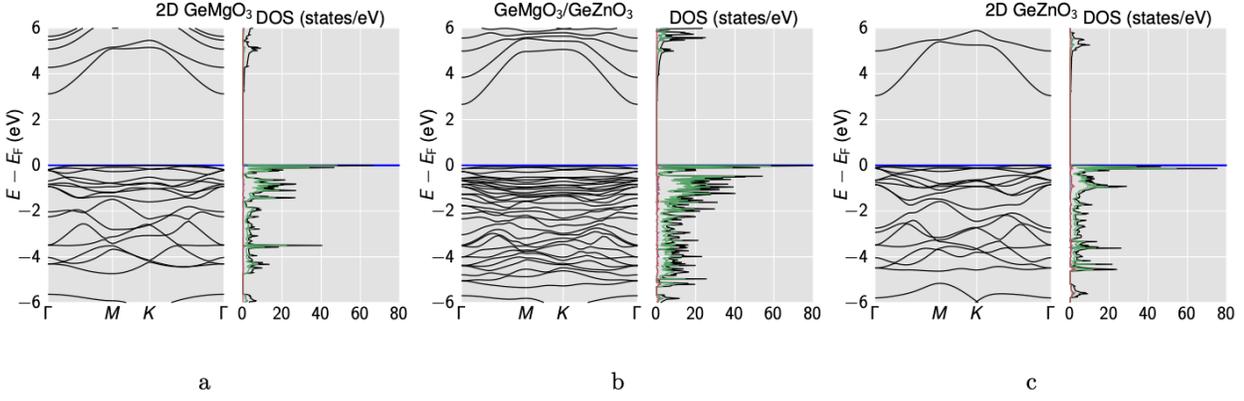

FIG. S15: Band structure and DOS for the individual 2D (a) GeMgO$_3$, (c) GeZnO$_3$ systems, and (b) the HS. The energies are aligned at the respective Fermi energy $E_F$.

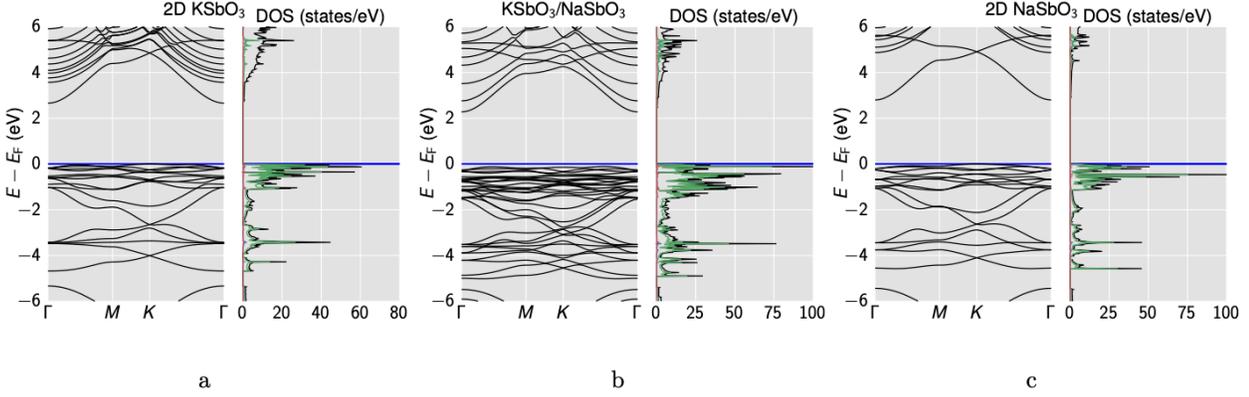

FIG. S16: Band structure and DOS for the individual 2D (a) KSbO$_3$, (c) NaSbO$_3$ systems, and (b) the HS. The energies are aligned at the respective Fermi energy $E_F$.

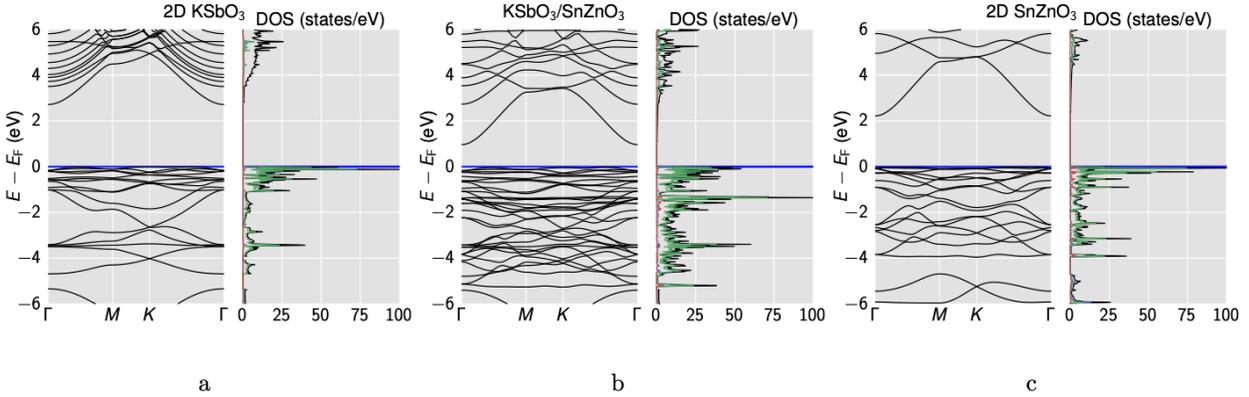

FIG. S17: Band structure and DOS for the individual 2D (a) KSbO$_3$, (c) SnZnO$_3$ systems, and (b) the HS. The energies are aligned at the respective Fermi energy $E_F$. Note that the band structures are computed for geometries optimized with D3 dispersion correction.

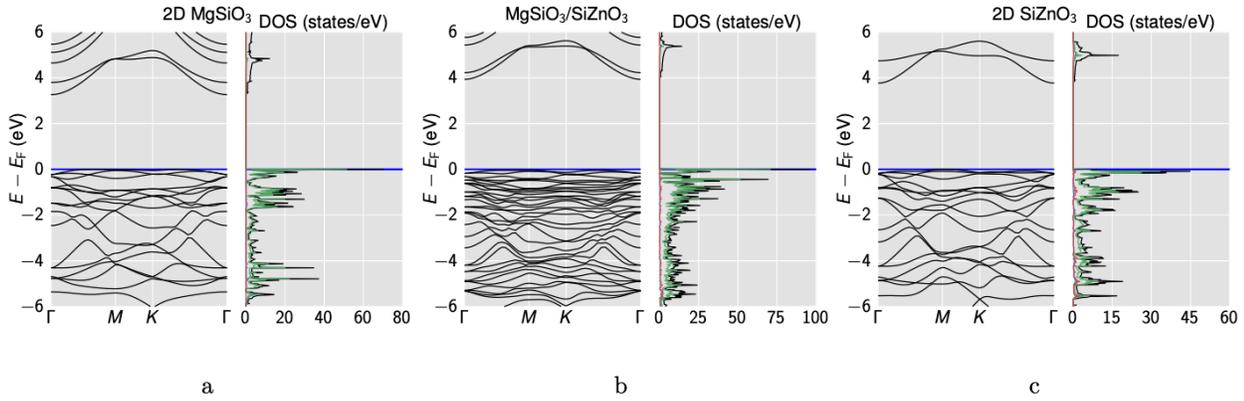

FIG. S18: Band structure and DOS for the individual 2D (a) MgSiO$_3$, (c) SiZnO$_3$ systems, and (b) the HS. The energies are aligned at the respective Fermi energy $E_\mathrm{F}$.

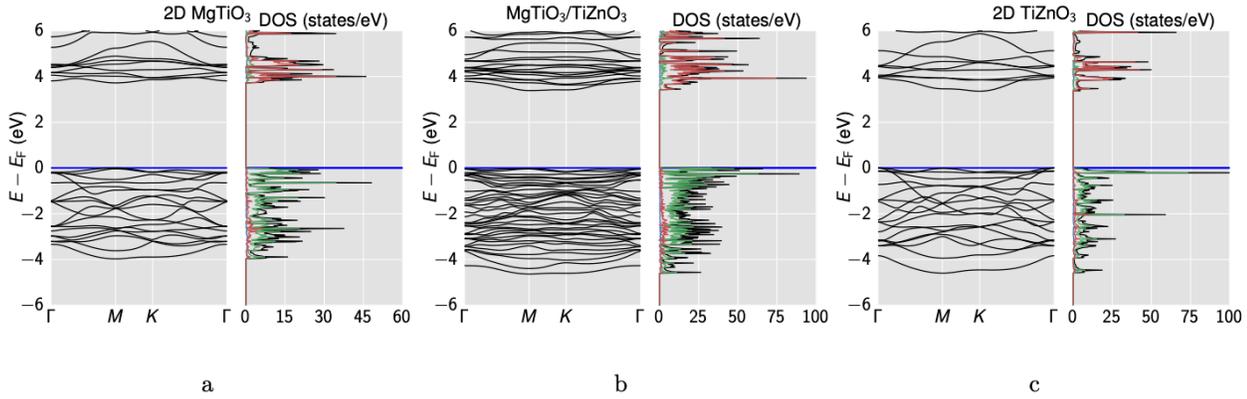

FIG. S19: Band structure and DOS for the individual 2D (a) MgTiO$_3$, (c) TiZnO$_3$ systems, and (b) the HS. The energies are aligned at the respective Fermi energy $E_\mathrm{F}$.

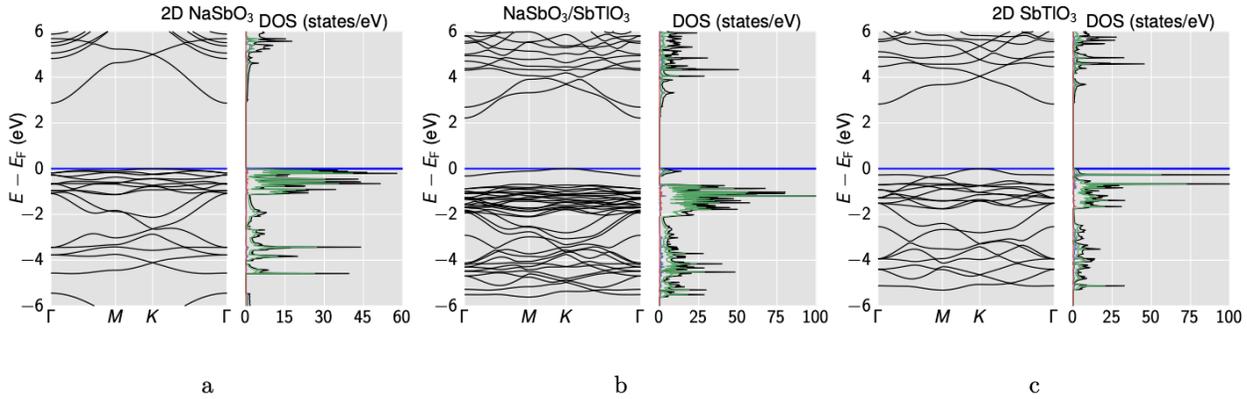

FIG. S20: Band structure and DOS for the individual 2D (a) NaSbO$_3$, (c) SbTlO$_3$ systems, and (b) the HS. The energies are aligned at the respective Fermi energy $E_\mathrm{F}$. Note that the band structures are computed for geometries optimized with D3 dispersion correction.

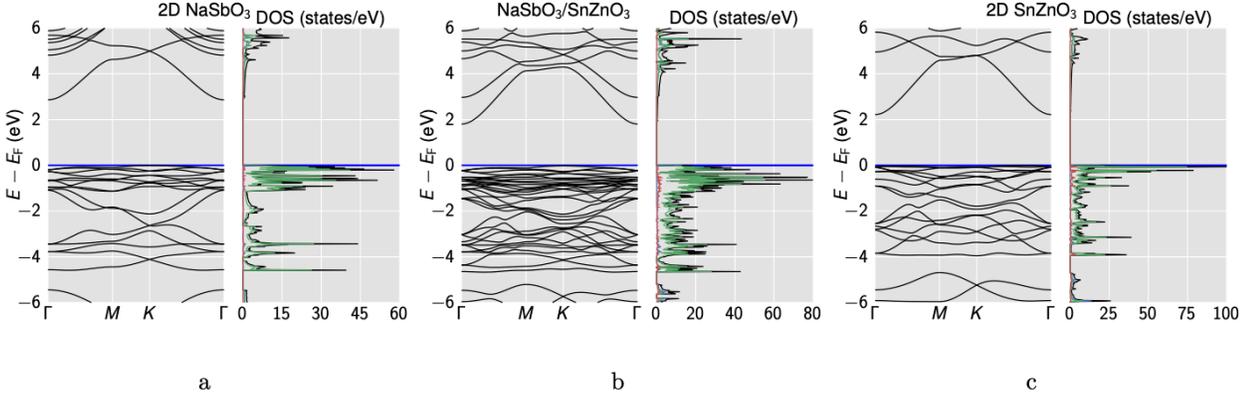

FIG. S21: Band structure and DOS for the individual 2D (a) NaSbO$_3$, (c) SnZnO$_3$ systems, and (b) the HS. The energies are aligned at the respective Fermi energy $E_F$. Note that the band structures are computed for geometries optimized with D3 dispersion correction.

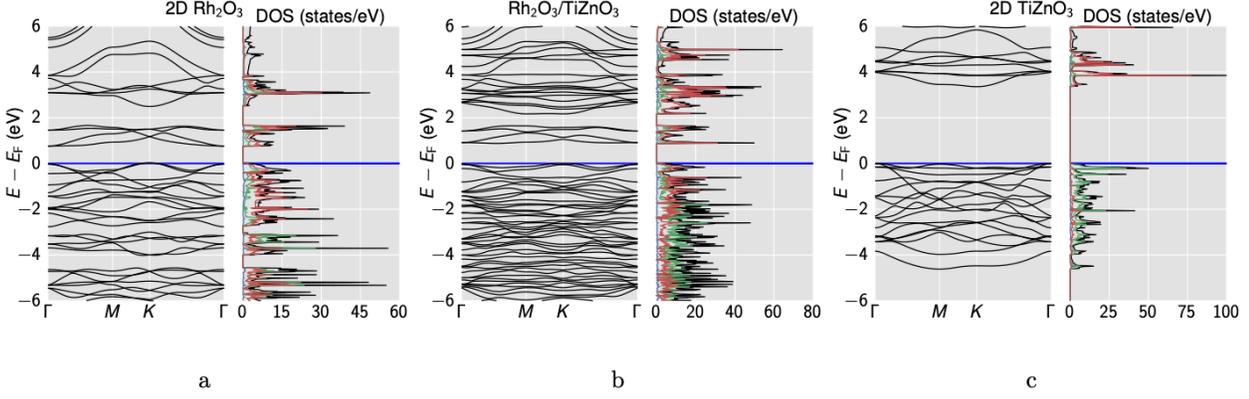

FIG. S22: Band structure and DOS for the individual 2D (a) Rh$_2$O$_3$, (c) TiZnO$_3$ systems, and (b) the HS. The energies are aligned at the respective Fermi energy $E_F$. Note that the band structures are computed for geometries optimized with D3 dispersion correction.

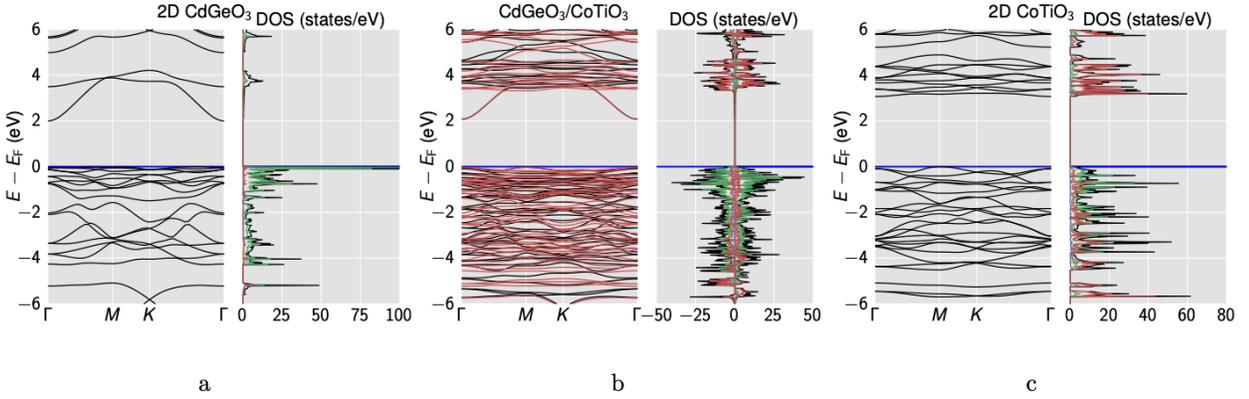

FIG. S23: Band structure and DOS for the individual 2D (a) CdGeO$_3$, (c) CoTiO$_3$ systems, and (b) the HS. The energies are aligned at the respective Fermi energy $E_F$. For the spin polarized band structure, majority spin bands (positive DOS) are indicated in black while minority spin bands (negative DOS) are in red.

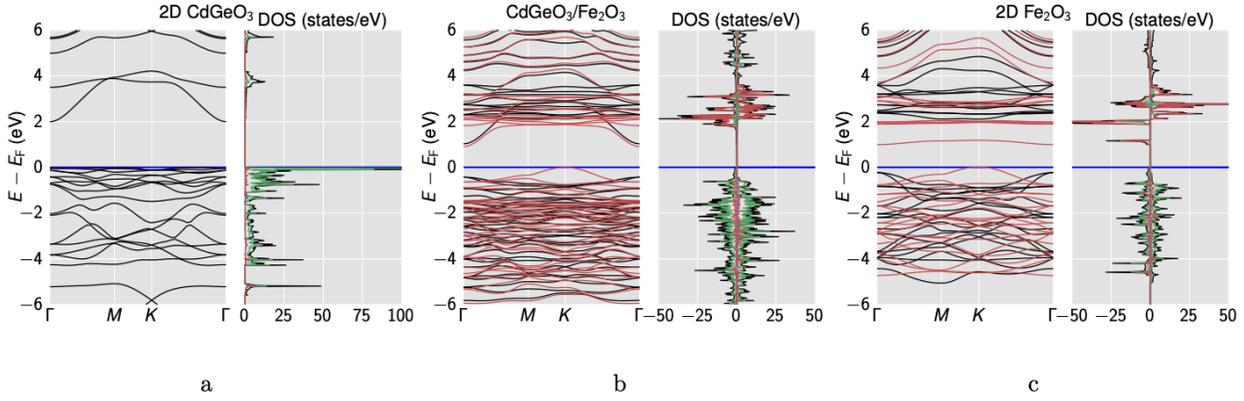

a

b

c

FIG. S24: Band structure and DOS for the individual 2D (a) CdGeO$_3$, (c) Fe$_2$O$_3$ systems, and (b) the HS. The energies are aligned at the respective Fermi energy $E_\mathrm{F}$.

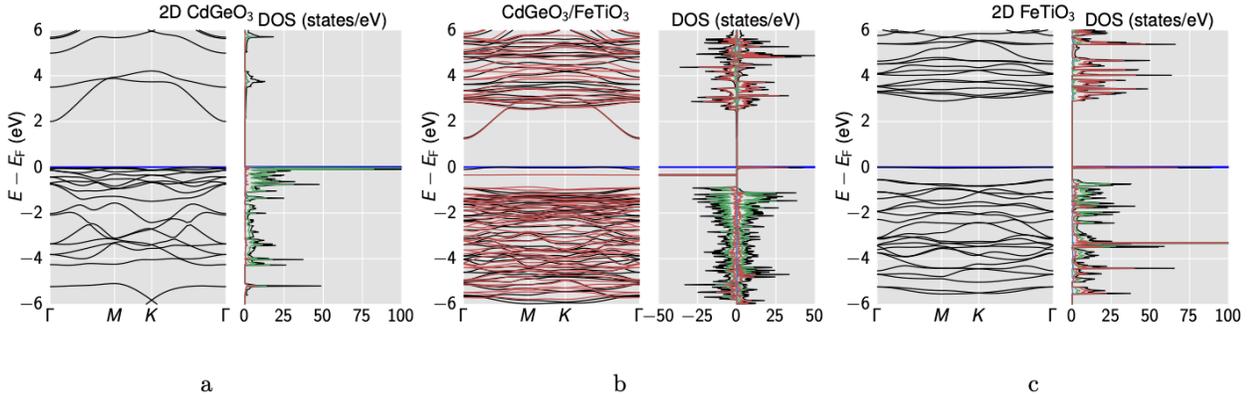

a

b

c

FIG. S25: Band structure and DOS for the individual 2D (a) CdGeO$_3$, (c) FeTiO$_3$ systems, and (b) the HS. The energies are aligned at the respective Fermi energy $E_\mathrm{F}$.

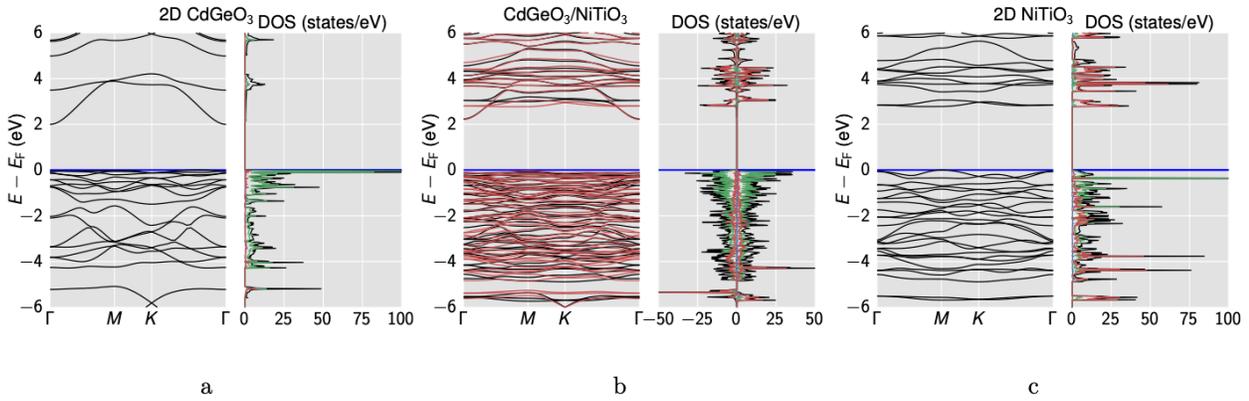

a

b

c

FIG. S26: Band structure and DOS for the individual 2D (a) CdGeO$_3$, (c) NiTiO$_3$ systems, and (b) the HS. The energies are aligned at the respective Fermi energy $E_\mathrm{F}$.

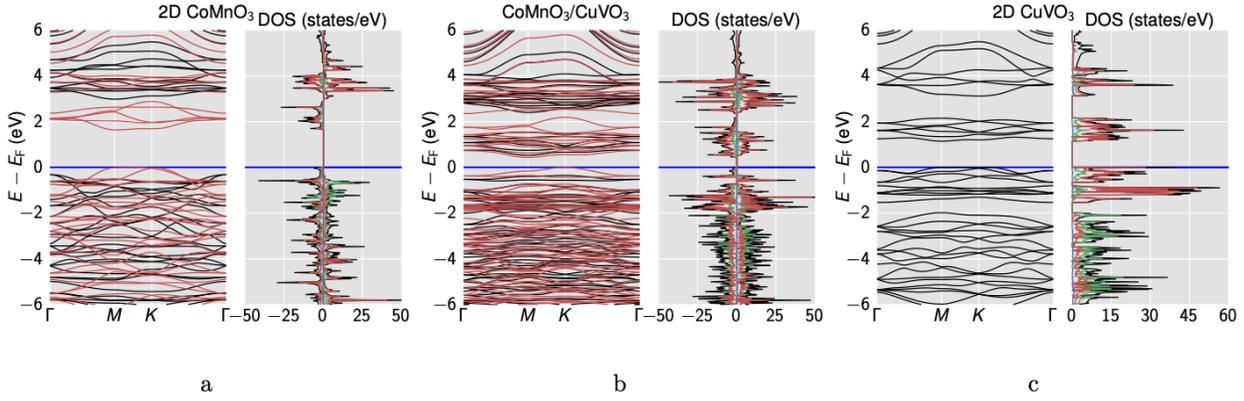

FIG. S27: Band structure and DOS for the individual 2D (a) CoMnO$_3$, (c) CuVO$_3$ systems, and (b) the HS. The energies are aligned at the respective Fermi energy $E_{\mathrm{F}}$.

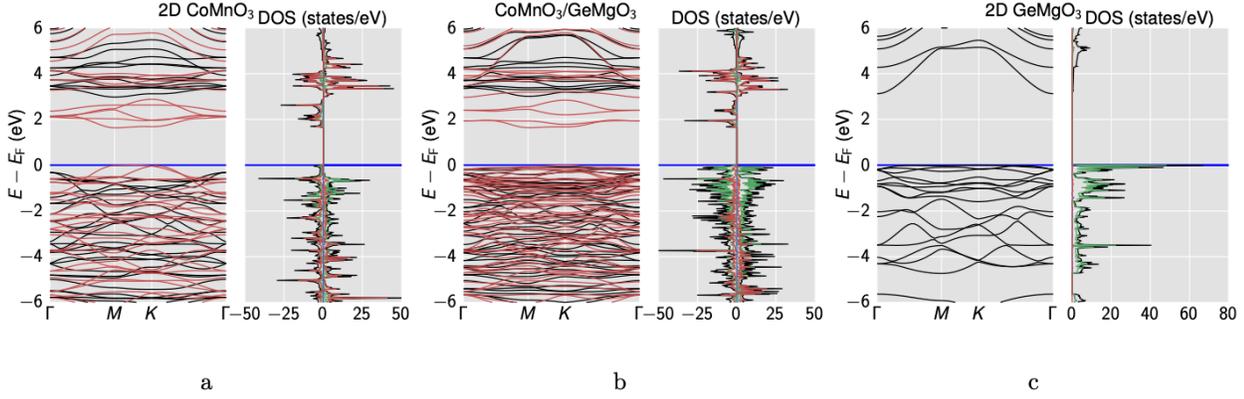

FIG. S28: Band structure and DOS for the individual 2D (a) CoMnO$_3$, (c) GeMgO$_3$ systems, and (b) the HS. The energies are aligned at the respective Fermi energy $E_{\mathrm{F}}$.

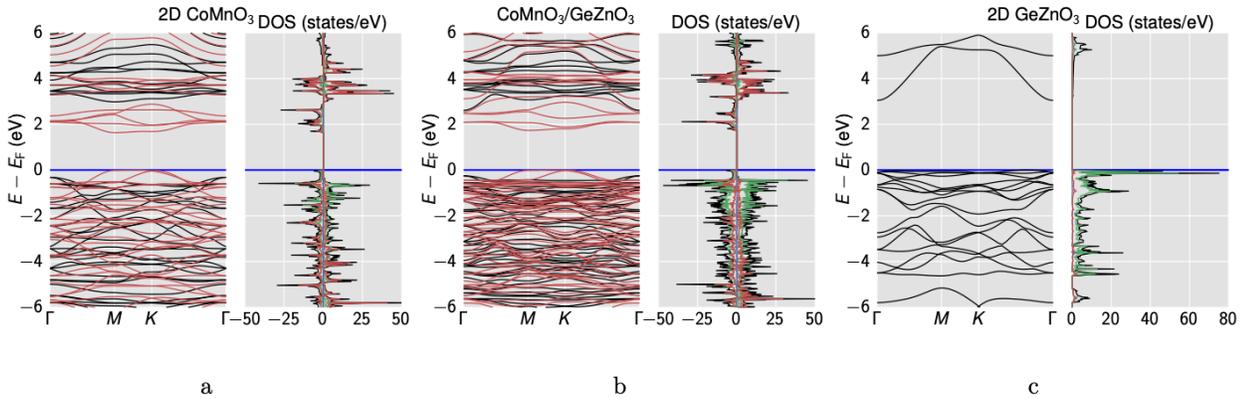

FIG. S29: Band structure and DOS for the individual 2D (a) CoMnO$_3$, (c) GeZnO$_3$ systems, and (b) the HS. The energies are aligned at the respective Fermi energy $E_{\mathrm{F}}$.

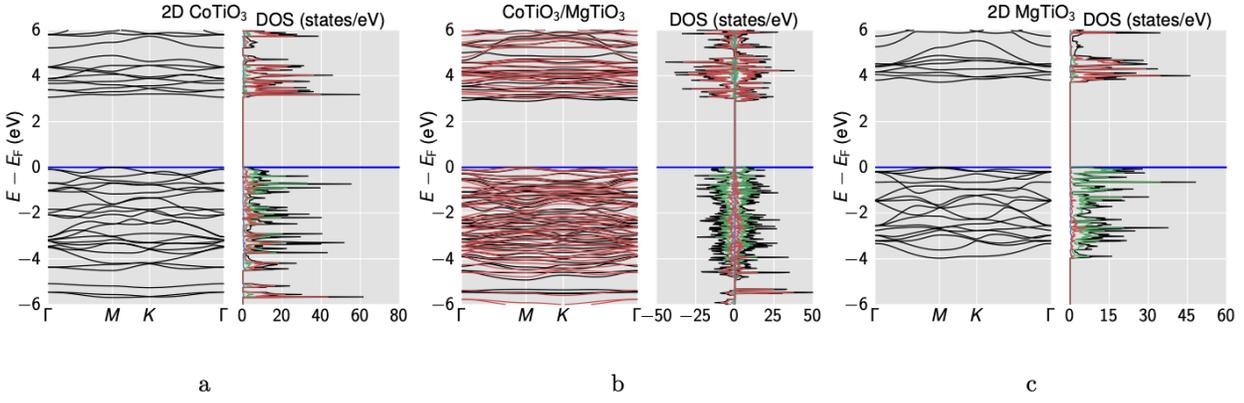

FIG. S30: Band structure and DOS for the individual 2D (a) CoTiO$_3$, (c) MgTiO$_3$ systems, and (b) the HS. The energies are aligned at the respective Fermi energy $E_\mathrm{F}$.

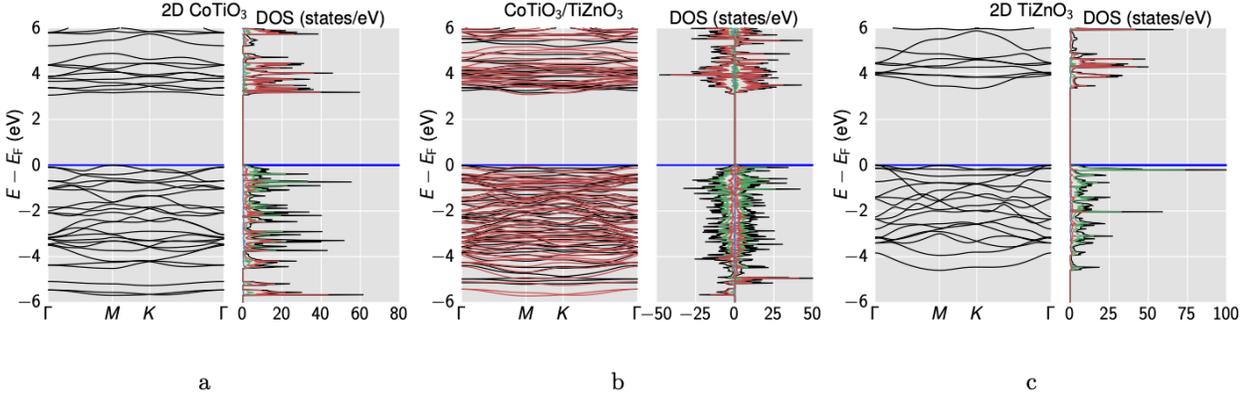

FIG. S31: Band structure and DOS for the individual 2D (a) CoTiO$_3$, (c) TiZnO$_3$ systems, and (b) the HS. The energies are aligned at the respective Fermi energy $E_\mathrm{F}$.

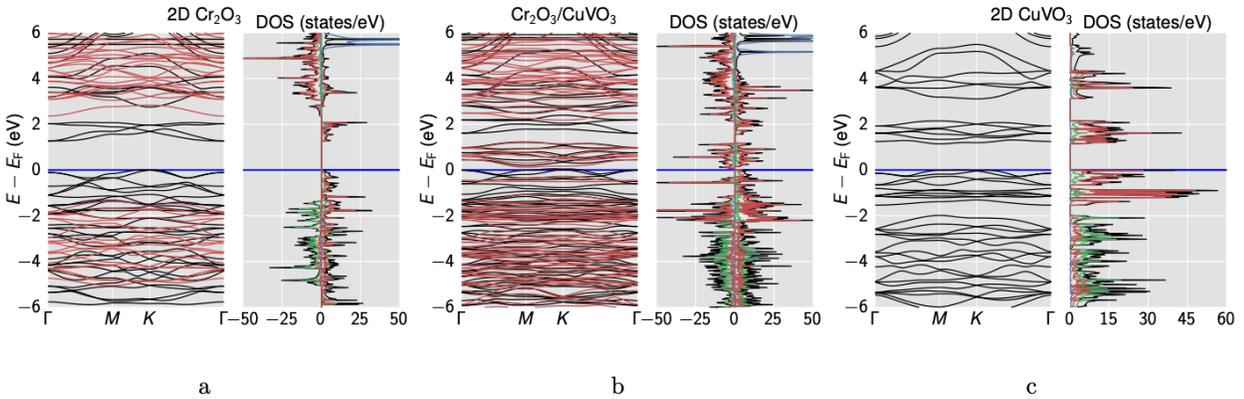

FIG. S32: Band structure and DOS for the individual 2D (a) Cr$_2$O$_3$, (c) CuVO$_3$ systems, and (b) the HS. The energies are aligned at the respective Fermi energy $E_\mathrm{F}$.

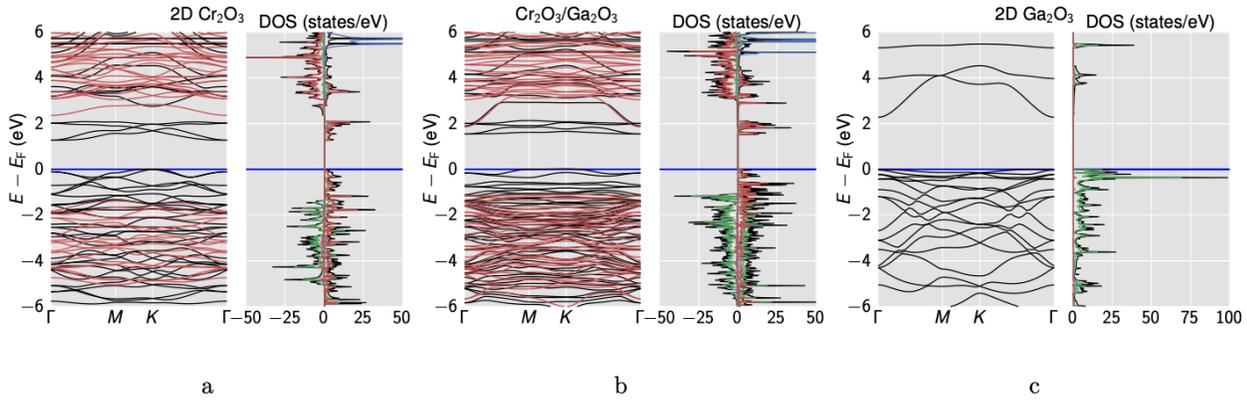

FIG. S33: Band structure and DOS for the individual 2D (a) $Cr_2O_3$, (c) $Ga_2O_3$ systems, and (b) the HS. The energies are aligned at the respective Fermi energy $E_F$.

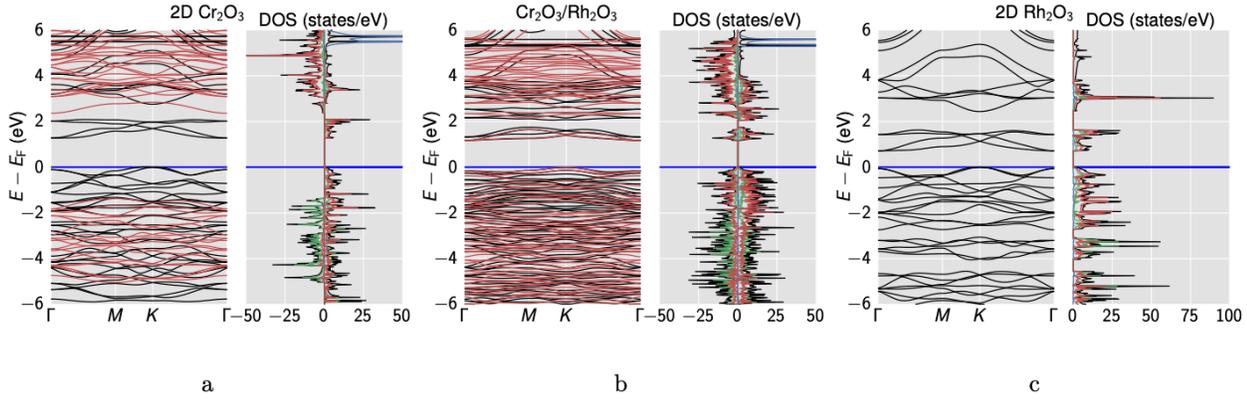

FIG. S34: Band structure and DOS for the individual 2D (a) $Cr_2O_3$, (c) $Rh_2O_3$ systems, and (b) the HS. The energies are aligned at the respective Fermi energy $E_F$.

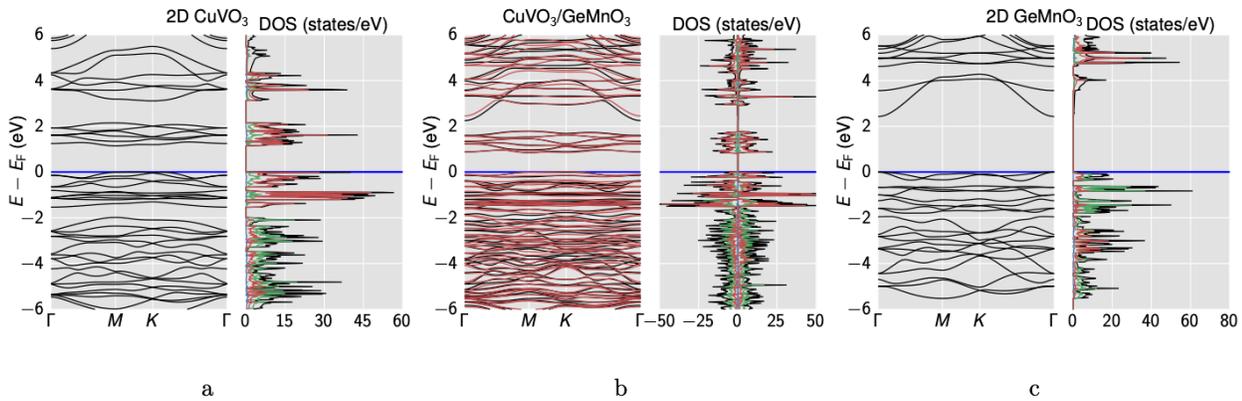

FIG. S35: Band structure and DOS for the individual 2D (a) $CuVO_3$, (c) $GeMnO_3$ systems, and (b) the HS. The energies are aligned at the respective Fermi energy $E_F$.

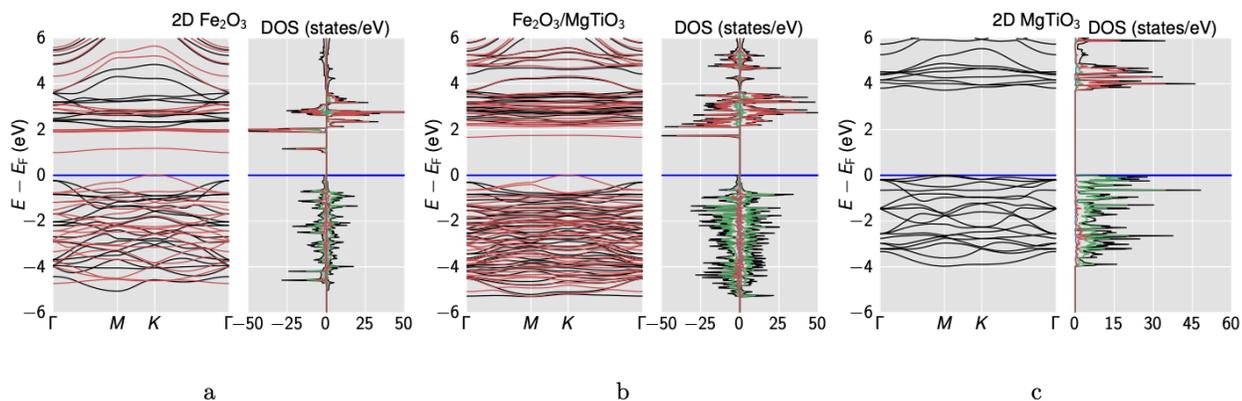

FIG. S36: Band structure and DOS for the individual 2D (a) $Fe_2O_3$, (c) $MgTiO_3$ systems, and (b) the HS. The energies are aligned at the respective Fermi energy $E_F$.

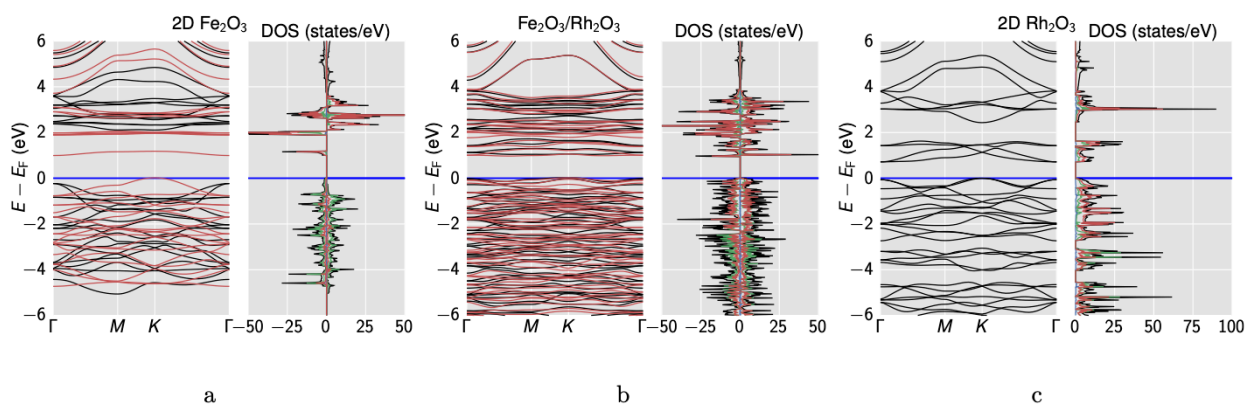

FIG. S37: Band structure and DOS for the individual 2D (a) $Fe_2O_3$, (c) $Rh_2O_3$ systems, and (b) the HS. The energies are aligned at the respective Fermi energy $E_F$.

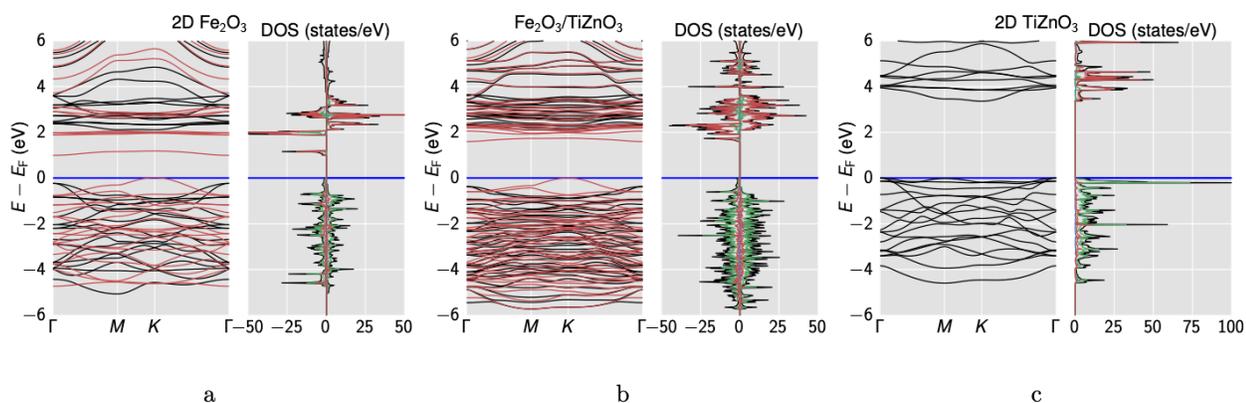

FIG. S38: Band structure and DOS for the individual 2D (a) $Fe_2O_3$, (c) $TiZnO_3$ systems, and (b) the HS. The energies are aligned at the respective Fermi energy $E_F$.

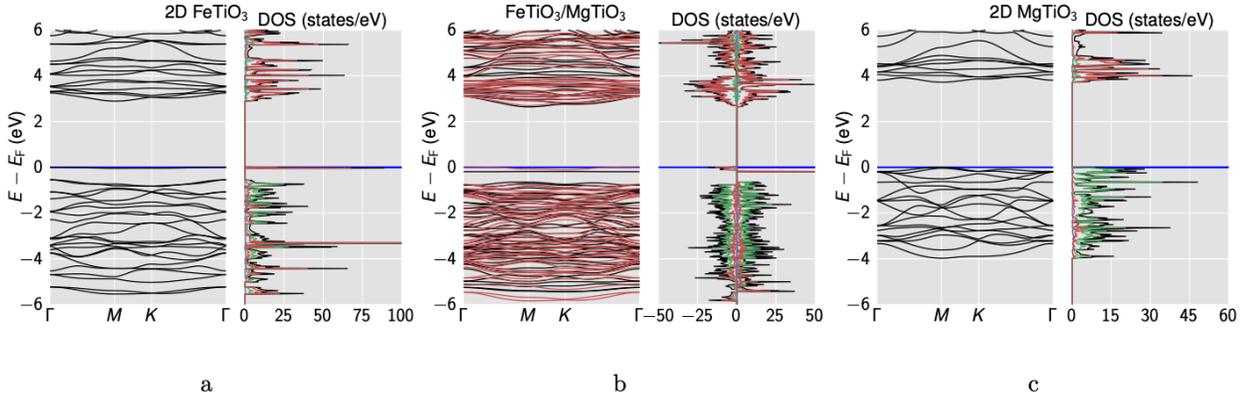

FIG. S39: Band structure and DOS for the individual 2D (a) FeTiO$_3$, (c) MgTiO$_3$ systems, and (b) the HS. The energies are aligned at the respective Fermi energy $E_{\mathrm{F}}$.

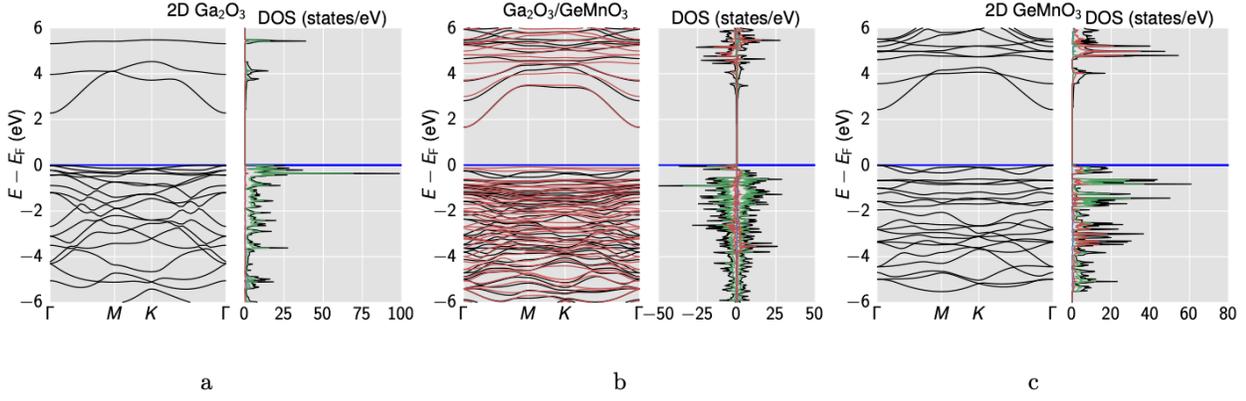

FIG. S40: Band structure and DOS for the individual 2D (a) Ga$_2$O$_3$, (c) GeMnO$_3$ systems, and (b) the HS. The energies are aligned at the respective Fermi energy $E_{\mathrm{F}}$.

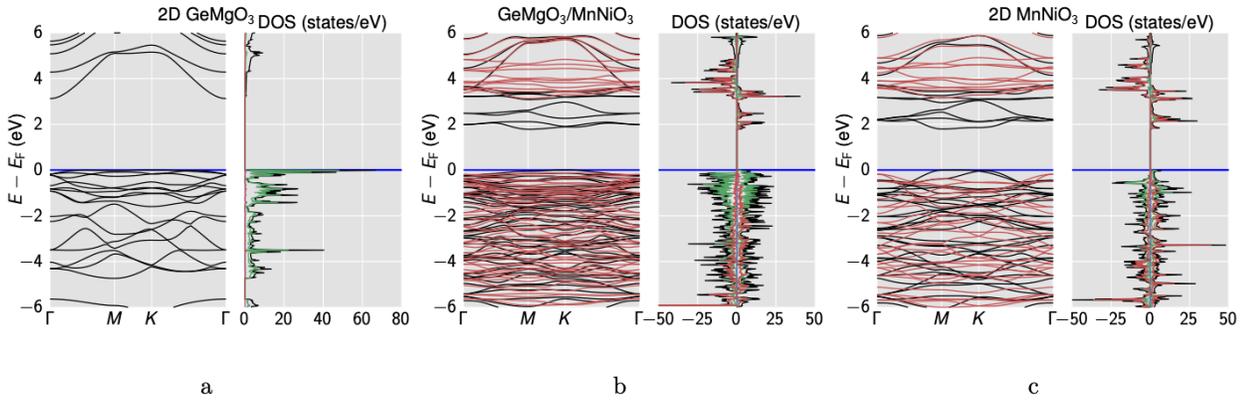

FIG. S41: Band structure and DOS for the individual 2D (a) GeMgO$_3$, (c) MnNiO$_3$ systems, and (b) the HS. The energies are aligned at the respective Fermi energy $E_{\mathrm{F}}$.

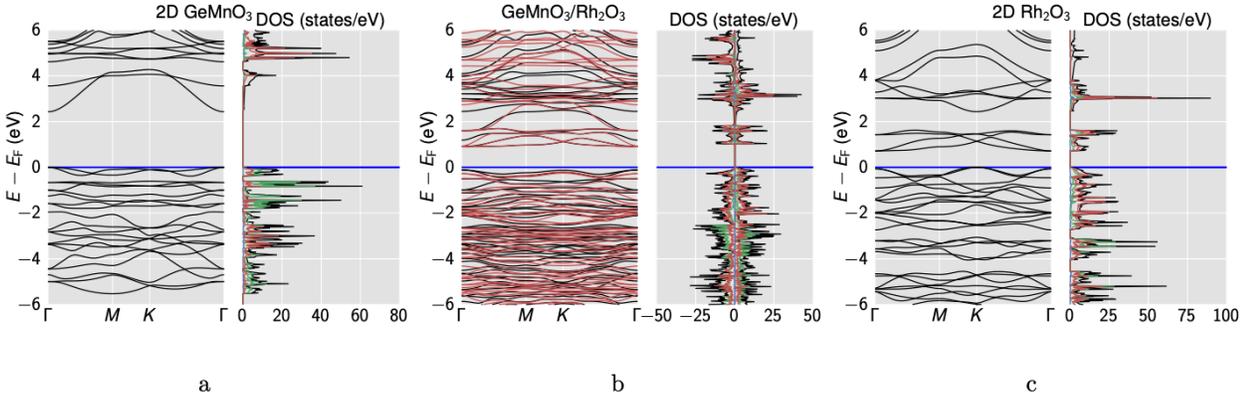

FIG. S42: Band structure and DOS for the individual 2D (a) GeMnO$_3$, (c) Rh$_2$O$_3$ systems, and (b) the HS. The energies are aligned at the respective Fermi energy $E_\mathrm{F}$.

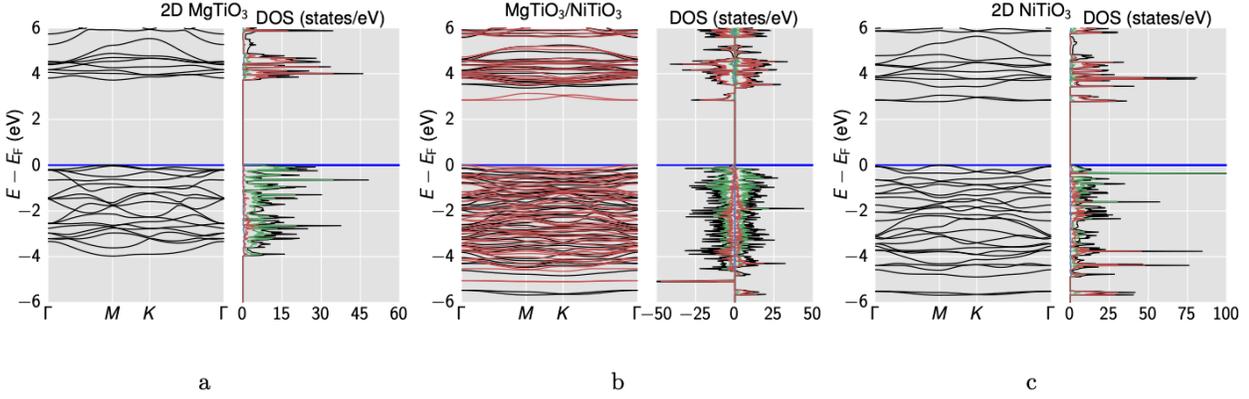

FIG. S43: Band structure and DOS for the individual 2D (a) MgTiO$_3$, (c) NiTiO$_3$ systems, and (b) the HS. The energies are aligned at the respective Fermi energy $E_\mathrm{F}$.

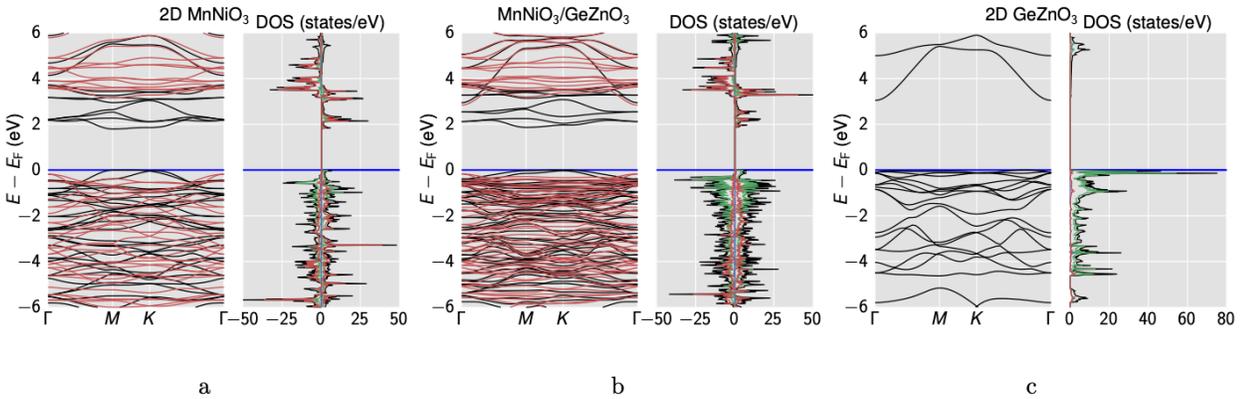

FIG. S44: Band structure and DOS for the individual 2D (a) MnNiO$_3$, (c) GeZnO$_3$ systems, and (b) the HS. The energies are aligned at the respective Fermi energy $E_\mathrm{F}$.

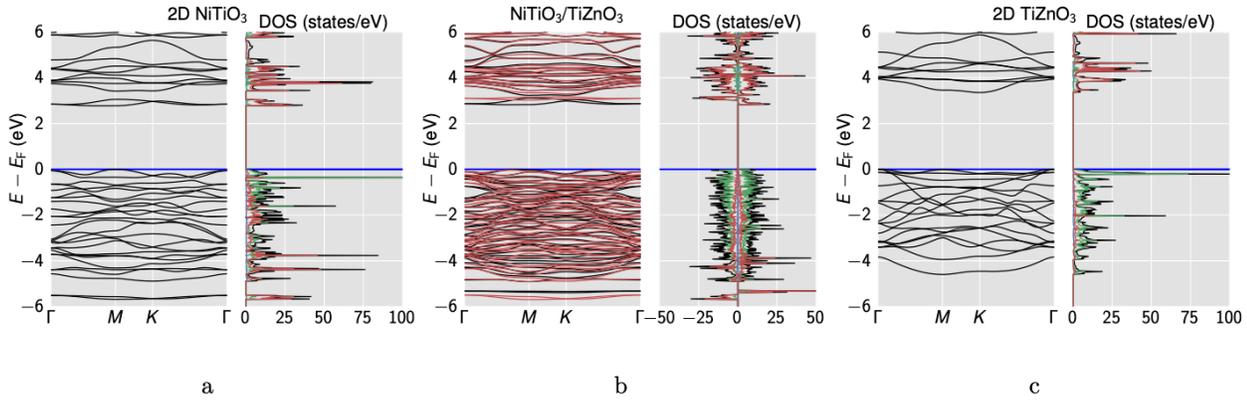

FIG. S45: Band structure and DOS for the individual 2D (a) NiTiO$_3$, (c) TiZnO$_3$ systems, and (b) the HS. The energies are aligned at the respective Fermi energy $E_F$.

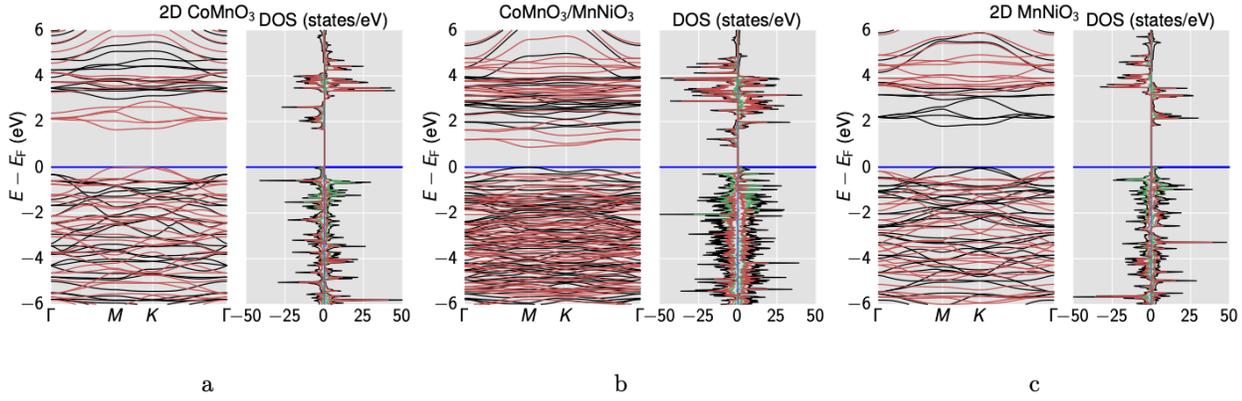

FIG. S46: Band structure and DOS for the individual 2D (a) CoMnO$_3$, (c) MnNiO$_3$ systems, and (b) the HS. The energies are aligned at the respective Fermi energy $E_F$.

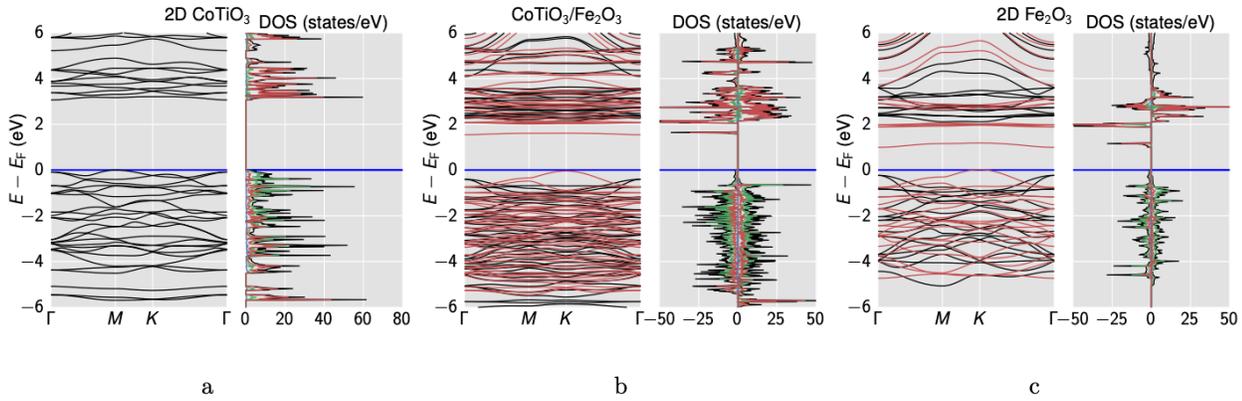

FIG. S47: Band structure and DOS for the individual 2D (a) CoTiO$_3$, (c) Fe$_2$O$_3$ systems, and (b) the HS. The energies are aligned at the respective Fermi energy $E_F$.

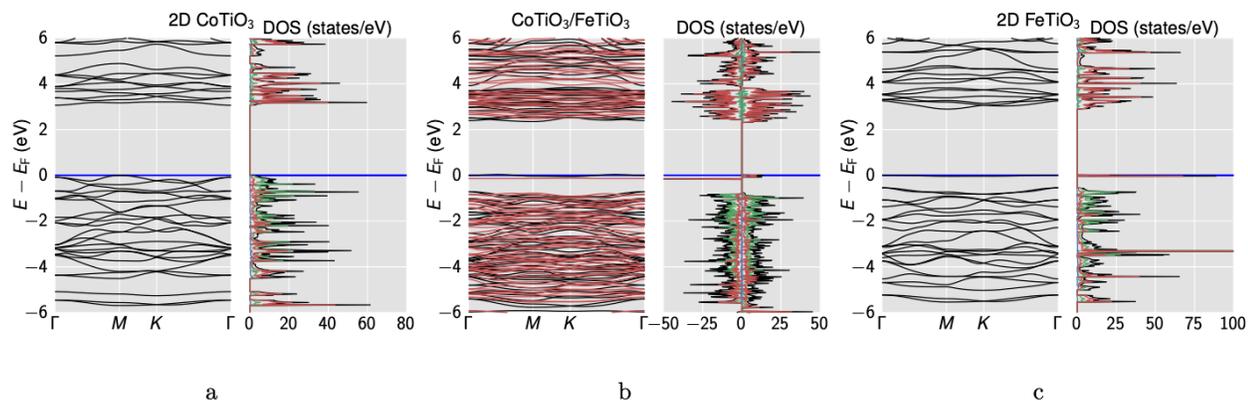

FIG. S48: Band structure and DOS for the individual 2D (a) CoTiO$_3$, (c) FeTiO$_3$ systems, and (b) the HS. The energies are aligned at the respective Fermi energy $E_F$.

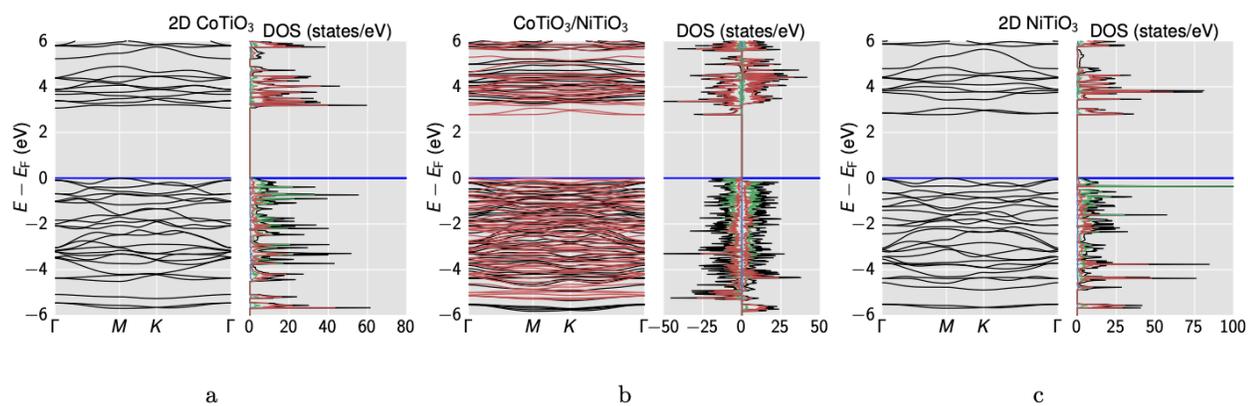

FIG. S49: Band structure and DOS for the individual 2D (a) CoTiO$_3$, (c) NiTiO$_3$ systems, and (b) the HS. The energies are aligned at the respective Fermi energy $E_F$.

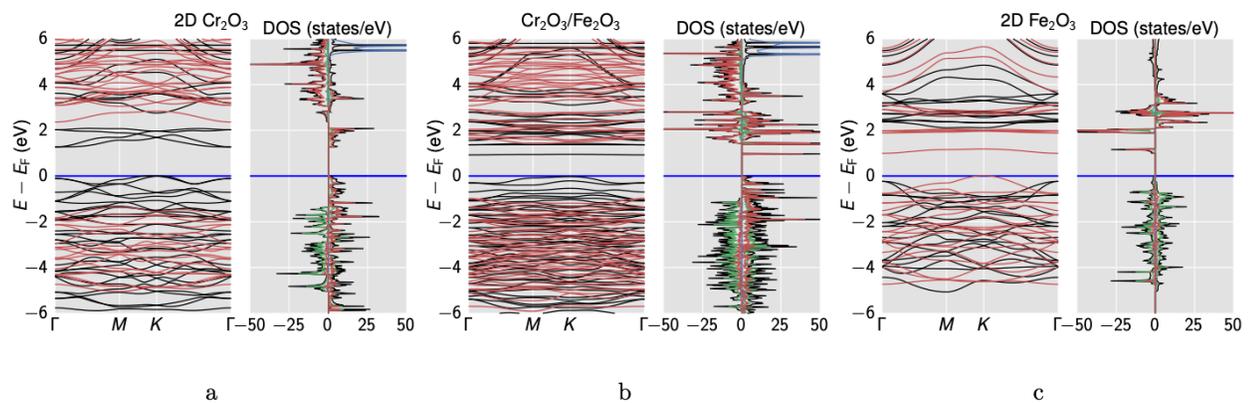

FIG. S50: Band structure and DOS for the individual 2D (a) Cr$_2$O$_3$, (c) Fe$_2$O$_3$ systems, and (b) the HS. The energies are aligned at the respective Fermi energy $E_F$.

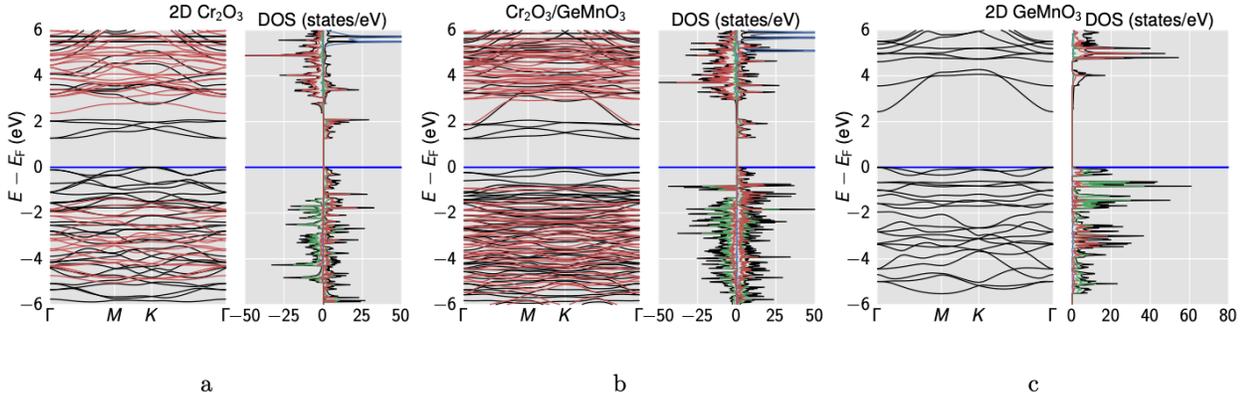

FIG. S51: Band structure and DOS for the individual 2D (a) $Cr_2O_3$, (c) $GeMnO_3$ systems, and (b) the HS. The energies are aligned at the respective Fermi energy $E_F$.

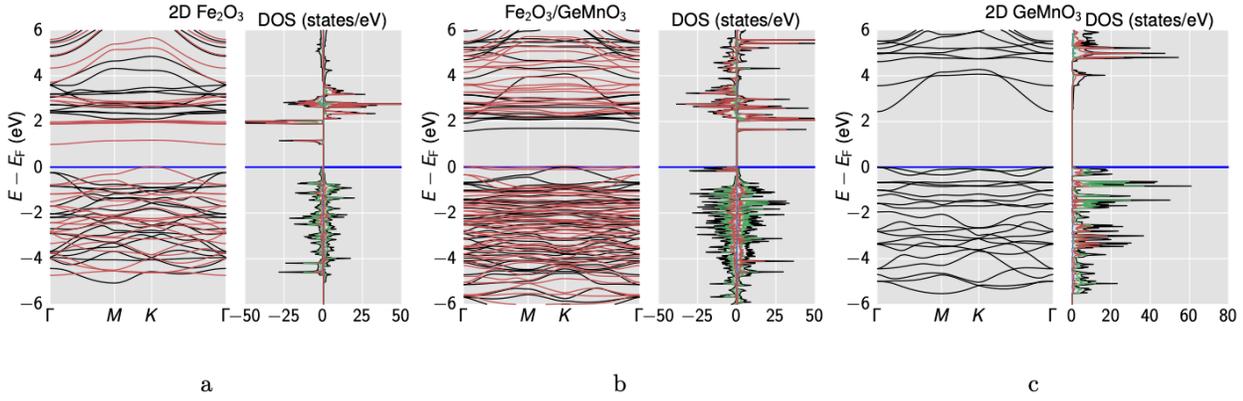

FIG. S52: Band structure and DOS for the individual 2D (a) $Fe_2O_3$, (c) $GeMnO_3$ systems, and (b) the HS. The energies are aligned at the respective Fermi energy $E_F$.

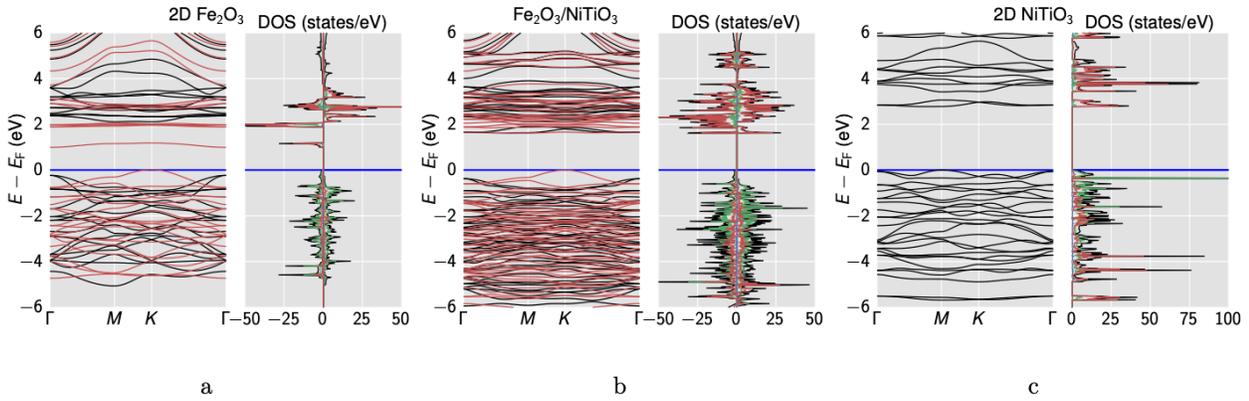

FIG. S53: Band structure and DOS for the individual 2D (a) $Fe_2O_3$, (c) $NiTiO_3$ systems, and (b) the HS. The energies are aligned at the respective Fermi energy $E_F$.

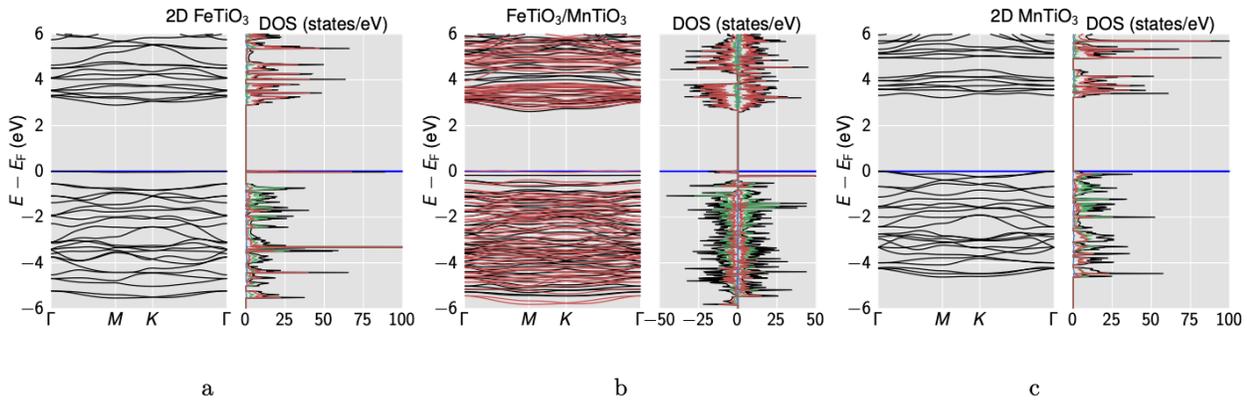

FIG. S54: Band structure and DOS for the individual 2D (a) FeTiO$_3$, (c) MnTiO$_3$ systems, and (b) the HS. The energies are aligned at the respective Fermi energy $E_{\text{F}}$.

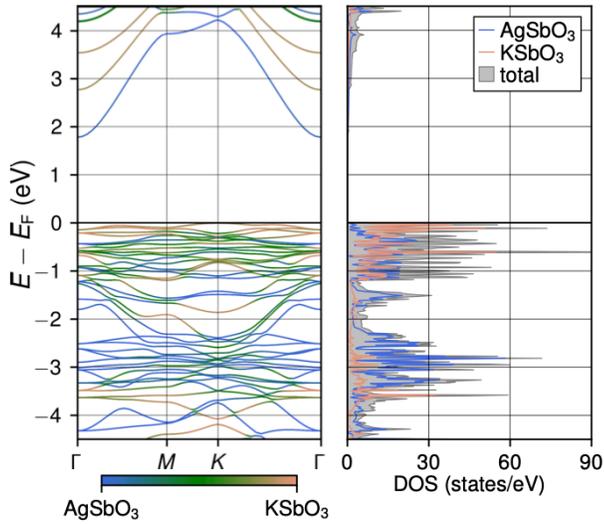

FIG. S55: Band structure and DOS with projection contributions from the atoms of each 2D sheet of AgSbO$_3$/KSbO$_3$. Note that the band structures are computed for geometries optimized with D3 dispersion correction.

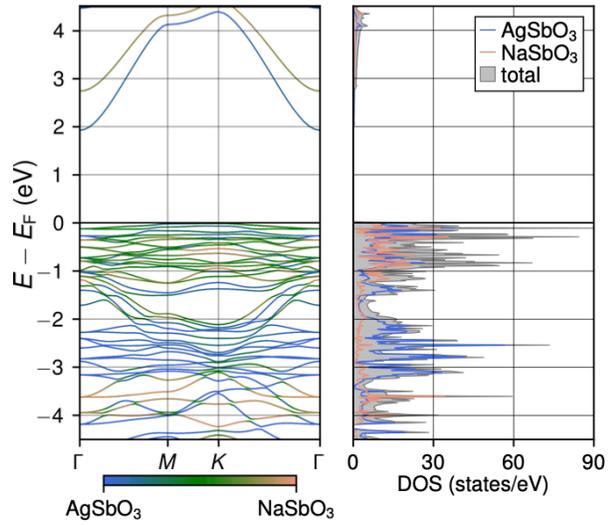

FIG. S56: Band structure and DOS with projection contributions from the atoms of each 2D sheet of AgSbO$_3$/NaSbO$_3$.

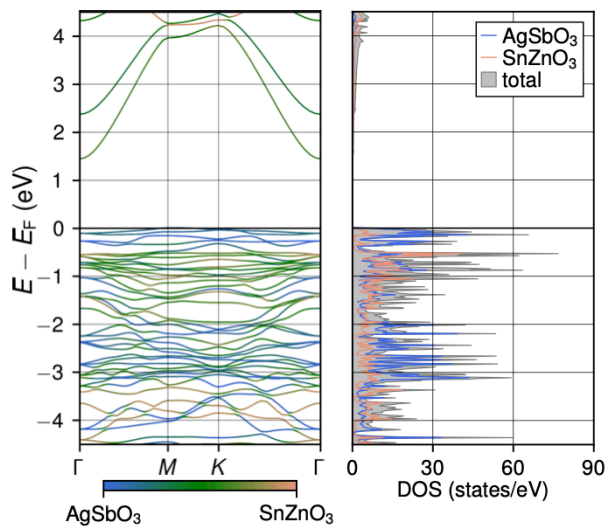

FIG. S57: Band structure and DOS with projection contributions from the atoms of each 2D sheet of $AgSbO_3$/ $SnZnO_3$. Note that the band structures are computed for geometries optimized with D3 dispersion correction.

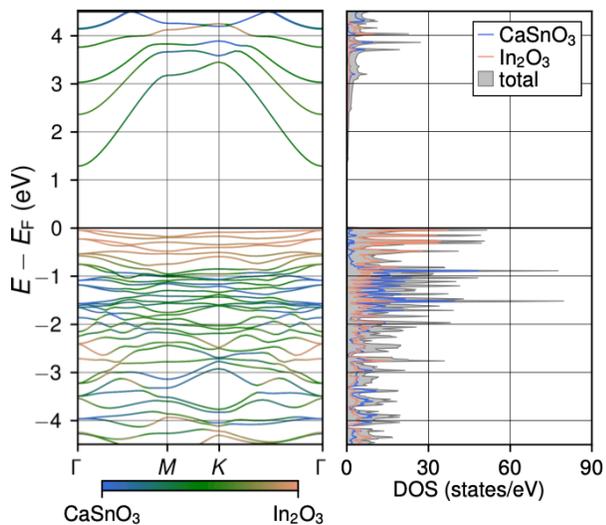

FIG. S59: Band structure and DOS with projection contributions from the atoms of each 2D sheet of $CaSnO_3$/ $In_2O_3$.

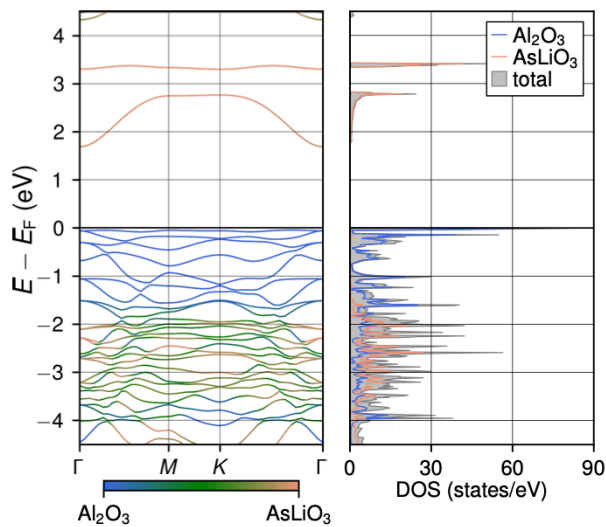

FIG. S58: Band structure and DOS with projection contributions from the atoms of each 2D sheet of $Al_2O_3$/ $AsLiO_3$.

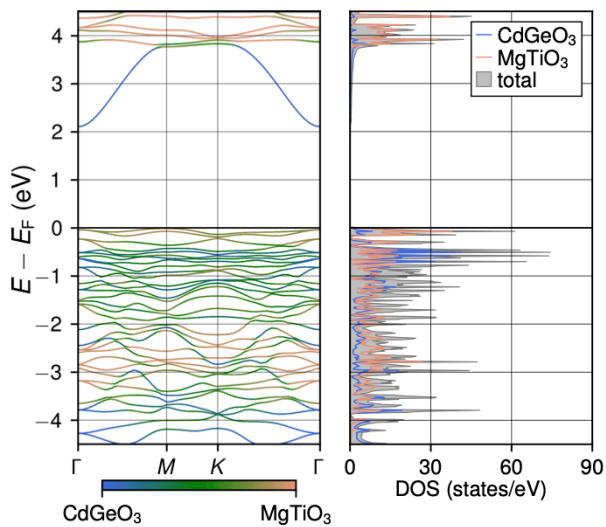

FIG. S60: Band structure and DOS with projection contributions from the atoms of each 2D sheet of $CdGeO_3$/ $MgTiO_3$.

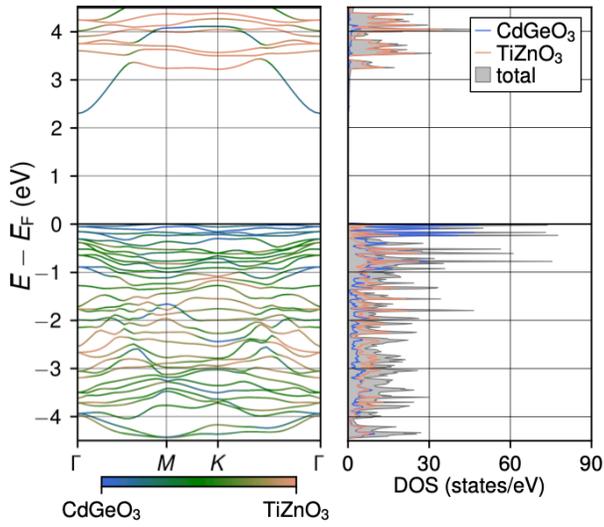

FIG. S61: Band structure and DOS with projection contributions from the atoms of each 2D sheet of CdGeO$_3$/TiZnO$_3$.

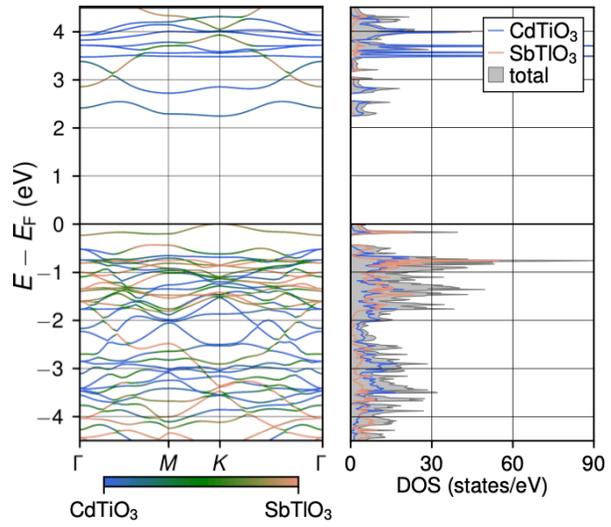

FIG. S63: Band structure and DOS with projection contributions from the atoms of each 2D sheet of CdTiO$_3$/SbTlO$_3$. Note that the band structures are computed for geometries optimized with D3 dispersion correction.

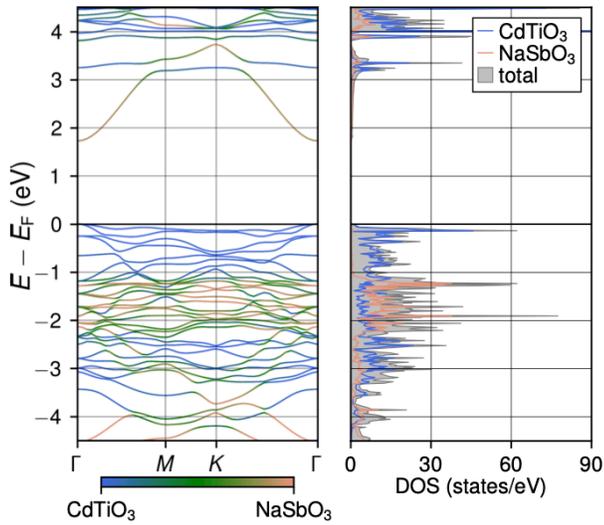

FIG. S62: Band structure and DOS with projection contributions from the atoms of each 2D sheet of CdTiO$_3$/NaSbO$_3$.

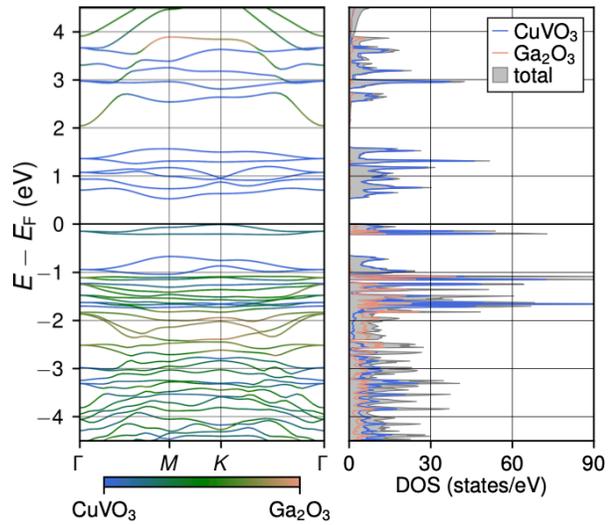

FIG. S64: Band structure and DOS with projection contributions from the atoms of each 2D sheet of CuVO$_3$/Ga$_2$O$_3$.

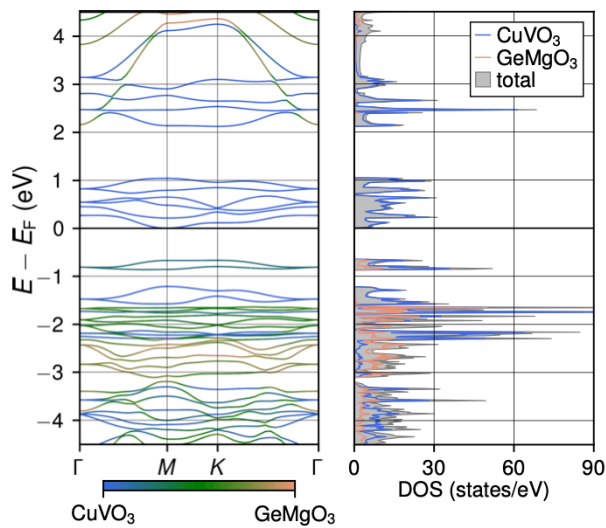

FIG. S65: Band structure and DOS with projection contributions from the atoms of each 2D sheet of CuVO$_3$/GeMgO$_3$.

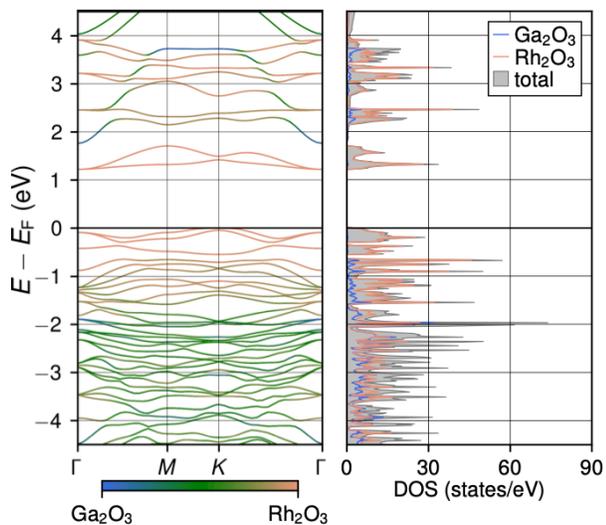

FIG. S67: Band structure and DOS with projection contributions from the atoms of each 2D sheet of Ga$_2$O$_3$/Rh$_2$O$_3$.

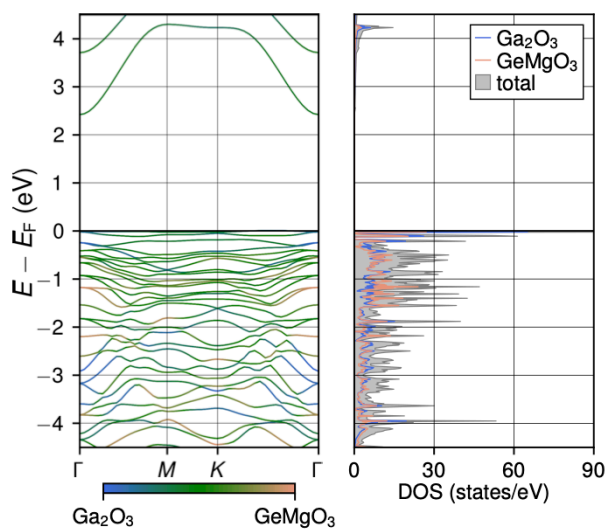

FIG. S66: Band structure and DOS with projection contributions from the atoms of each 2D sheet of Ga$_2$O$_3$/GeMgO$_3$.

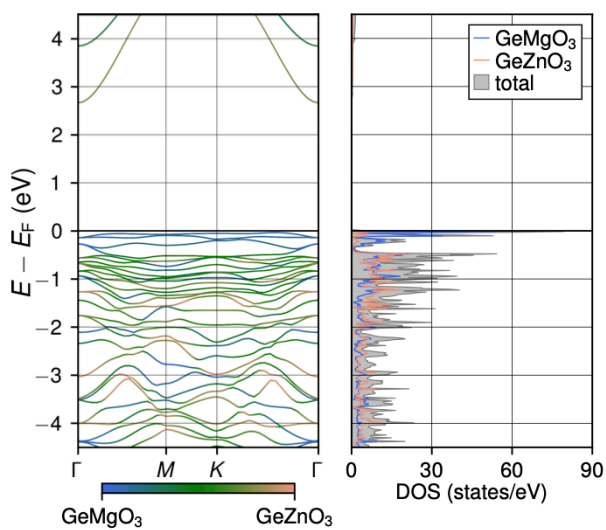

FIG. S68: Band structure and DOS with projection contributions from the atoms of each 2D sheet of GeMgO$_3$/GeZnO$_3$.

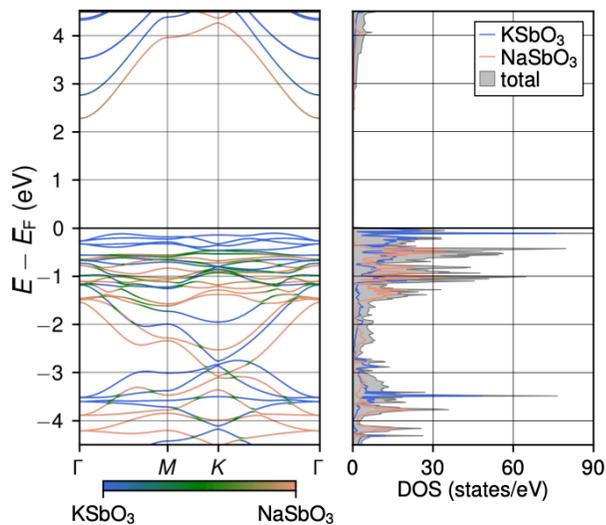

FIG. S69: Band structure and DOS with projection contributions from the atoms of each 2D sheet of $KSbO_3$/ $NaSbO_3$.

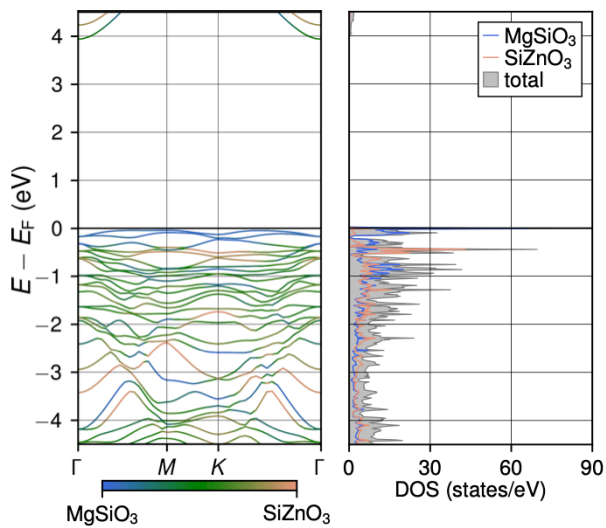

FIG. S71: Band structure and DOS with projection contributions from the atoms of each 2D sheet of $MgSiO_3$/ $SiZnO_3$.

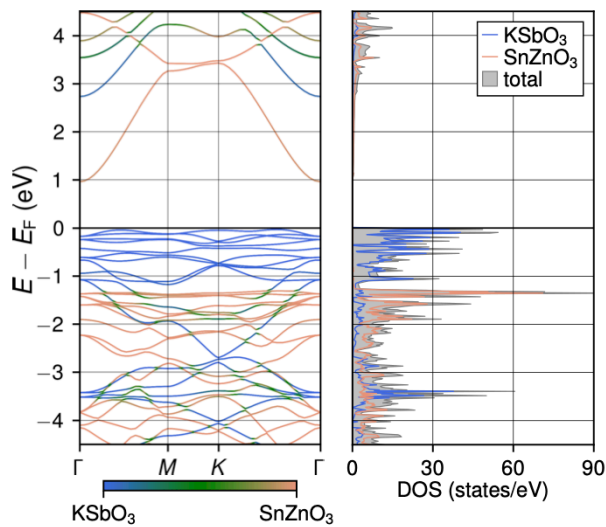

FIG. S70: Band structure and DOS with projection contributions from the atoms of each 2D sheet of $KSbO_3$/ $SnZnO_3$. Note that the band structures are computed for geometries optimized with D3 dispersion correction.

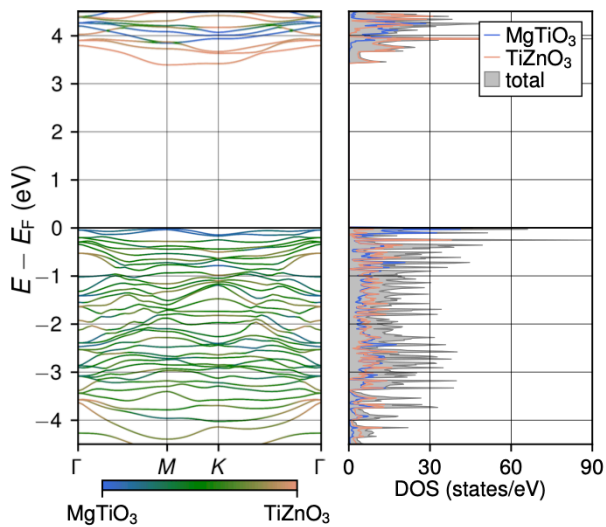

FIG. S72: Band structure and DOS with projection contributions from the atoms of each 2D sheet of $MgTiO_3$/ $TiZnO_3$.

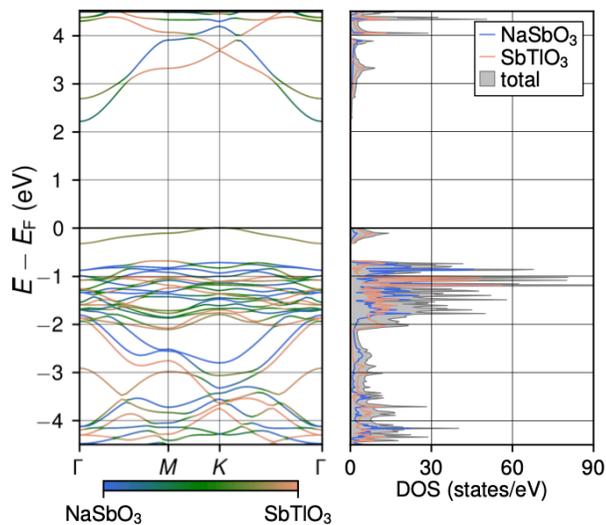

FIG. S73: Band structure and DOS with projection contributions from the atoms of each 2D sheet of $NaSbO_3$/$SbTlO_3$. Note that the band structures are computed for geometries optimized with D3 dispersion correction.

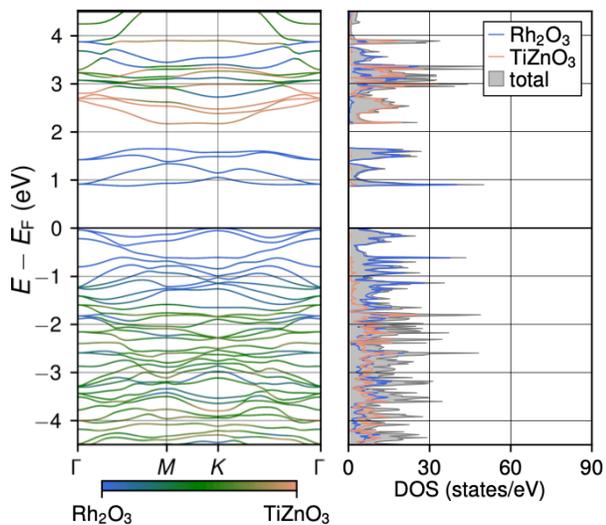

FIG. S75: Band structure and DOS with projection contributions from the atoms of each 2D sheet of $Rh_2O_3$/$TiZnO_3$. Note that the band structures are computed for geometries optimized with D3 dispersion correction.

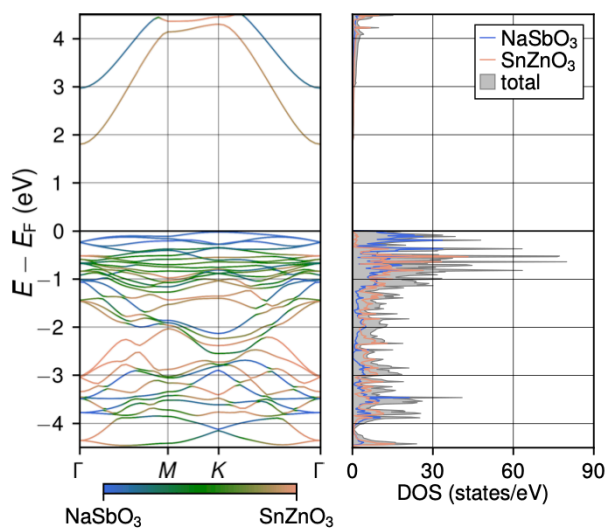

FIG. S74: Band structure and DOS with projection contributions from the atoms of each 2D sheet of $NaSbO_3$/$SnZnO_3$. Note that the band structures are computed for geometries optimized with D3 dispersion correction.

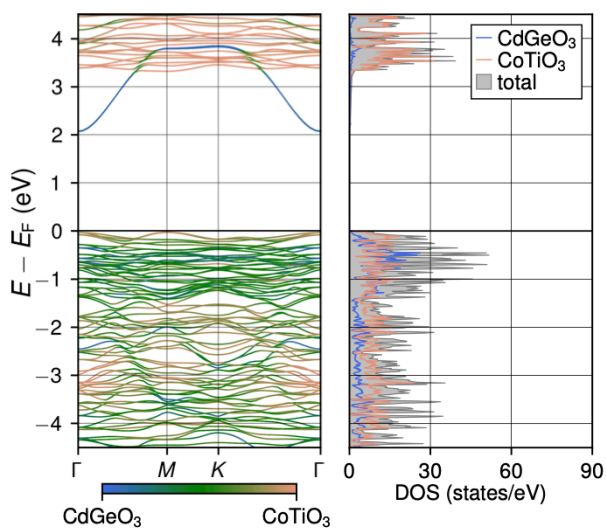

FIG. S76: Band structure and DOS with projection contributions from the atoms of each 2D sheet of $CdGeO_3$/$CoTiO_3$.

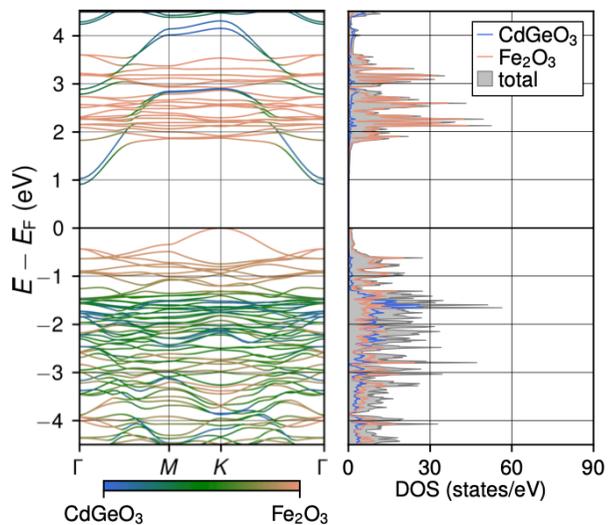

FIG. S77: Band structure and DOS with projection contributions from the atoms of each 2D sheet of CdGeO$_3$/Fe$_2$O$_3$.

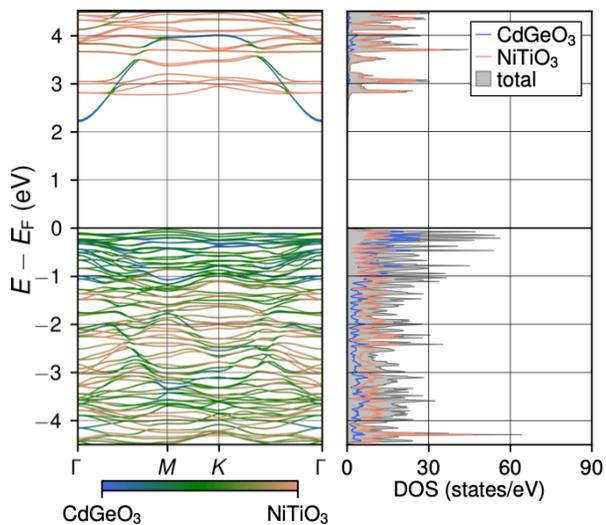

FIG. S79: Band structure and DOS with projection contributions from the atoms of each 2D sheet of CdGeO$_3$/NiTiO$_3$.

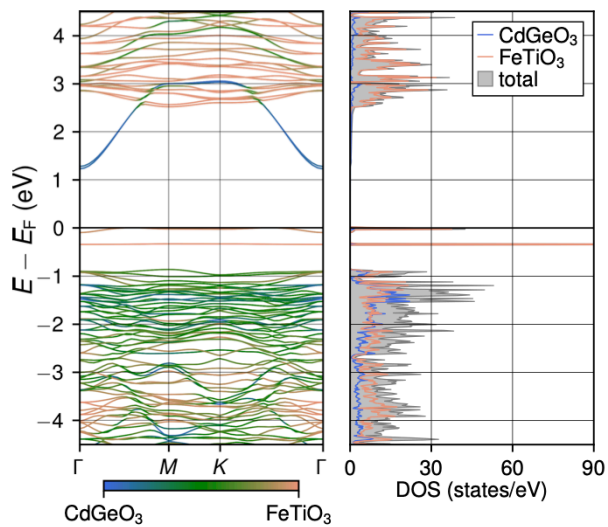

FIG. S78: Band structure and DOS with projection contributions from the atoms of each 2D sheet of CdGeO$_3$/FeTiO$_3$.

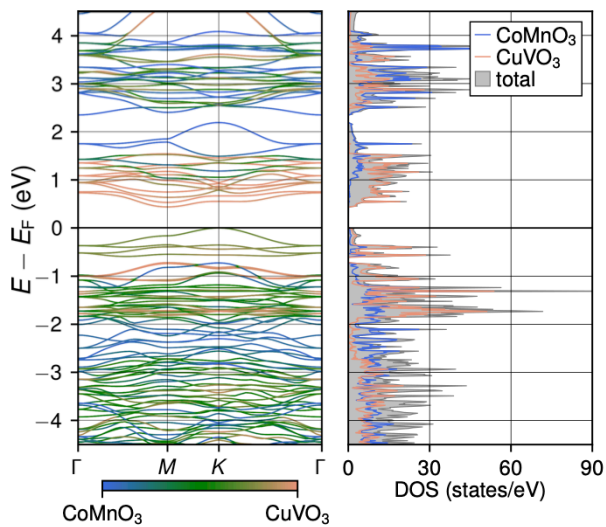

FIG. S80: Band structure and DOS with projection contributions from the atoms of each 2D sheet of CoMnO$_3$/CuVO$_3$.

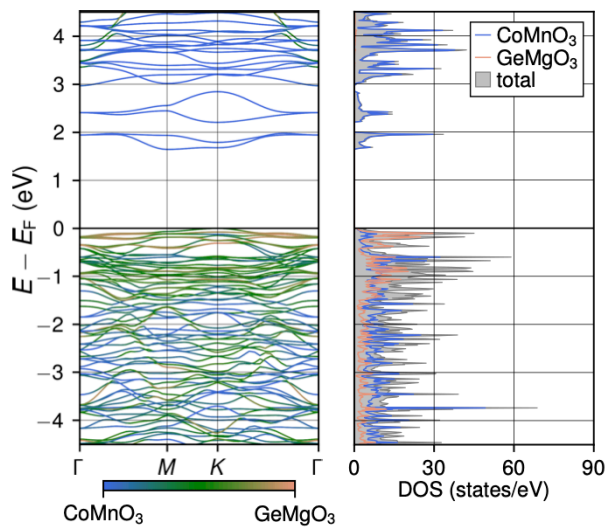

FIG. S81: Band structure and DOS with projection contributions from the atoms of each 2D sheet of CoMnO$_3$/GeMgO$_3$.

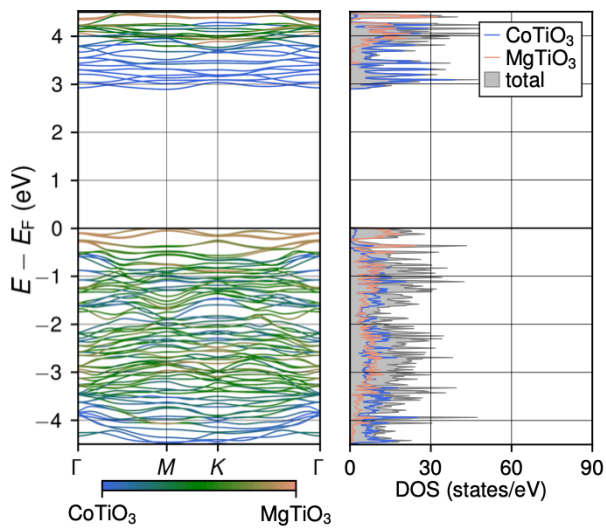

FIG. S83: Band structure and DOS with projection contributions from the atoms of each 2D sheet of CoTiO$_3$/MgTiO$_3$.

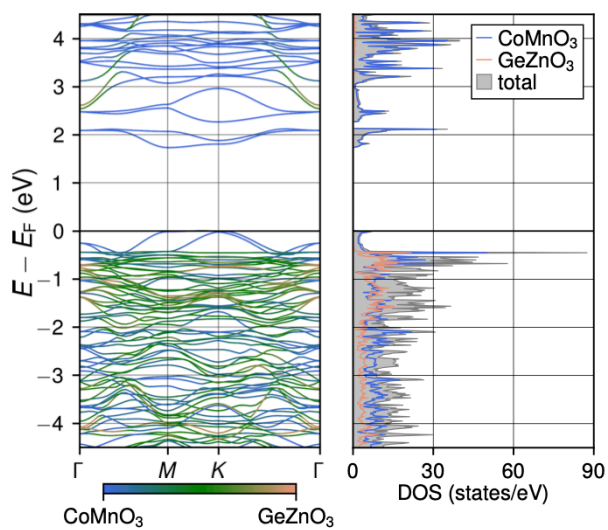

FIG. S82: Band structure and DOS with projection contributions from the atoms of each 2D sheet of CoMnO$_3$/GeZnO$_3$.

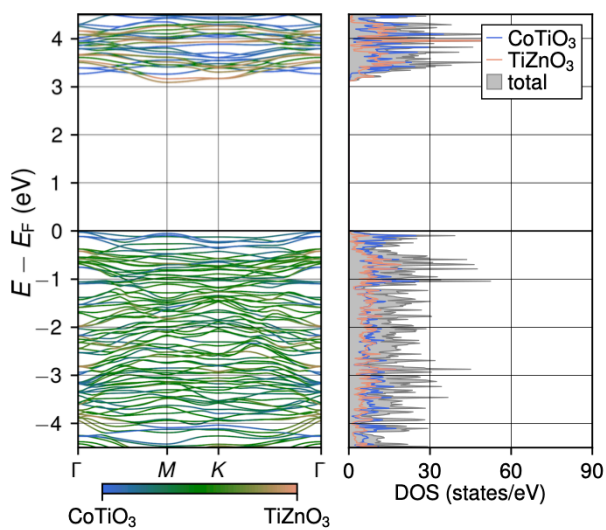

FIG. S84: Band structure and DOS with projection contributions from the atoms of each 2D sheet of CoTiO$_3$/TiZnO$_3$.

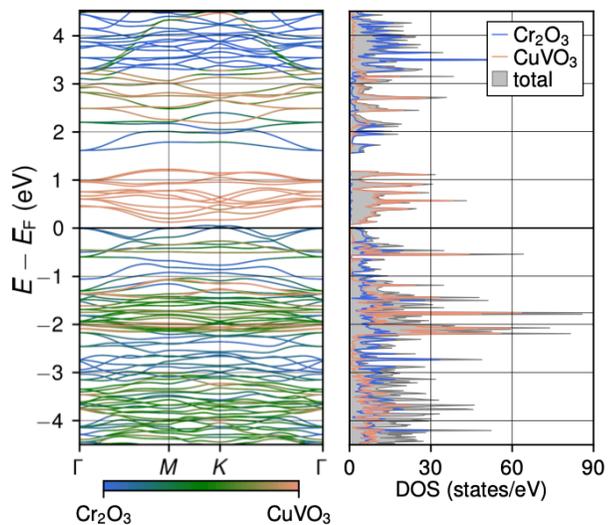

FIG. S85: Band structure and DOS with projection contributions from the atoms of each 2D sheet of Cr₂O₃/CuVO₃.

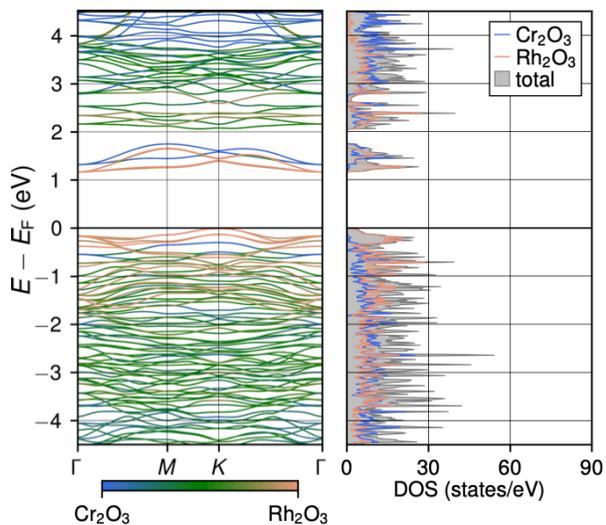

FIG. S87: Band structure and DOS with projection contributions from the atoms of each 2D sheet of Cr₂O₃/Rh₂O₃.

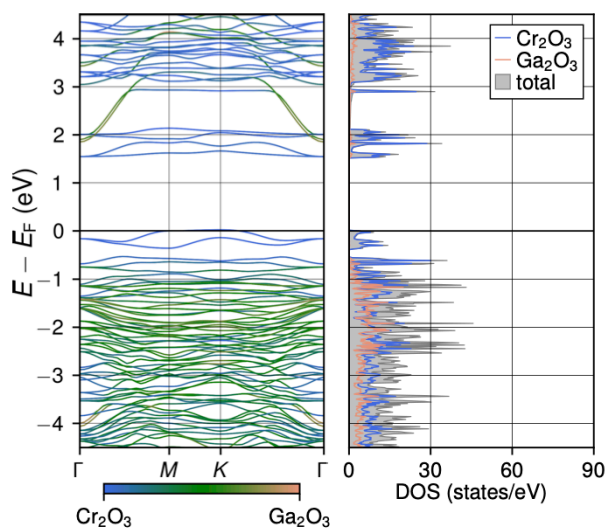

FIG. S86: Band structure and DOS with projection contributions from the atoms of each 2D sheet of Cr₂O₃/Ga₂O₃.

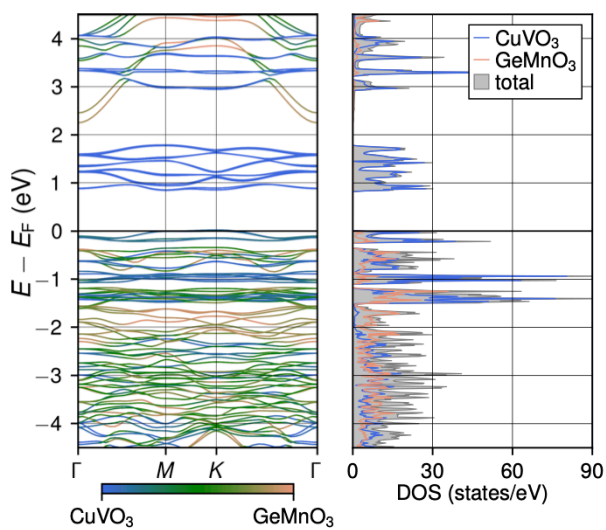

FIG. S88: Band structure and DOS with projection contributions from the atoms of each 2D sheet of CuVO₃/GeMnO₃.

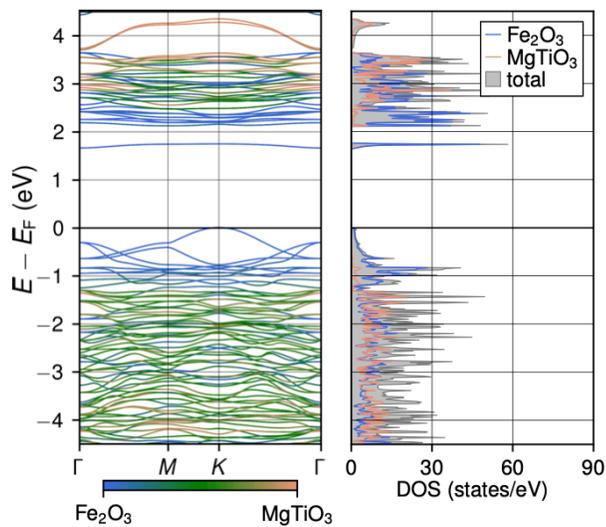

FIG. S89: Band structure and DOS with projection contributions from the atoms of each 2D sheet of Fe$_2$O$_3$/MgTiO$_3$.

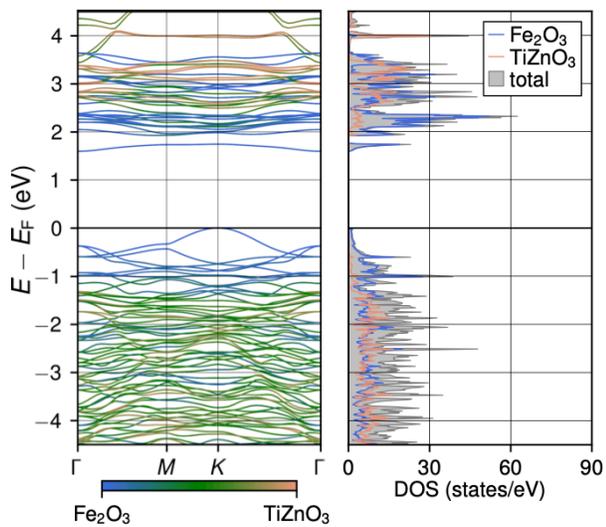

FIG. S91: Band structure and DOS with projection contributions from the atoms of each 2D sheet of Fe$_2$O$_3$/TiZnO$_3$.

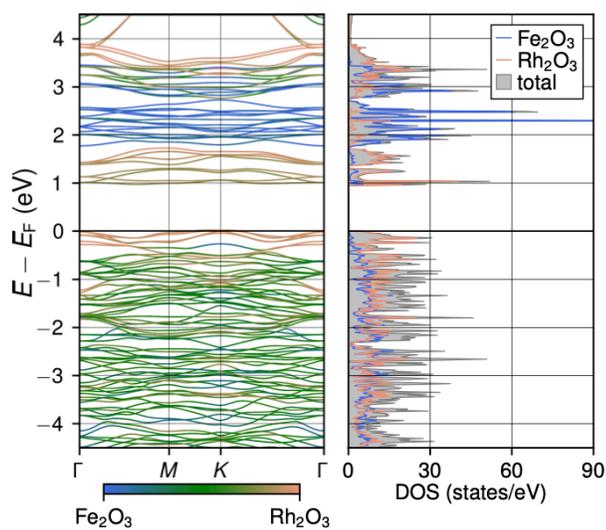

FIG. S90: Band structure and DOS with projection contributions from the atoms of each 2D sheet of Fe$_2$O$_3$/Rh$_2$O$_3$.

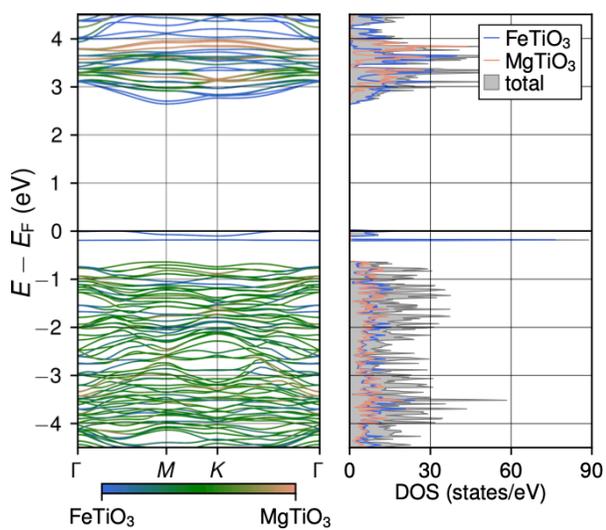

FIG. S92: Band structure and DOS with projection contributions from the atoms of each 2D sheet of FeTiO$_3$/MgTiO$_3$.

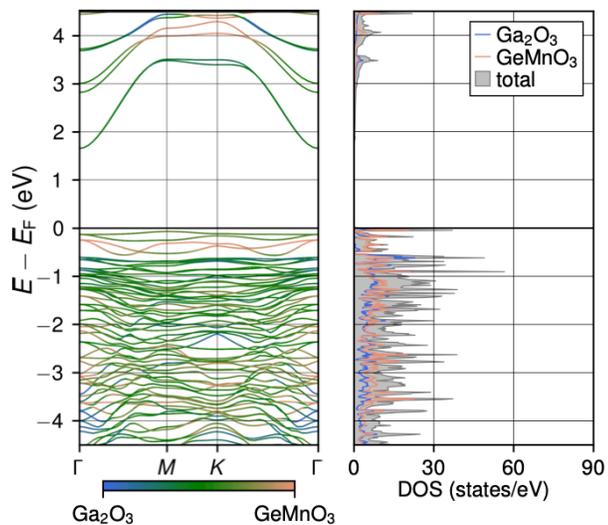

FIG. S93: Band structure and DOS with projection contributions from the atoms of each 2D sheet of $Ga_2O_3/$ $GeMnO_3$.

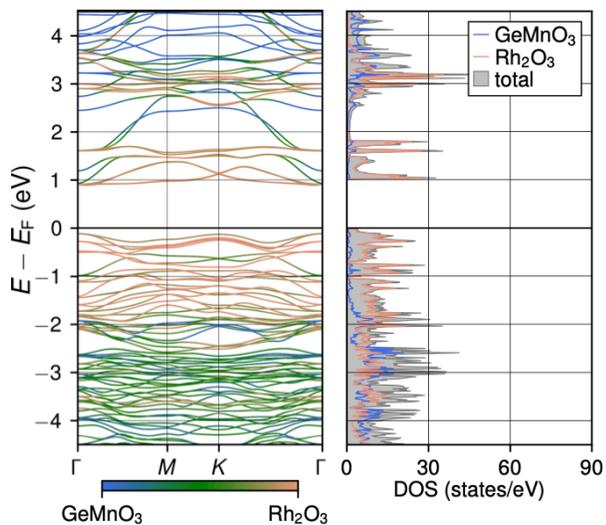

FIG. S95: Band structure and DOS with projection contributions from the atoms of each 2D sheet of $GeMnO_3/$ $Rh_2O_3$.

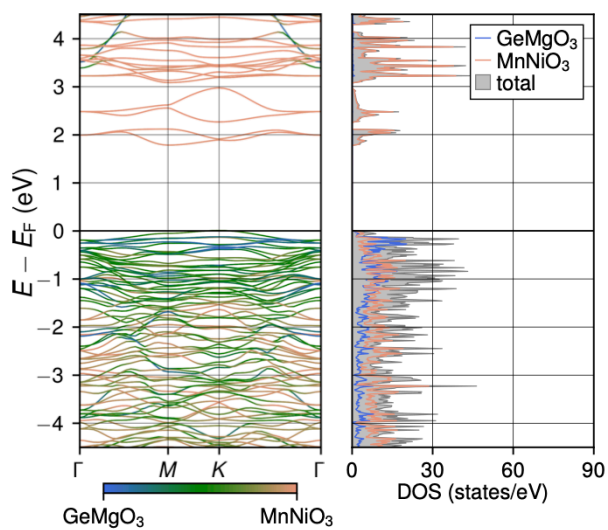

FIG. S94: Band structure and DOS with projection contributions from the atoms of each 2D sheet of $GeMgO_3/$ $MnNiO_3$.

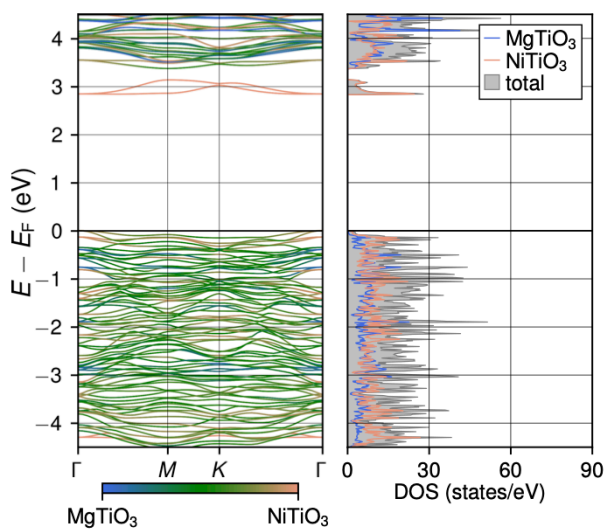

FIG. S96: Band structure and DOS with projection contributions from the atoms of each 2D sheet of $MgTiO_3/$ $NiTiO_3$.

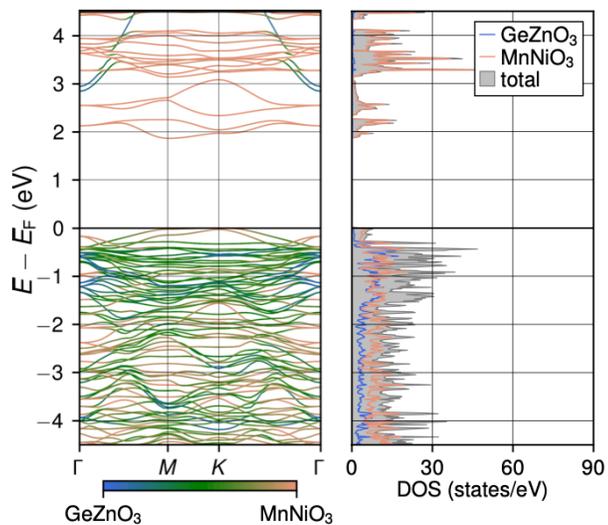

FIG. S97: Band structure and DOS with projection contributions from the atoms of each 2D sheet of MnNiO$_3$/GeZnO$_3$.

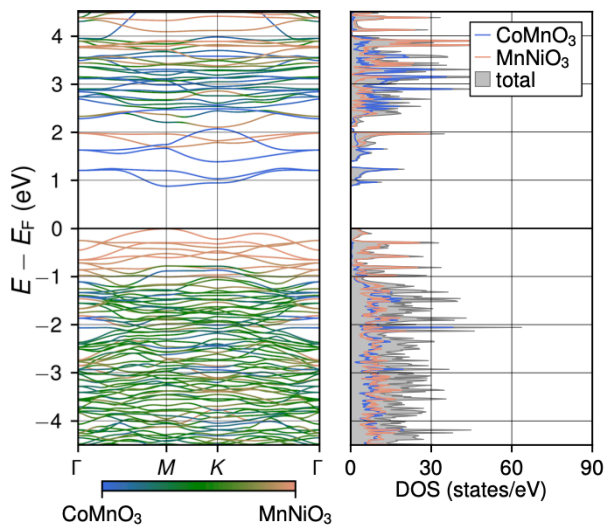

FIG. S99: Band structure and DOS with projection contributions from the atoms of each 2D sheet of CoMnO$_3$/MnNiO$_3$.

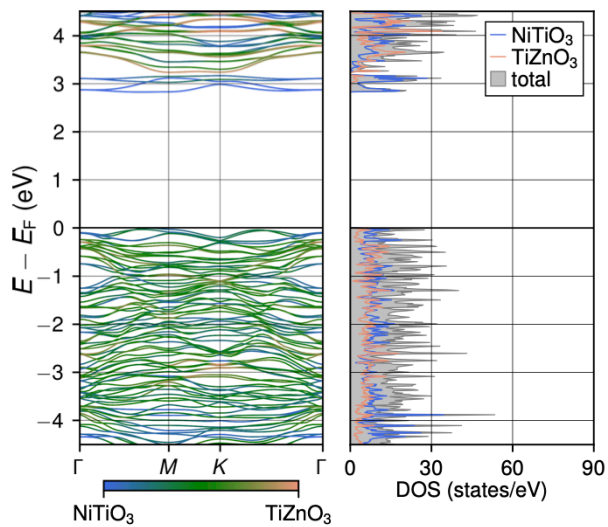

FIG. S98: Band structure and DOS with projection contributions from the atoms of each 2D sheet of NiTiO$_3$/TiZnO$_3$.

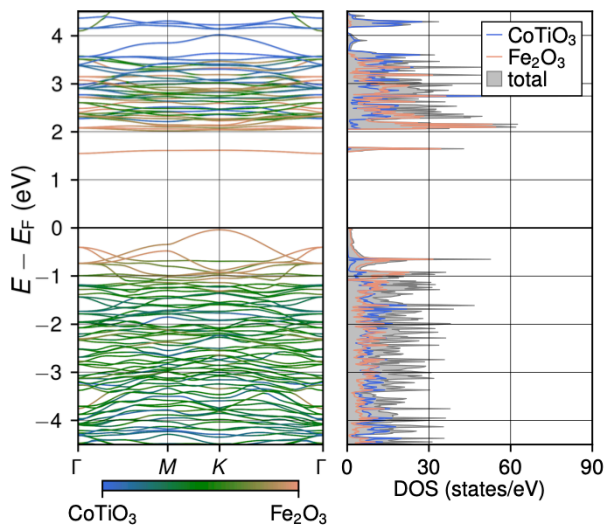

FIG. S100: Band structure and DOS with projection contributions from the atoms of each 2D sheet of CoTiO$_3$/Fe$_2$O$_3$.

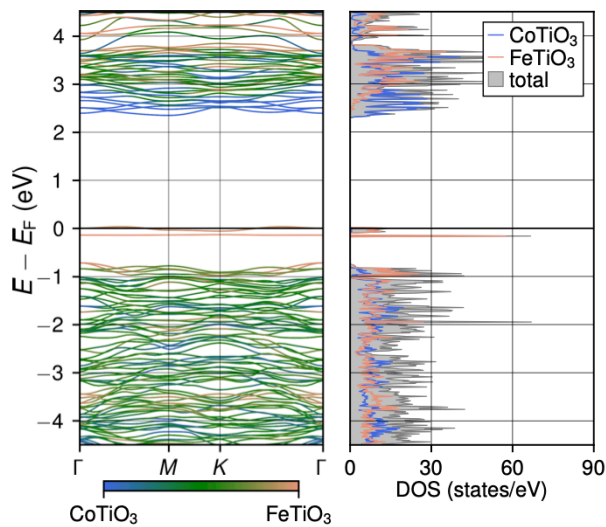

FIG. S101: Band structure and DOS with projection contributions from the atoms of each 2D sheet of CoTiO$_3$/FeTiO$_3$.

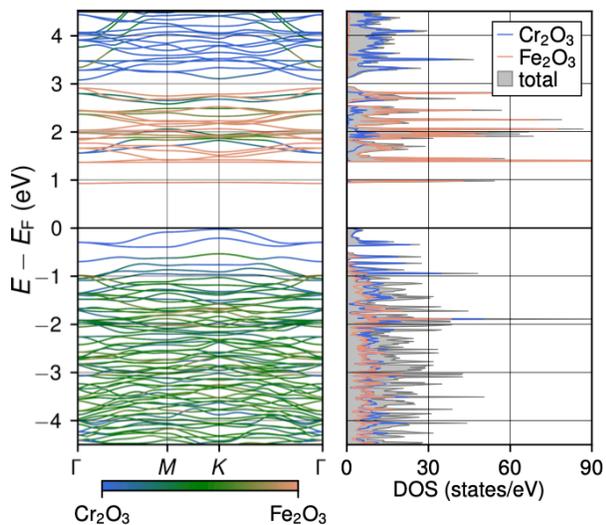

FIG. S103: Band structure and DOS with projection contributions from the atoms of each 2D sheet of Cr$_2$O$_3$/Fe$_2$O$_3$.

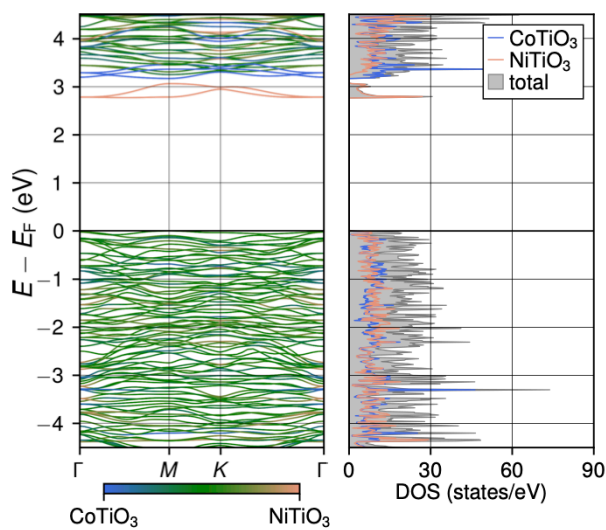

FIG. S102: Band structure and DOS with projection contributions from the atoms of each 2D sheet of CoTiO$_3$/NiTiO$_3$.

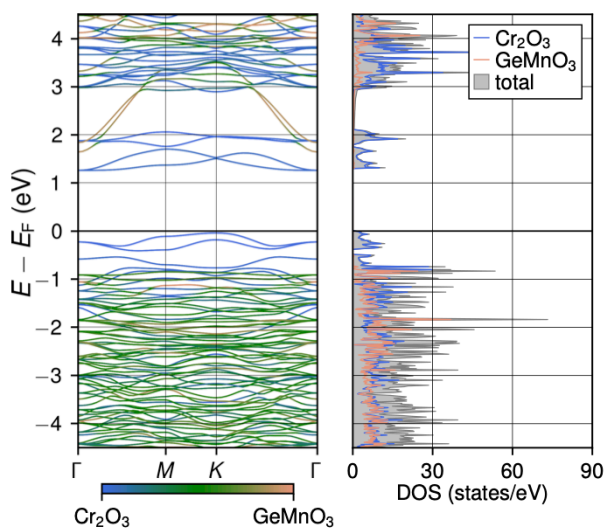

FIG. S104: Band structure and DOS with projection contributions from the atoms of each 2D sheet of Cr$_2$O$_3$/GeMnO$_3$.

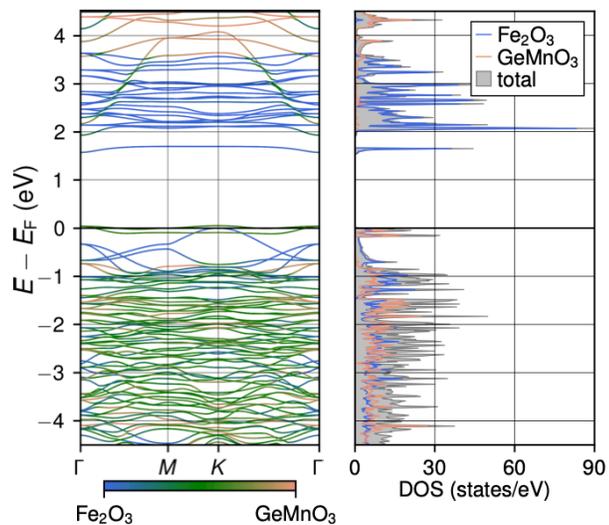

FIG. S105: Band structure and DOS with projection contributions from the atoms of each 2D sheet of Fe$_2$O$_3$/GeMnO$_3$.

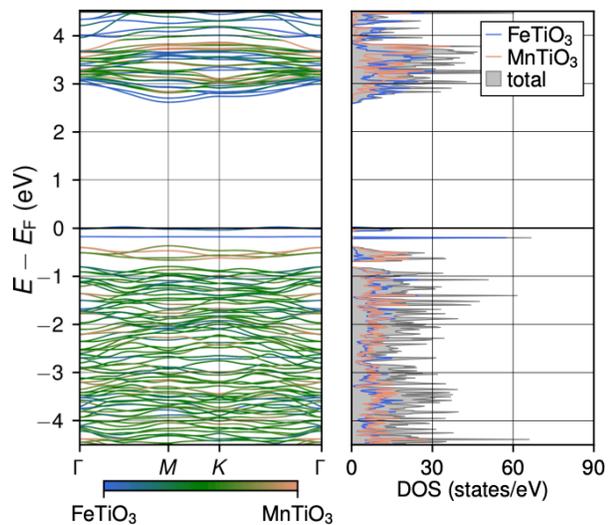

FIG. S107: Band structure and DOS with projection contributions from the atoms of each 2D sheet of FeTiO$_3$/MnTiO$_3$.

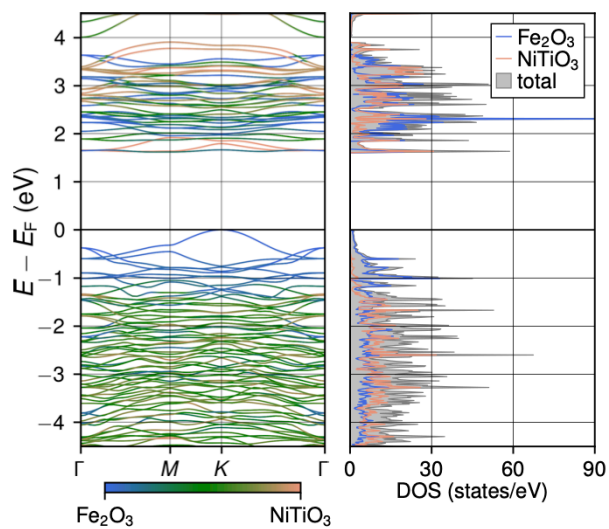

FIG. S106: Band structure and DOS with projection contributions from the atoms of each 2D sheet of Fe$_2$O$_3$/NiTiO$_3$.

## IV. Charge density on the plane above the top surface for twisted HSs

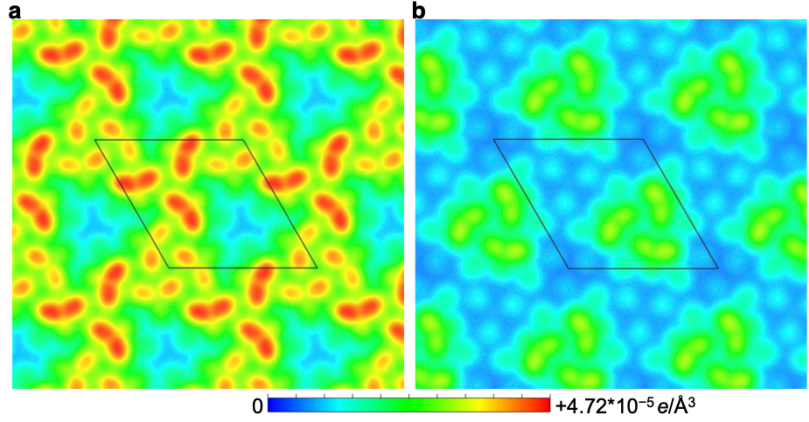

FIG. S108: (**a/b**) Top view of the charge density variation on a plane 3.18/3.23 Å above the top surface (above MgTiO$_3$) of Fe$_2$O$_3$/MgTiO$_3$ for 22/38°. The unit cell is indicated by the black frame.

## V. Magnetisation density on the plane below the bottom surface for twisted HSs

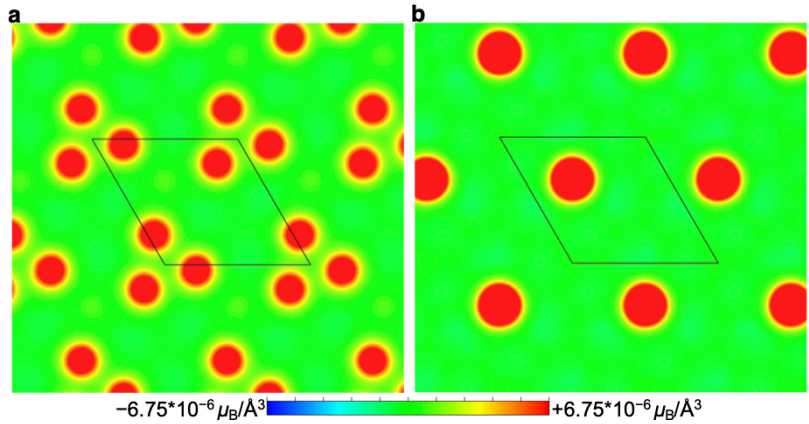

FIG. S109: (**a/b**) Top view of the magnetisation density variation on a plane 3.69/3.15 Å below the bottom surface (below Fe$_2$O$_3$) of Fe$_2$O$_3$/MgTiO$_3$ for 22/38°. The unit cell is indicated by the black frame.


\end{document}